\documentclass[fleqn,usenatbib]{mnras}

\usepackage{newtxtext,newtxmath}

\usepackage[T1]{fontenc}

\DeclareRobustCommand{\VAN}[3]{#2}
\let\VANthebibliography\thebibliography
\def\thebibliography{\DeclareRobustCommand{\VAN}[3]{##3}\VANthebibliography}

\usepackage{graphicx}	
\usepackage{amsmath}	
\usepackage{lscape}
\usepackage{xspace}
\usepackage{url}

\newcommand{\gaia}{{\it Gaia}\xspace}

\newcommand{\kepler}{{\it Kepler}\xspace}
\newcommand{\tess}{{\it TESS}\xspace}
\newcommand{\SNR}{SNR\xspace}

\newcommand{\edit}[1]{#1}
\newcommand{\editt}[1]{#1}

\title[Circumbinary planets with Gaia]{Predicting Gaia astrometry's ability to constrain the populations of circumbinary planets}

\author[T. A. Baycroft et al.]{
Thomas A. Baycroft,$^{1,2}$\thanks{E-mail: tbaycroft@sjtu.edu.cn}
Amaury H.M.J. Triaud,$^{1}$
Johannes Sahlmann$^{3}$
\\
$^{1}$School of Physics and Astronomy, University of Birmingham, Edgbaston, Birmingham B15 2TT, UK\\
$^{2}$Tsung-Dao Lee Institute, Shanghai Jiao Tong University, 1 Lisuo Road, Shanghai 201210, China\\
$^{3}$European Space Agency (ESA), European Space Astronomy Centre (ESAC), Camino Bajo del Castillo s/n, 28692, Villanueva de la Ca\~nada, Madrid, Spain
}

\date{Accepted 2026 February 26. Received 2026 February 20; in original form 2025 November 11}

\pubyear{\the\year{}}

\begin{document}
\label{firstpage}
\pagerange{\pageref{firstpage}--\pageref{lastpage}}
\maketitle

\begin{abstract} 
The coming data releases of \gaia are expected to result in an upheaval of exoplanet science, in particular for long period giant planets (\(0.2\,{\rm M_{J}}\,\leq M\leq25\,{\rm M_{J}}\)). One class of exoplanets which \gaia will help investigate is circumbinary planets.  Using the current knowledge of the circumbinary exoplanet population as well as expectations for the \gaia sensitivity, we investigate the impact \gaia will have on our understanding of circumbinary planets. We compare our results to a pre-launch estimate, the main differences arising from a better understanding of the circumbinary planet population, which result in a lower expected yield than previously predicted, though still significant compared to the known population. We make a rough yield estimate, with conservative detection criteria and parameter-space cuts, predicting in the 10s - 100s of detections in \gaia DR4. More importantly, we show how the yield estimate varies strongly with different assumptions on the injected circumbinary population, showing \gaia\editt{'s} sensitivity to the mass and orbital period distribution of circumbinary planets. We find that \gaia circumbinary exoplanet detections will be biased towards planets closer to the instability zone surrounding the binary, due to the larger number of binaries on wider orbits and the limited timespan of \gaia. We also assess the impact \gaia will have on known circumbinary systems, one being that it may resolve the question of reliability of the claimed planets orbiting post-common-envelope binaries, with \gaia DR5 being sensitive to between 3 and 11 out of 32 such planet candidates.
\end{abstract}




\section{Introduction}

Circumbinary planet \edit{and brown dwarf\footnote{\edit{We will refer to both true planets and low-mass brown dwarfs \(M\leq25\,{\rm M_{J}}\) \citep{schneider_defining_2011} as "planets" for simplicity.}}} detections number in the few dozen so far\edit{, with 35 included in the exoplanet.eu \citep{thebault_complete_2025} catalogue and 53 in the NASA exoplanet archive \citep{christiansen_nasa_2025}}. These come from a variety of techniques, such as transits \citep[e.g.][]{doyle_kepler-16_2011}, radial velocities \citep[e.g.][]{standing_radial-velocity_2023}, microlensing \citep[e.g.][]{bennett_first_2016}, direct imaging \citep[e.g.][]{delorme_direct-imaging_2013}, and eclipse timing variations \citep[e.g.][]{beuermann_two_2010,goldberg_5mjup_2023}. Astrometry has been involved in the analysis for three circumbinary systems, it was used to study the inclination of HD202206 \citep[][see Section \ref{sec:known_ms}]{benedict_hd_2017}, and the proper-motion anomaly method \citep{kervella_stellar_2022} has implied a circumbinary companion orbiting HD155555 (HIP 84586) \citep{gratton_stellar_2024}, and provided slight evidence for a circumbinary companion orbiting HW Virginis \citep{baycroft_new_2023}. 

Of these circumbinary planets, the transiting sample from \textit{Kepler} (12 planets orbiting 10 binaries) is the only one that has been suitable for population statistics \citep[e.g.][]{armstrong_abundance_2014,martin_planets_2014,li_uncovering_2016}. The distribution of mutual inclinations between binary and circumbinary planets was found to mean that the majority of the circumbinary planet population should lie close to coplanar \citep{armstrong_abundance_2014,martin_planets_2014}. In addition, even accounting for the transit method's bias towards short period planets, the fact that the transiting sample from \textit{Kepler} were almost all found as close to the binary as is dynamically possible has been interpreted as a real feature. The innermost planets of the circumbinary systems form a "piled-up" population \citep{martin_planets_2014,li_uncovering_2016} just outside the instability region that surrounds the inner binary \citep{holman_long-term_1999,georgakarakos_empirical_2024}. 

In contrast to these results, ground-based radial velocity surveys - while demonstrating the ability to detect the planets discovered in transit such as Kepler-16b \citep{doyle_kepler-16_2011,triaud_bebop_2022} and TIC172900988b \citep{kostov_tic_2021,sairam_new_2024} - have not detected many circumbinary planets near the stability limit, instead finding planets and candidates at longer orbital periods \citep{baycroft_progress_2024,baycroft_bebop_2025}. There are either poorly understood biases in one or both of the transit and radial velocity surveys, or some other physical reason why these two different methods are probing different sub-populations of circumbinary planets. The mass distribution of the circumbinary planet population is also poorly understood, with half of the transiting sample from \textit{Kepler} having weakly constrained masses. The masses of the circumbinary planets from \textit{Kepler} are mostly around the mass of Saturn, but a handful of other planets have larger masses than this. TIC\,172900988b has a mass \(\approx 2\,{\rm M_{Jup}}\) \citep{sairam_new_2024}, Kepler-1660b a mass \(\approx 5\,{\rm M_{Jup}}\) \citep{goldberg_5mjup_2023}, and BEBOP-4b a mass \(\approx 20\,{\rm M_{Jup}}\) \citep{triaud_bebop_2025}. The set of circumbinary planet/candidate masses detected with radial velocities, while containing some super-Jupiter mass planets, is on the whole lower mass than the general population orbiting single-stars \citep{baycroft_progress_2024}.

In independent sets of observations and research programmes, many circumbinary planets have been claimed to orbit post-common-envelope binaries, with detections based on eclipse timing variations \citep[e.g.][]{beuermann_two_2010,basturk_eclipse_2023}. These are typically massive super-Jupiters (\(M\sin{i}>{\rm M_{Jup}}\)) on long orbital periods (multiple years). There is a debate as to whether such eclipse timing variation signals are indeed caused by  orbiting circumbinary companions \citep{hardy_first_2015,pulley_eclipse_2022}, but also some astrometric evidence that such planets might exist \citep{baycroft_new_2023}. Since circumbinary planets do exist orbiting main sequence binaries, as those found in transit and radial-velocity are, it stands to reason some might survive the common-envelope process their host binary is very likely to follow \citep{columba_statistics_2023}. \edit{It has also been proposed that these could arise in a second-generation of planet formation from ejected material \citep[e.g.][]{ledda_quest_2023}.}

\gaia provides a new opportunity for a population-level analysis of circumbinary exoplanets, with the potential for a large sample size of optimal close binary systems, among which many circumbinary planets can be found. \citet{sahlmann_gaias_2015} estimated the expected yield of circumbinary planets from \gaia. \edit{Mostly based on the \kepler results \citep{armstrong_abundance_2014,martin_planets_2014,li_uncovering_2016},} a number of assumptions were made about the circumbinary planet population, namely:
\begin{itemize}
    \item The occurrence rate of circumbinary gas giants is \(\approx 10\%\)
    \item The mass distribution is the same as that of planets orbiting single-stars
    \item All circumbinary planets are in the "pile-up" at \(6\times P_{\rm bin}\)
\end{itemize}
Of these assumptions, the first is still consistent with the current understanding, but the other two are not (see \edit{the discussion} above \edit{and in \citealt{baycroft_progress_2024,baycroft_tools_2025}}). We therefore revisit this problem, partly with updated knowledge of the \gaia sensitivity, but especially of the circumbinary planet population properties. We also assess the ability of \gaia to weigh-in on known circumbinary systems, in particular the post-common-envelope systems mentioned above.

We present our constructed samples of circumbinary planets, our criteria for circumbinary detectability in \gaia, and the calculation method for expected yields in Section \ref{sec:methods}. In Section \ref{sec:results} we present the results of different injected circumbinary planet populations on the yield expected from \gaia, as well as the expectations for \gaia relating to known circumbinary planets orbiting main-sequence binaries and post-common-envelope binaries. We conclude in Section \ref{sec:conclusions}.

\section{Methods}\label{sec:methods}

\subsection{Synthetic Population}\label{sec:synthpop}

We generate a synthetic population of binaries, using \gaia sources and the binary star population from \citet{raghavan_survey_2010}, and then inject a circumbinary planet population inspired by the detected population. The use of \citet{raghavan_survey_2010} for the binary fractions and distributions is arbitrary to a certain extent, other similar studies exist \citep[e.g.][]{duquennoy_multiplicity_1991,duchene_stellar_2013,offner_origin_2023}. These various studies are in broad agreement over the multiplicity rates, and \citet{raghavan_survey_2010} gives the clearest distributions of binary mass ratios and eccentricities in a graphical format that we could artificially reproduce (see Figure \ref{fig:gaia_bins}). While the population of binary stars used will likely impact the exact numbers predicted, the conclusions about how changing the planet distribution affects the yields are not strongly affected (see Section \ref{sec:yielddisc}).

\subsubsection{Synthetic population of binary stars}

To assemble the binary star population we use the sample of \gaia DR3 sources within 200pc \citep{gaia_collaboration_gaia_2016,gaia_collaboration_gaia_2023}, use\editt{ing} the following requirements \editt{to obtain the distance cut to ensure that we only select sources with significant parallaxes and good photometry:} 
\begin{itemize}
    \item \editt{1000/parallax \(\leq 200\)}
    \item parallax\_over\_error \(\geq10\)
    \item phot\_g\_mean\_flux\_over\_error \(\geq{\bf 50}\)
    \item phot\_bp\_mean\_flux\_over\_error \(\geq20\)
    \item \editt{phot\_rp\_mean\_flux\_over\_error \(\geq20\)}
\end{itemize}
\editt{We find that the exact value used for these selections has little impact, reducing each threshold to 6 only increases the predicted yields by less than 5\%.}

We only consider the yield of circumbinary planets within 200pc, similarly to \citet{sahlmann_gaias_2015}. This will lead to an underestimate of the total number of detections that \gaia will provide, but those detections beyond 200pc will likely be dominated by the highest mass planets. While considering binaries to further distances from the solar system would allow for comparisons with predictions of planets orbiting single stars \citep[e.g.][]{perryman_astrometric_2014,lammers_exoplanet_2025}, we chose to restrict to 200pc which allows for comparison with \citet{sahlmann_gaias_2015}. 

We also restrict ourselves to main sequence stars for the moment. Disregarding that some systems are already known binaries, their parameters are assumed to be that of a single star. \edit{The mass of each star is calculated from from the \gaia bp-rp colour \citep[following][]{pecaut_intrinsic_2013}}\footnote{Updated table taken from: \url{https://www.pas.rochester.edu/~emamajek/EEM_dwarf_UBVIJHK_colors_Teff.txt}}.

A population of binary stars is injected following \citet{raghavan_survey_2010}\edit{, using the existing stars from \gaia as primary stars}. The binary fraction based on spectral type is listed in Table \ref{tab:binarityfrac}; these values are the lower bounds on the \textit{multiplicity fraction} from \citet{raghavan_survey_2010}. 

\begin{table}
    \centering
    \caption{Fraction of stars in the synthetic sample to which a binary companion is assigned, divided by spectral type. Distribution based on \citet{raghavan_survey_2010}.}
    \begin{tabular}{lc}
    \hline
    Spectral type & Binary fraction \\
    \hline
    B & \(75\%\) \\
    A & \(70\%\) \\
    F & \(\edit{45}\%\) \\
    G & \(\edit{40}\%\) \\
    K & \(\edit{35}\%\) \\
    M & \(\edit{30}\%\) \\
    \hline
    \end{tabular}
    \label{tab:binarityfrac}
\end{table}

\begin{figure*}
    \centering
    \includegraphics[width=0.8\columnwidth]{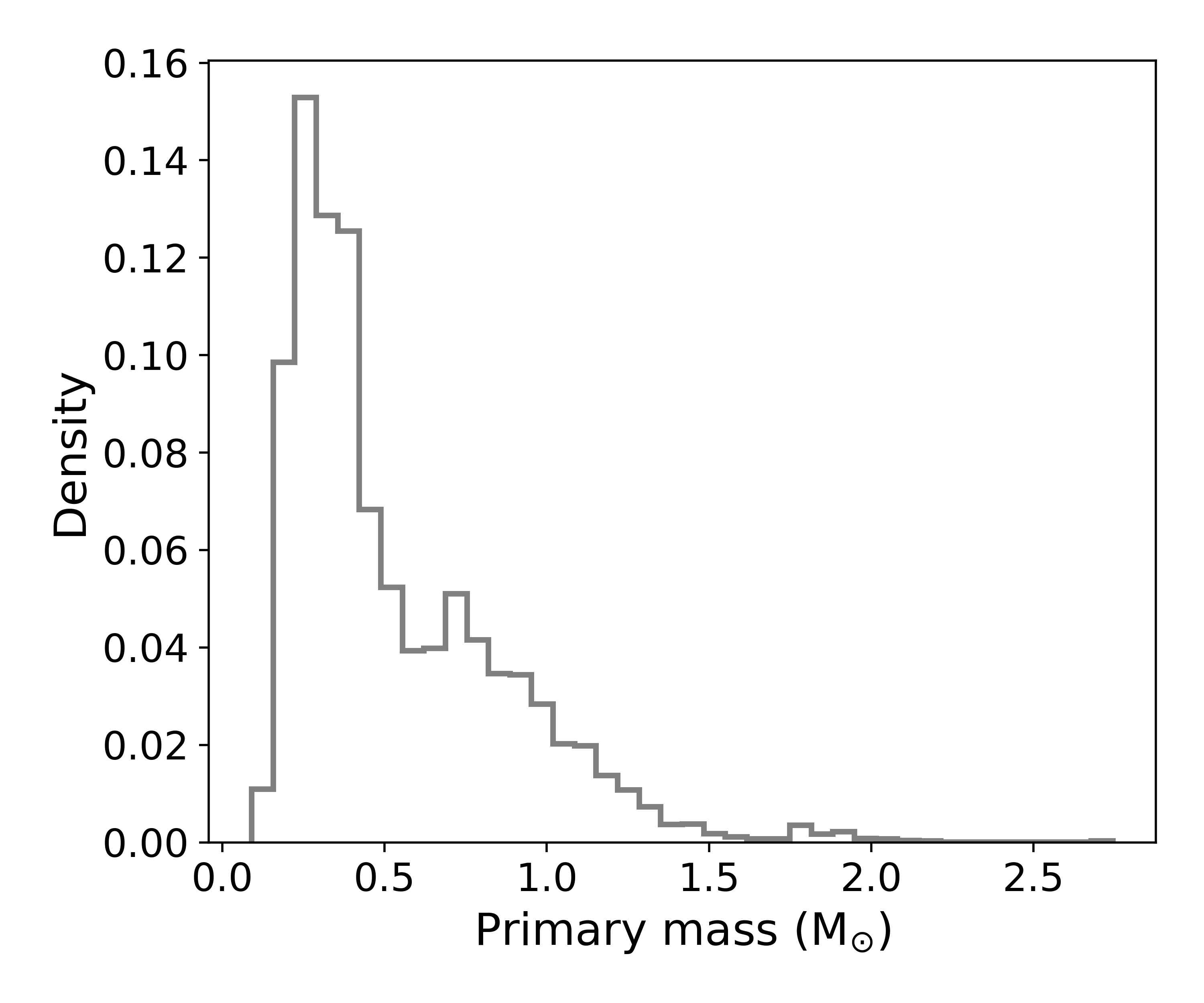}
    \includegraphics[width=0.8\columnwidth]{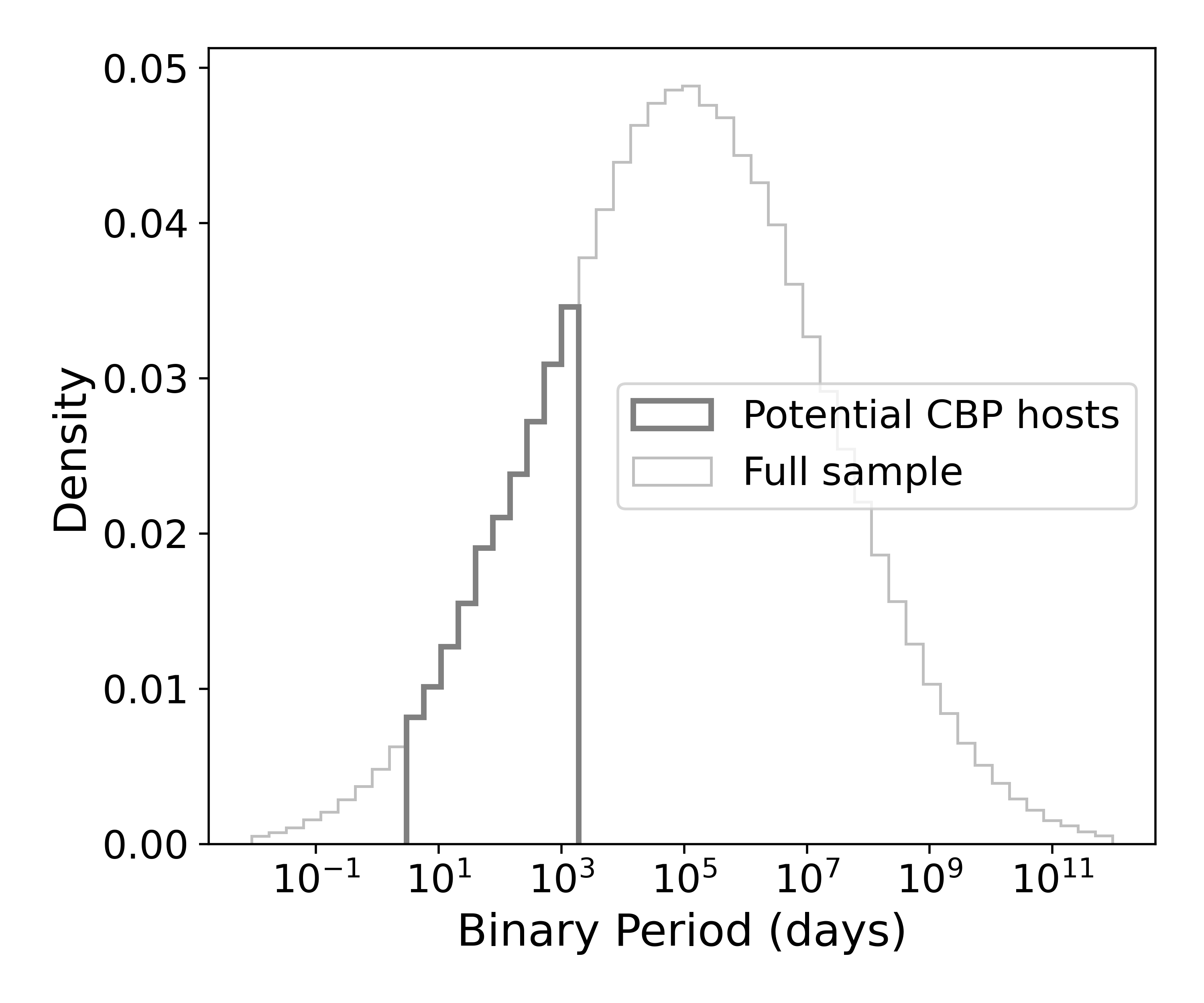}\\
     \includegraphics[width=0.8\columnwidth]{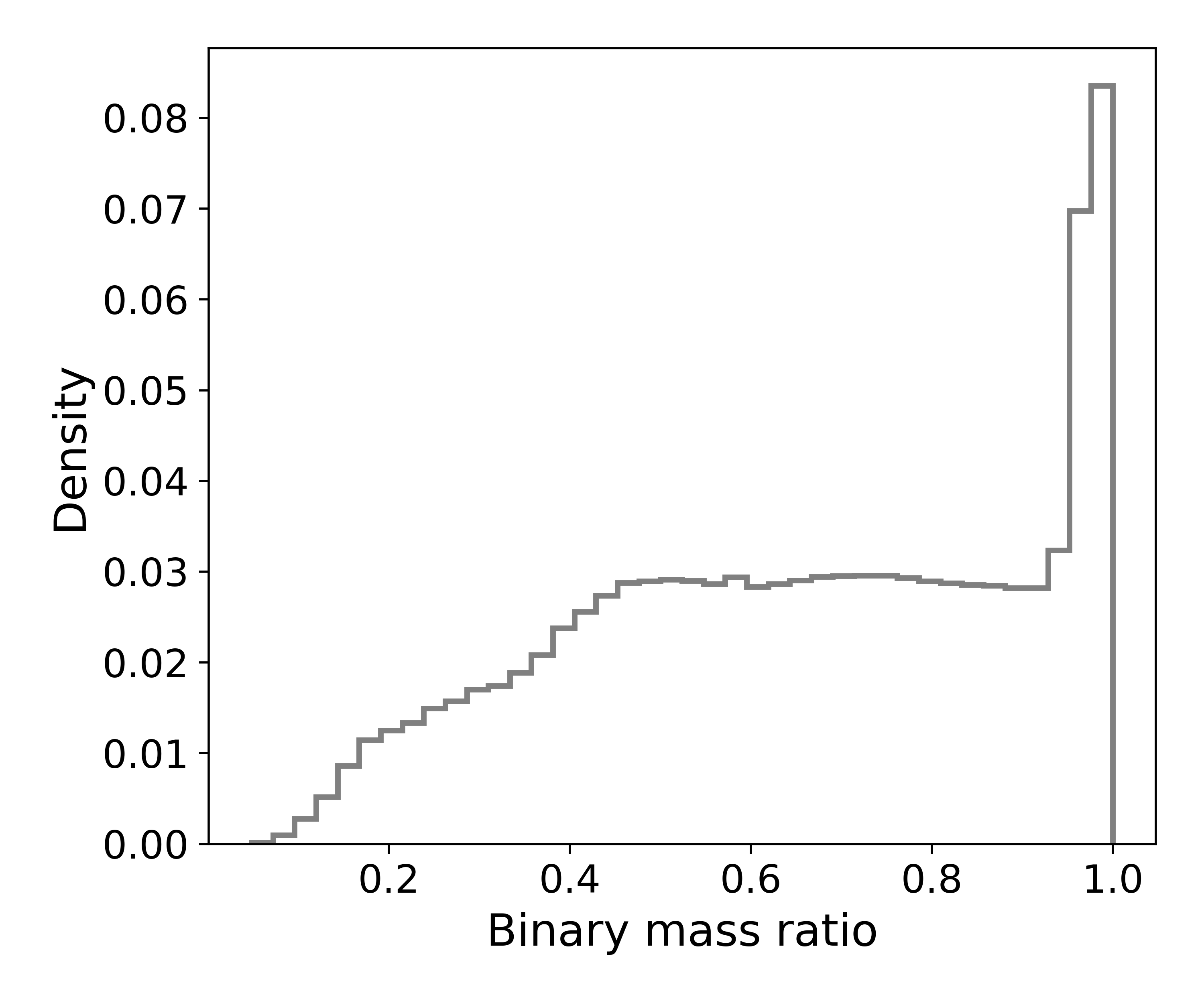}
    \includegraphics[width=0.8\columnwidth]{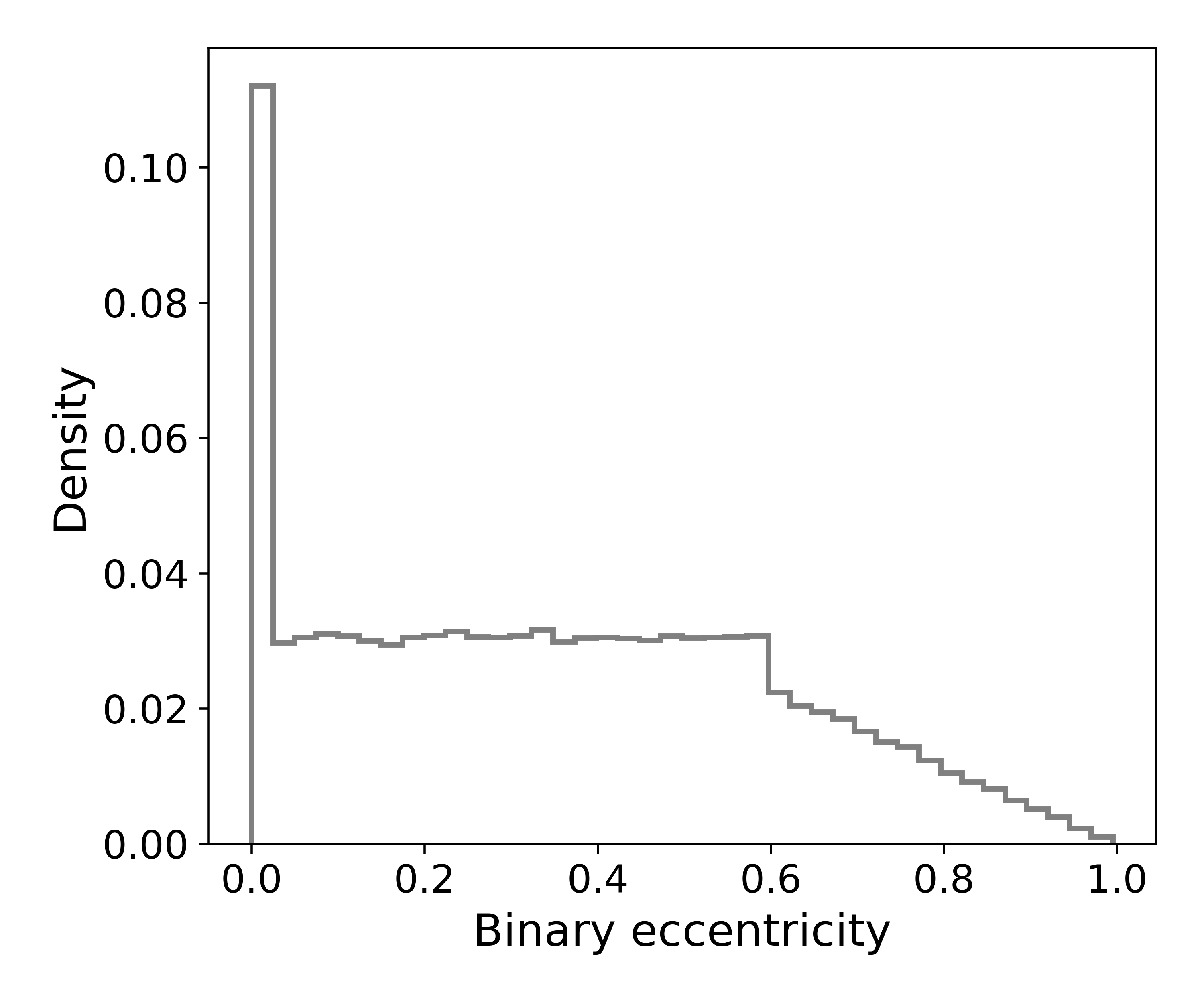}
    \caption{Distributions of parameters for the synthetic population of binary stars which will be used for simulation for \gaia yield of circumbinary planets. \edit{Orbital period distribution shows the full simulated sample in the background while specifying the range that is actually considered for potential circumbinary planet hosts.}}
    \label{fig:gaia_bins}
\end{figure*}

The injected binary parameter distributions are presented in Figure \ref{fig:gaia_bins}. 
The period (\(P_{\rm bin}\)), mass ratio (\(q\)), and eccentricity (\(e\)), distributions are taken from \citet{raghavan_survey_2010}, where approximations are made to reproduce the \(q\) and \(e\) distributions. 

\edit{For \(P\) the distribution is log-normal (\(\log P \sim \mathcal{N}(5.03,2.28)\)), and we remove binaries with \(P\leq3\,{\rm days}\). The \(P\) distribution shown in Figure \ref{fig:gaia_bins} shows the full generated distribution, as well as the section of it which which is considered in this work, which includes the 3 day cut and also a line at half the timespan of \gaia DR5 since planets should not exist on stable orbits near enough longer-period binaries than this to be detected in \gaia data.} For \(q\), we start from a uniform distribution in the range \([0,1]\). \edit{Two} change\edit{s are} performed to mimic the observed overabundance of equal mass binaries, to recreate a decreasing binary fraction  from \(q = 0.2\) down to 0\edit{, and to account for for the mass ratio distribution being more skewed towards q=1 for binaries with \(P\leq100\,{\rm days}\) \citep{raghavan_survey_2010}}. For \(e\) we use an initially uniform distribution in the range \([0,1]\), and perform two operations to reproduce the distribution seen in \citet{raghavan_survey_2010} where high eccentricities are less common. The total of \edit{four} changes are as follows:

\begin{enumerate}
    \item For binaries with \(q<0.182\), with probability \(1-5.5q\) the mass ratio for that binary is instead chosen uniformly between \([0.95,1]\).
    \item \edit{For binaries with \(P\leq100\,{\rm days}\), \(\frac{1-q}{3}\) is added to the values of q}
    \item All binaries with \(P_{\rm bin} < 12\) days are made circular since they would likely have tidally circularised.
    \item For binaries with \(e>0.6\), with probability \(2.5(e-0.6)\) the eccentricity of that binary is instead chosen uniformly between \([0,0.6]\).
\end{enumerate}

\edit{We also remove any systems where the companion star is a brown dwarf \(M<0.08\,{\rm M_{\odot}}\), except if the primary star is M type, at which point we place a threshold in secondary mass at \(M=0.06\,{\rm M_{\odot}}\) since such systems would warrant the status of binary star.}

These result in the distributions in eccentricity and mass ratio seen in the bottom panels of Figure \ref{fig:gaia_bins}.

\subsubsection{Discussion of binary population}

\edit{The main decision taken is the choice of binary fraction. \editt{This is not as simple as taking the fraction of lone binaries "binarity fraction" since some known circumbinary systems have distant stellar companions such as \kepler-64 \citep{schwamb_planet_2013}, a circumbinary planet within a hierarchical quadruple star system. Nor is it suitable to take the "multiplicity fraction" of all non-single stars, since many tertiaries would be close enough to inhibit circumbinary planet formation, or impact the \gaia astrometry. The better solution lies somewhere in between. The distance to the outer binary in \kepler-64 is \(\sim1000\) AU which would put it around \(10^5\) days, at the peak of the binary period distribution (see Figure \ref{fig:gaia_bins}). \gaia will resolve sources with separation \(\geq 400 \) mas \citep{gaia_collaboration_gaia_2022,holl_gaia_2023-1}, which at 200 pc corresponds to 80 AU, therefore most wide triples which would have space for a circumbinary planet to the inner binary will be resolved in \gaia. We therefore use a binary fraction around halfway between the binarity and multiplicity fractions.} In \citet{raghavan_survey_2010} the multiplicity fraction for FGK stars\footnote{line 2 of their Table 16 (including candidates)} is \(46\%\) and the binar\editt{ity} fraction is \(34\%\), the rest being triples and quadruples. The value we chose was \(40\%\) (accounting for the removed systems with \(P\leq3\,{\rm days}\) this drops to \(39\%\)). \editt{We note that the choice of 3 days as a cut is not based on a physical expectation, but purely an empirical observation that below this approximate boundary almost all (96\% of) binaries have a tertiary companion \citep{tokovinin_tertiary_2006}. Comparatively few binaries have short orbital periods so the exact location of this cut has little impact on our results.}

To assess the effect that \editt{our method for dealing with tertiaries} has, we also repeat the ensuing analysis with lower binary fraction\editt{,} aiming to \editt{only include lone binaries with no extra components. A} summary of the results is presented in Appendix \ref{sec:fewerbinaries}, where very little difference is found except for the yields are \(75-80\%\) of the original values.} \editt{As it stands, the results in the main text do include systems that will be triples (and above) which includes an implicit assumption that binaries with a wide tertiary companion host circumbinary planets at the same rate as lone binaries. Even if the tertiary is wide enough to allow planets to form, the rate may still be different. Since this rate is not known, we chose to accept this limitation which could lead to an overestimate of the yield by up to 20-25\% . In a very optimistic scenario, \gaia might even be able to constrain this rate.}

\edit{While the \(P\) distribution is very wide, spanning orders of magnitude, only the range between \(3\) days and \(\sim 1000\) days will be able to host detectable circumbinary planets in our setup. The \(q\) distribution is used to obtain the luminosity ratio used to calculate if the binary is detected astrometrically, but its main impact on the calculated yields is through the total mass of the binary which is used to calculate the astrometric signal of the planets. If the true \(q\) were higher (lower) than what we simulate the yield would decrease (increase). The \(e\) distribution is only used to calculate the circumbinary stability limit beyond which we place the planets.}

\edit{When calculating the astrometric SNR (see Section \ref{sec:detectability} we take the luminosity of the secondary star into account for calculating the size of the photocentre orbit around the centre-of-mass (i.e. the size is 0 for an equal brightness binary). However, we do not take into account any extra luminosity of the secondary for the total brightness of the system, which is used to calculate the astrometric precision. Extra luminosity would mean no change or an improvement in precision for the majority of systems (see Figure \ref{fig:prec_approx}) and this would likely only be a small effect, therefore not including this is a conservative choice so as to not potentially bias towards larger yields.}

\edit{The synthetic population as a whole contains \(\sim450\,000\) binaries, though most of these are not be suitable for circumbinary planets detected by \gaia (see top-right panel of Figure \ref{fig:gaia_bins}.} The simulated binary population used in \citet{sahlmann_gaias_2015} was generated using mass distributions for primary and secondary stars from a combination of \textit{Kepler} and radial velocity surveys; assuming the local density of FGK stars  remains constant out to 200 pc; and the including a flat binarity rate of \(13.5^{+1.8}_{-1.6}\,\%\) \citep[\edit{for binaries with \(P\leq10\,{\rm years}\);}][]{halbwachs_multiplicity_2003}. They also restrict to binaries with \(P<304\,{\rm days}\) and \(0.6\,{\rm M_{\odot}} \leq M_{\rm A} \leq 1.3\,{\rm M_{\odot}}\), to
obtain a sample of \(\approx 38\,000\) binaries. Applying the same cuts to our sample (which includes longer period binaries and a wider range of primary masses) we obtain \(\approx 53\,000\) binaries when applying the period cut, and obtain  \(\approx 27\,000\) binaries when we also apply the primary mass cut. The set of accessible binaries over which we will simulate circumbinary exoplanet in this paper is 40\% larger than in \citet{sahlmann_gaias_2015} but within the parameter range common with \citet{sahlmann_gaias_2015} we obtain 30\% fewer systems. We expect that will not have a drastic impact on the number of circumbinary planets we obtain in our yields. The effect of the number of binaries is much smaller than the effect of changing the circumbinary planet population, as discussed below.

\subsubsection{Injected populations of circumbinary planets}

For the circumbinary planet population, we only inject gas giants, as these are the only planets that \gaia will realistically be sensitive to, and they are the circumbinary population for which we have a better understanding. We inject a planet around 10\% of the binaries \citep[the approximate occurrence rate of circumbinary gas giants;][]{armstrong_abundance_2014,martin_planets_2014,baycroft_progress_2024}. Similarly to \citet{sahlmann_gaias_2015} who included companions up to a mass of \(M = 30{\rm \, M_{Jup}}\), we include companions up to a mass of  \(M = 25{\rm \, M_{Jup}}\) as "planets" in our samples.This includes low mass brown dwarfs below the so called "brown-dwarf desert" \citep{grether_how_2006,sahlmann_search_2011}, which is located between \(\approx 25-45\,{\rm M_{Jup}}\), as planets, despite them being above the deuterium-fusing limit. With the eventual aim to understand planet formation around binaries, we chose to include companions up to this mass as ones that likely would have formed in the same way as the rest of the super-Jupiter planet population, and refer to all of these as "planets" throughout this work \citep[as recommended in][]{schneider_definition_2018}.

The mass and orbital period distributions of circumbinary planets are still not very well known. In the previous study on \gaia's circumbinary planet yield \citep{sahlmann_gaias_2015}, planet masses were drawn from the population detected orbiting single-stars, and the orbital periods were all fixed at \(6\times\) the binary orbital period (\(P_{\rm bin}\)). The \kepler circumbinary planets known at the time were all clustered around \(6\times P_{\rm bin}\) so this was a good assumption then. Additional detections in \kepler and \tess as well as preliminary results from the BEBOP\footnote{Binaries Escorted By Orbiting Planets} radial velocity survey have changed this picture \citep[e.g.][]{kostov_kepler-1647b_2016,baycroft_progress_2024,baycroft_bebop_2025}. Particularly, the radial-velocity survey finds very few planets detected close to the binary, and their first published planets are all at relatively longer orbital periods relative to the binary. 

In comparison with the injected population used by \citet{sahlmann_gaias_2015}, fewer high-mass circumbinary objects have been found with most of the confirmed planets being between the mass of Saturn and the mass of Jupiter. However, there are still circumbinary super-Jupiters and low-mass brown dwarfs claimed. The \(5\,{\rm M_{Jup}}\) planet orbiting Kepler-1660, detected based on eclipse timing variation is typically considered confirmed \citep{goldberg_5mjup_2023}. Directly imaged circumbinary companions also exists in this mass range, two super-Jupiters (masses \(\approx 4\) and \(\approx 10\,{\rm M_{Jup}}\)) were observed in WISPIT 1 \citep{van_capelleveen_wide_2025}, and a planet/low-mass brown dwarf\footnote{Although with a companion:binary mass ratio of \(\approx 0.1\) this could potentially be classed as a triple rather than a bona-fide circumbinary system} (mass \(\approx 12\) or  \(\approx16\,{\rm M_{Jup}}\)) in VHS 1256 \citep{dupuy_masses_2023}, additionally a \(\approx 20\,{\rm M_{Jup}}\) circumbinary companion has been detected from radial velocities \citep{triaud_bebop_2025}. Hence massive circumbinary objects with \(M>{\rm M_{Jup}}\) do exist, if rarely, so we inject populations that extend up to high masses but with a reduced rate compared to the population used in \citet{sahlmann_gaias_2015}. 

We generate 7 different setups with different mass and period distributions for the circumbinary planets. These 7 use different combinations of mass distributions as well as relative orbital period distributions. The Setup 1 is our fiducial setup, and we then produce three setups maintaining the fiducial period distribution and varying the mass distribution, and three the other way around. We perform these 7 simulations to assess the impact of the assumptions behind each injected distribution on the recovered yields. Since the true circumbinary planet population is unknown, this experiment allows us to see which parameters are most important to the recovered planet yield, and thus, what \gaia will constrain best. The different setups are summarised in Table \ref{tab:GaiaYieldSetup}.

\begin{figure}
    \centering
    \includegraphics[width=0.8\columnwidth]{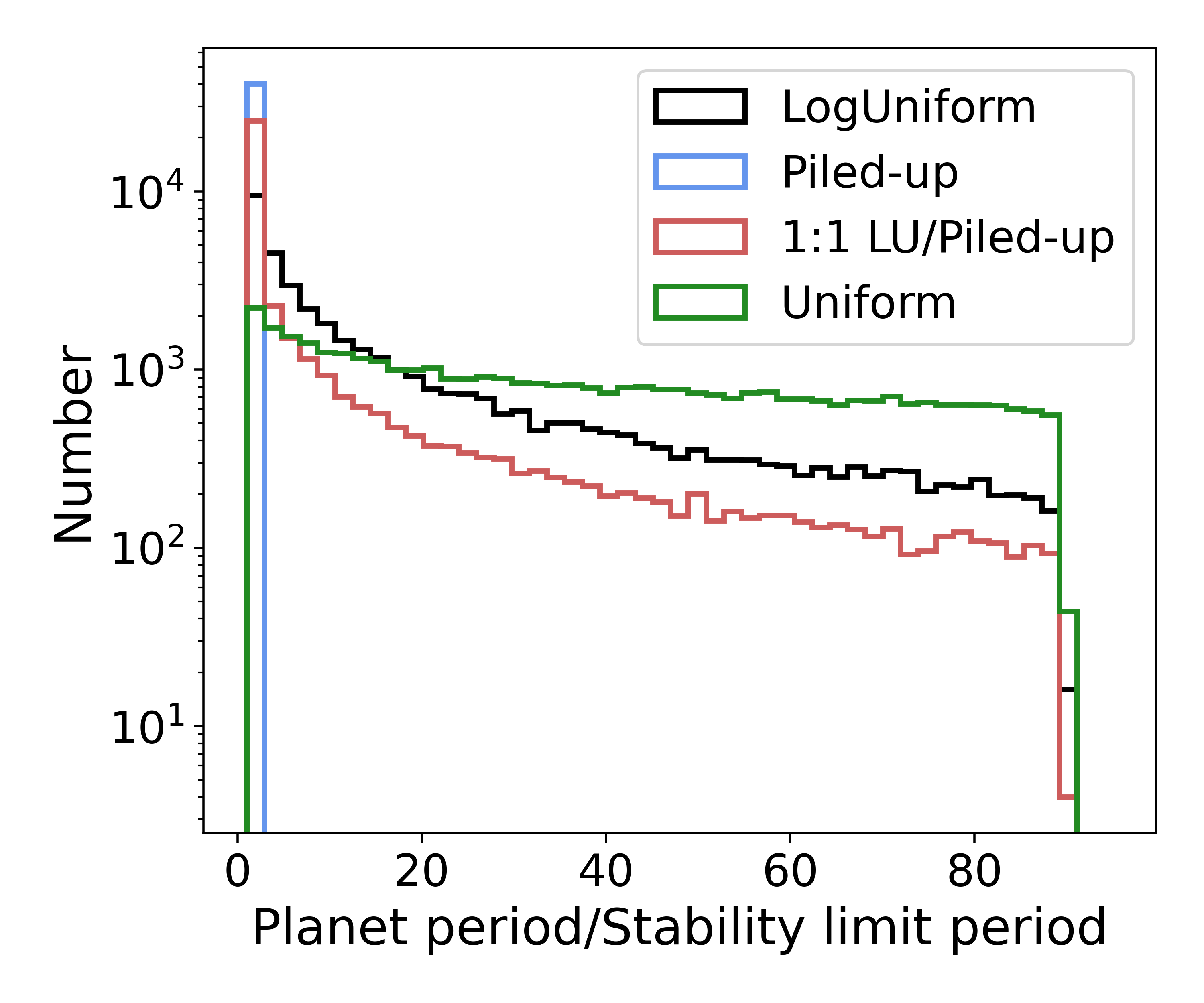}
    \includegraphics[width=0.8\columnwidth]{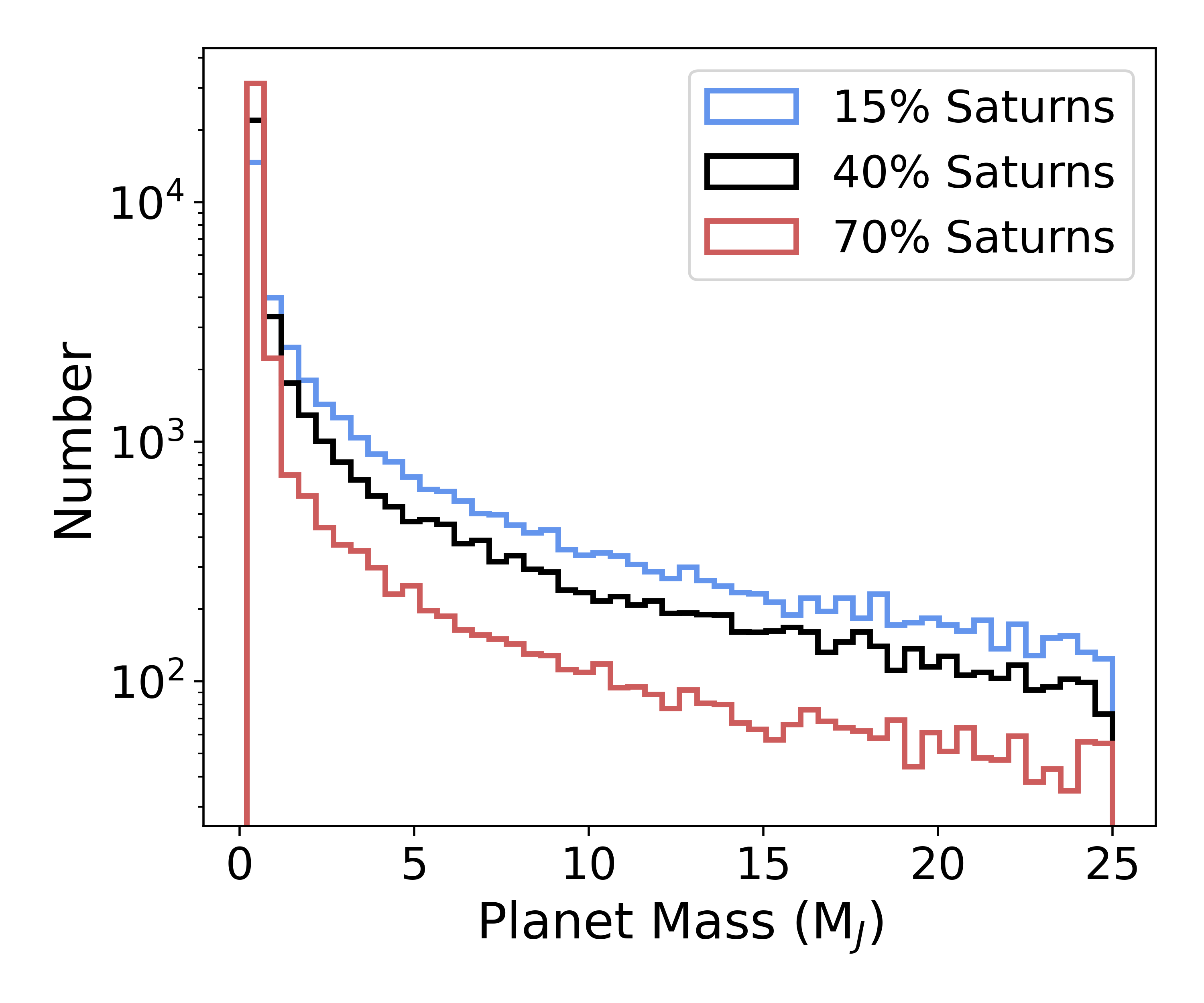}
    \caption{Planet mass and scaled orbital period distributions used in circumbinary planet injections to test \gaia yield. There are four different scaled period distributions used (top), and three different mass distributions used  (bottom). The middle distribution (in black) in each case is our fiducial one, it is kept the same when varying the other parameter.}
    \label{fig:gaia_planet_injections}
\end{figure}

We use 4 different orbital period distributions, more specifically the distributions are generated in the space of \(a_{\rm sc} = a_{\rm pl}/r_{\rm stab}\) \citep{baycroft_bebop_2025}, the semi-major axis of the circumbinary planet divided by the orbital radius of the stability limit \citep[calculated using][]{holman_long-term_1999}. In an improvement over placing all the planets at \(6\times P_{\rm bin}\) these steps take into account the binary eccentricity and mass ratio to calculate the instability limit \edit{If wanting to take into account the eccentricity of the injected planets as well, the limit defined in \citet{georgakarakos_empirical_2024} would be more appropriate, but here we assume circular orbits for the planets. We only include circular orbits since the \gaia SNR does not depend on eccentricity (see Section \ref{sec:detectability}), this shortcoming of the method is compensated by the fact we chose a conservative SNR threshold meaning any small impacts that eccentricity (or inclination) might have on the detectability should not matter}. The four distributions we use are: 
\begin{itemize}
    \item \(a_{\rm sc} \sim \mathcal{LU}(1,20)\)
    \item \(a_{\rm sc} = 1.2 \)
    \item \(a_{\rm sc} \sim \mathcal{U}(1,20)\)
    \item half \(a_{\rm sc} \sim \mathcal{LU}(1,20)\) and half \(a_{\rm sc} = 1.2 \)
\end{itemize}
where \(\mathcal{U}\) means a Uniform distribution, and \(\mathcal{LU}\) a Log-Uniform distribution. These distributions are plotted in the top panel of \ref{fig:gaia_planet_injections}. Note that the distributions are generated in the space of scaled semi-major axis (\(a_{\rm sc}\), but shown in scaled orbital period. Among those four, when we set \(a_{\rm sc} = 1.2\) we create a population of planets piled-up outside the stability limit, analogous to \(6\times P_{\rm bin}\), in order to compare to \citet{sahlmann_gaias_2015}. 

We also produce 3 different mass distributions. Each simulation is a combination of two functions:
\begin{itemize}
    \item \(M_{\rm pl} \sim \mathcal{N}(0.3\,M_{\rm Jup},0.2\,M_{\rm Jup})\), cut to only include planets with mass \(M>0.2\,M_{\rm Jup}\)
    \item \(M_{\rm pl} \sim \mathcal{LU}(0.2\,M_{\rm Jup},25\,M_{\rm Jup})\)
\end{itemize}
where \(\mathcal{N}\) is a Normal (or Gaussian) distribution. The first distribution peaks around the mass of Saturn (the typical mass of the circumbinary planets from \kepler). The combinations are constructed such that the percentage of planets drawn from the Normally distributed "Saturns distribution" are approximately \(15\%\), \(40\%\), and \(70\%\). The fiducial case is that with 40\% Saturns. These distributions are shown in the bottom panel of \ref{fig:gaia_planet_injections}. We also test a fourth mass distribution, taking the fiducial \(40\%\) Saturn mixture, and scaling the mass of the planet linearly with the mass of the inner binary. The rationale for this test is that the protoplanetary disc mass is expected to correlate with the central  mass \citep[be it single or binary;][]{andrews_mass_2013} which implies more material available to form more massive planets in higher mass binaries.

Setup 1 is the fiducial case (shown in black in Figure \ref{fig:gaia_planet_injections}). Setups 4, 5, and 6 keep the mass distributions as in Setup 1 and vary the distribution of \(a_{\rm sc}\). Setups 2, 3, and 7 maintain the distribution in \(a_{\rm sc}\) as in Setup 1 and vary the mass distribution with Setup 7 being that where the masses are scaled by the total mass of the binary.

\begin{table}
    {\centering
    \caption{Injected distributions in Mass and scaled semi-major axis of circumbinary planets for the 7 different setups.}
    \begin{tabular}{lll}
    \hline
    Setup number & \(M_{\rm pl}\) distribution & \(a_{\rm sc}\) distribution\\
    \hline
    1 & 40\% \(\mathcal{N}\) and 60\% \(\mathcal{LU}\) &  \(\mathcal{LU}\) \\
    2 & 15\% \(\mathcal{N}\) and 85\% \(\mathcal{LU}\) &  \(\mathcal{LU}\) \\
    3 & 70\% \(\mathcal{N}\) and 30\% \(\mathcal{LU}\) &  \(\mathcal{LU}\) \\
    4 & 40\% \(\mathcal{N}\) and 60\% \(\mathcal{LU}\) & \( = 1.2\)\\
    5 & 40\% \(\mathcal{N}\) and 60\% \(\mathcal{LU}\) & 1:1 \(\mathcal{LU}\) and \(=1.2\)\\
    6 & 40\% \(\mathcal{N}\) and 60\% \(\mathcal{LU}\) &  \(\mathcal{U}\) \\
    7   & 40\% \(\mathcal{N}\) and 60\% \(\mathcal{LU}\) \(^{*}\) &  \(\mathcal{LU}\) \\
    \hline
    \end{tabular}\\}
    \(^{*}\)scaled by \(M_{\rm bin}\)
    \label{tab:GaiaYieldSetup}
\end{table}

\subsection{Criteria for Detectability}\label{sec:detectability}

To detect an orbit with \gaia , the photocentre must trace a wide enough path to be noticed in a short enough time for good data coverage. We use conservative criteria, and require that an orbit has to be fully covered by \gaia's dataspan to classify it as a detection. This is conservative because, especially for larger amplitude signals, detections can in principle be made on partially covered orbits \citep[e.g.][]{gaia_collaboration_discovery_2024}. We use a \SNR\, (SNR) to assess each detection \citep[as in][]{sahlmann_gaias_2015}, with \({\rm SNR}>20\) chosen as the threshold beyond which an orbit can be  detected. The astrometric SNR as defined in \citet{sahlmann_gaias_2015} is:

\begin{equation}
    {\rm SNR} = \alpha \frac{\sqrt{N_{\rm obs}}}{\sigma},
\end{equation}

where \(N_{\rm obs}\) is the number of \gaia observations, \(\sigma\) their average uncertainty, and \(\alpha\) is the semi-major axis of the photocentre orbit. \edit{\(\alpha\) being in angular units will depend inversely on the distance to the system and will increase for a larger mass ratio, but decrease for a larger luminosity ratio (equations can be found in e.g. \citealt{sahlmann_gaias_2015}).} 

This SNR cut is a simplified detection criterion that neglects any influence of potential systematic errors. The true ability of \gaia to detect orbits has more aliases and caveats, such as dependence on eccentricity and sky position \citep{el-badry_generative_2024}. In the case of a yield estimate such as this, the simplified criterion is acceptable since the uncertainty in the circumbinary planet population properties it is the limiting factor of our study.

The average single-epoch astrometric uncertainty \editt{for stars brighter than Gmag} of 12 at the time of \gaia's launch was expected to be \(\approx 30\,{\rm \mu as}\) \editt{with each of the up-to 9 individual CCD astrometric measurements at each epoch having an expected uncertainty of \(\approx 90\,{\rm \mu as}\)} \citep{de_bruijne_gaia_2014}. At the most recent data release (DR3) the measured precision was different, in particular, the precision is worse for bright stars \citep[Gmag\(<9\); ][]{holl_gaia_2023}. This will likely still improve going forwards to DR4 and DR5, but in this work we use an approximation to the function in \citet{holl_gaia_2023}, shown in Figure \ref{fig:prec_approx} to estimate the precision of the \gaia epoch astrometry for each star\footnote{While a visual of the updated estimate of the DR4 astrometric uncertainty is available in \url{https://great.ast.cam.ac.uk/Greatwiki/GreatMeet-PM18?action=AttachFile&do=view&target=EAS2025-S1-Brown.pdf}, this is not so different from that of \citet{holl_gaia_2023}, so we use the published estimate}. \editt{We then multiply this by 3 to obtain the expected individual CCD measurement uncertainty, since that is the type of measurement we are using for the \SNR calculation.}

To classify as a detectable circumbinary planet in this work, the planet must be detectable in the \gaia astrometry, for which the constraints used are: \({\rm SNR} > 20\) and \(P_{\rm pl} < T_{\rm span}\). \(T_{\rm span}\) is the coverage of the astrometric data, which is \(\approx 2000\) days for DR4 and \(\approx 3800\) days for DR5. On top of this, the binary should also be detectable in the \gaia astrometry (i.e. \({\rm SNR} > 20\)). \edit{To class as detectable, the orbital coverage criterion relies only on the orbital period of the planet (since that of the binary must be smaller than that).}

We will use this definition of detectability on synthetic samples of circumbinary planets. For known systems we will not impose that the binary is detectable in the \gaia astrometry since it is already known. There is also the possibility that there would be a sample of circumbinary planets astrometrically detectable with \gaia data, but where the binary itself is not astrometrically detectable, such as short period \edit{(\(\lesssim 10\) days)} eclipsing binaries \edit{that are too distant for \gaia to detect, or twin binaries which would have almost no photocentre motion since the photocentre and centre of mass would be co-located}. Investigating these showed yields much lower than those for binaries detected astrometrically such that we chose not to include these in our synthetic "detections".

\subsection{Yield Estimates}

For a given setup, we simulate the population of circumbinary planets as detailed above, and record which planets are detectable and what their properties as well as that of the binaries they are orbiting. The "yield" is the number of detected planets from this; on top of the yield, the distribution of the various parameters (planetary and binary) can be studied. To smooth out the small number statistics, we repeat the simulation and yield calculations 100 times and combine/average the results.

We test the 7 different planet population scenarios, both for an expected DR4 yield and an expected DR5 yield. \editt{To calculate the expected number of CCD observations in DR4 and DR5 we take the 'astrometric\_n\_good\_obs\_al' (number of good CCD measurements taken in DR3) for each target, and extrapolate this to obtain the number of data that will be used in DR4 and DR5}. When comparing the results from the different injections, we mostly use the DR5 yield as this is the prediction of what the full \gaia dataset will produce and with a larger sample of planets, trends are clearer.

In Section \ref{sec:results} we present the distributions of various parameters for the detected planets in our synthetic sample. We also show the distribution for planets where the binary orbital period \(P_{\rm bin} < 50 \,{\rm days}\). This is an approximate threshold below which the currently known population of binaries hosting circumbinary planets lies. The comparative rarity of circumbinary planets from transits or radial velocities orbiting longer period binaries is due to there being fewer long period binaries in eclipsing binary catalogues \edit{(from which most dedicated searches have originated, e.g. \citealt{armstrong_abundance_2014,martin_bebop_2019,martin_searching_2021}), the transit geometry affecting longer period planets,} and due to the longer baselines of data then required to detect the circumbinary planets. To investigate how the yield of circumbinary planets from \gaia is affected by the orbital period distribution of circumbinary planets involves: the injected distribution itself (which is relative to the orbital period of the binary); the sensitivity of \gaia to longer period orbits; and the binary frequency increasing with longer binary period. Giving attention to this sub-sample of "short-period" binaries allows us to investigate how these different effects can be disentangled. The distributions shown in Section \ref{sec:results}, show the results on both the whole sample and this "short-period" sub-sample of binaries.

\section{Results}\label{sec:results}

\subsection{Gaia yield prediction across the injections}

\begin{figure}
    \centering
    \includegraphics[width=0.8\columnwidth]{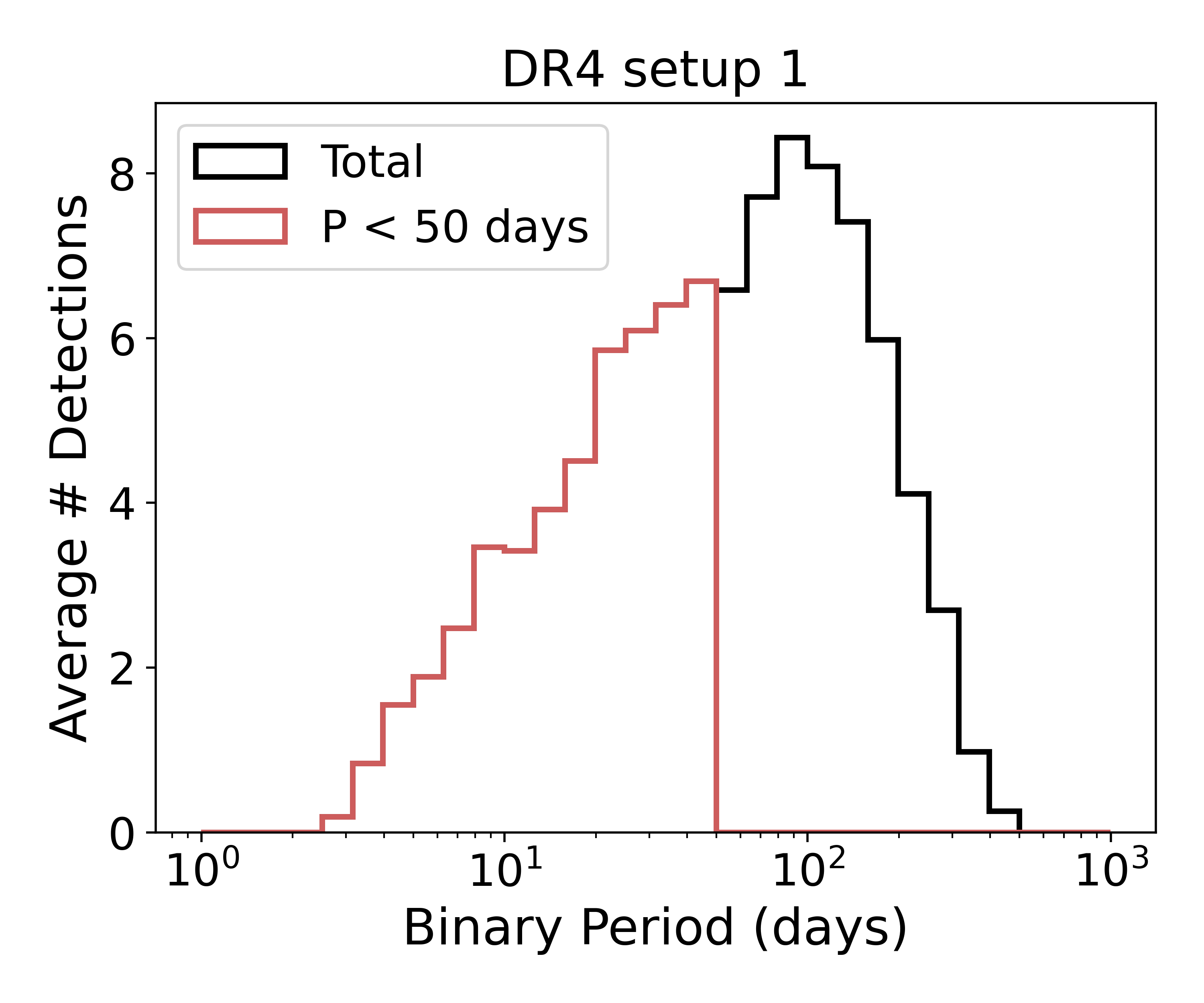}
    \includegraphics[width=0.8\columnwidth]{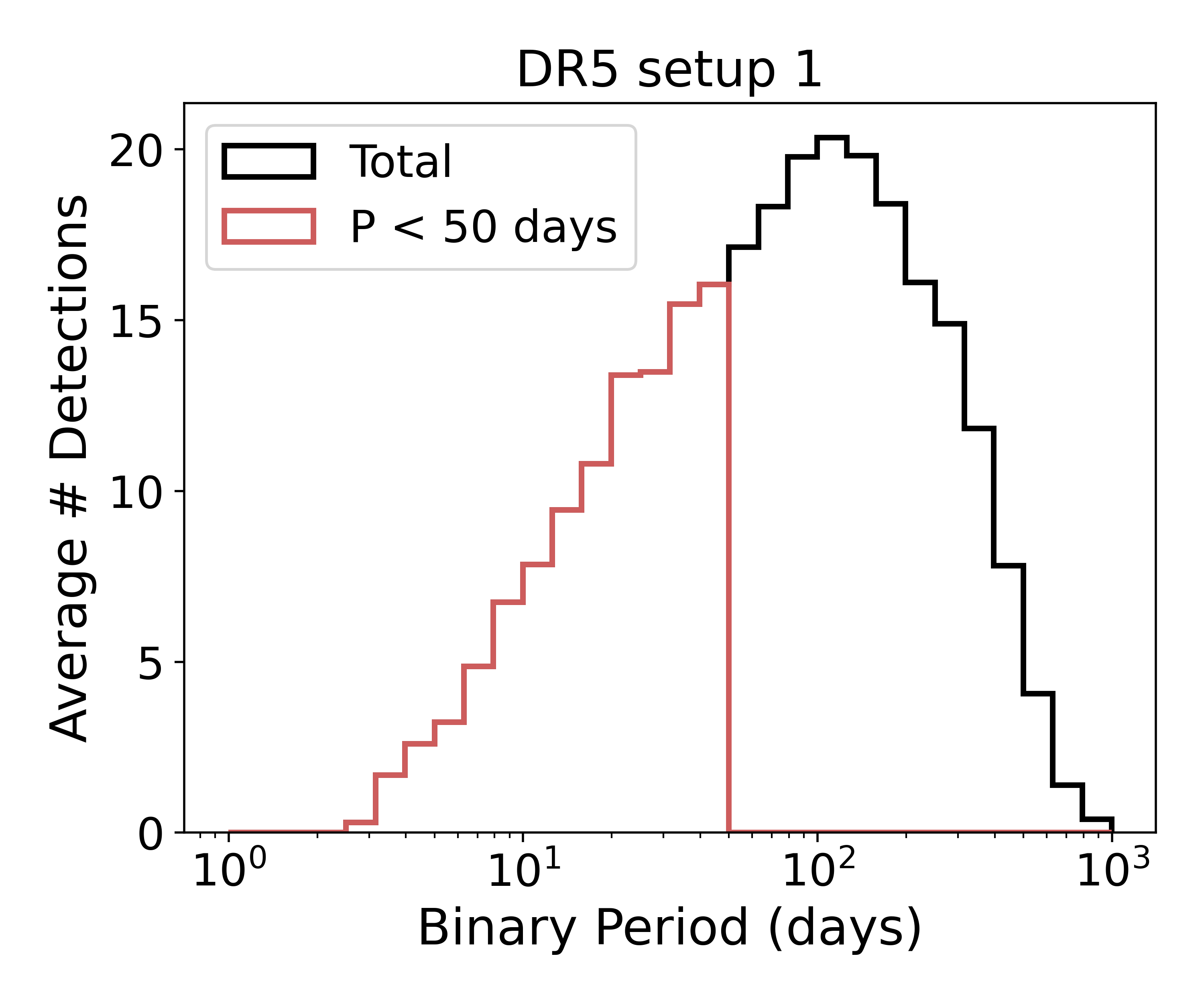}
    \caption{Distributions of the binary orbital periods around which circumbinary planets are expected to be detected in \edit{S}etup 1 (the \edit{fiducial} setup see Table \ref{tab:GaiaYieldSetup}). Top for DR4 and bottom for DR5, the other distributions are those restricted to binaries with \(P_{\rm bin}<50\,{\rm days}\).}
    \label{fig:Gaia_Pbins}
\end{figure}

The orbital periods of the binaries around which planets are detected in Setup 1 (our fiducial case) are drawn as histograms in Figure \ref{fig:Gaia_Pbins} comparing DR4 and DR5. Between DR4 and DR5 the total number of expected detections increases. Additionally, detections extend to planets orbiting longer binary periods as they become detectable due to the increased timespan covered by \gaia DR5.

\begin{figure}
    \centering
    \includegraphics[width=0.8\columnwidth]{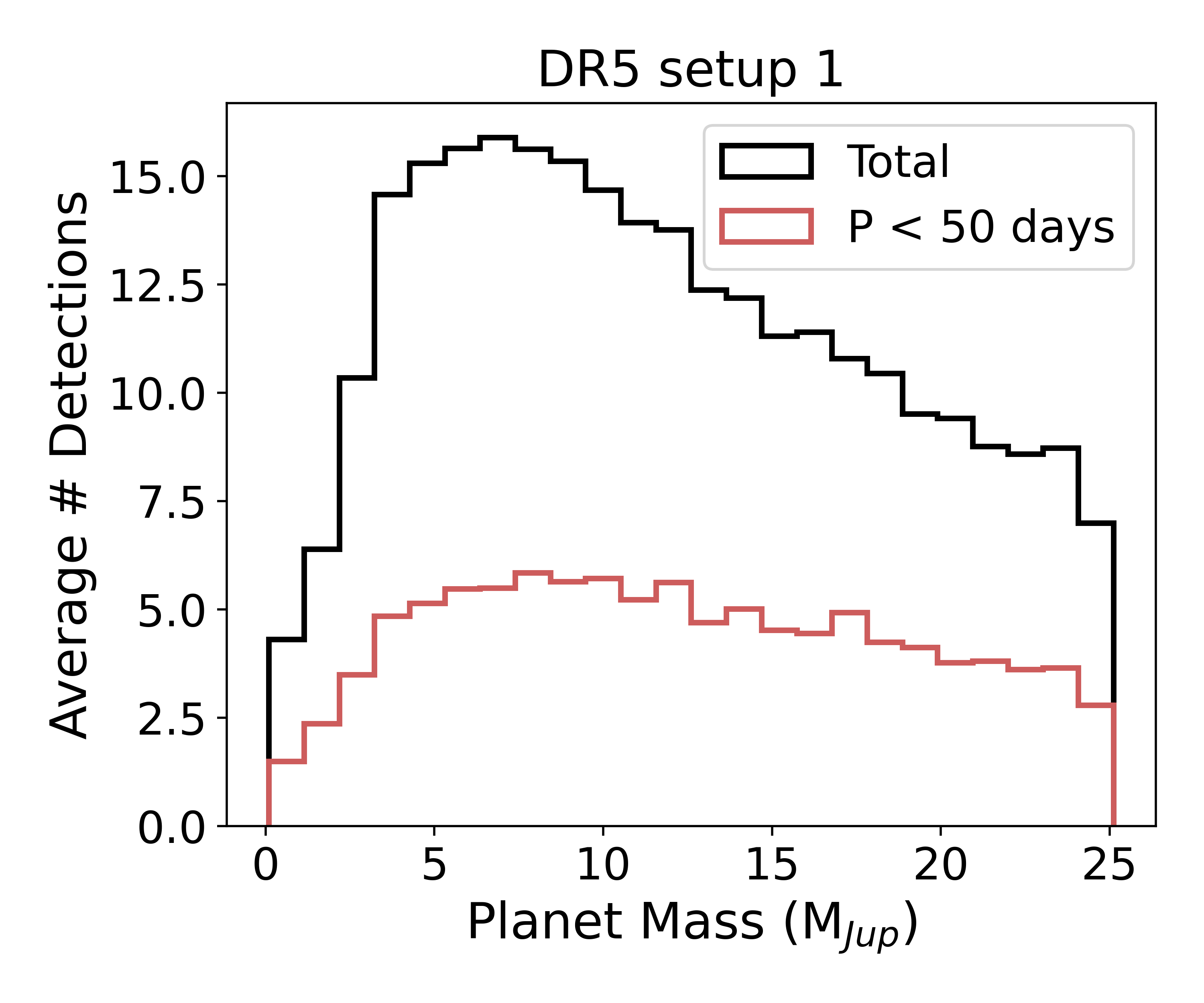}
    \includegraphics[width=0.8\columnwidth]{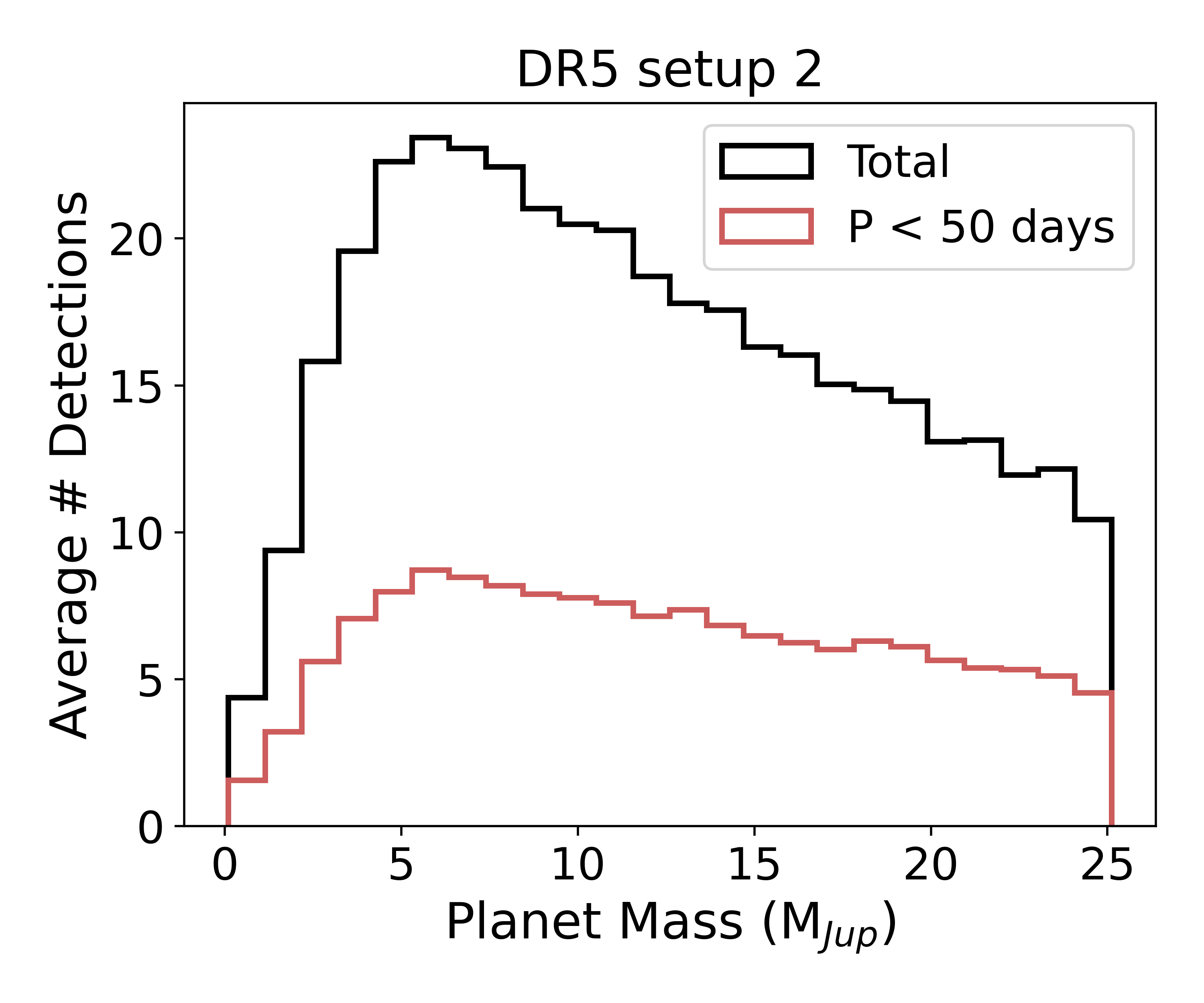}\\
    \includegraphics[width=0.8\columnwidth]{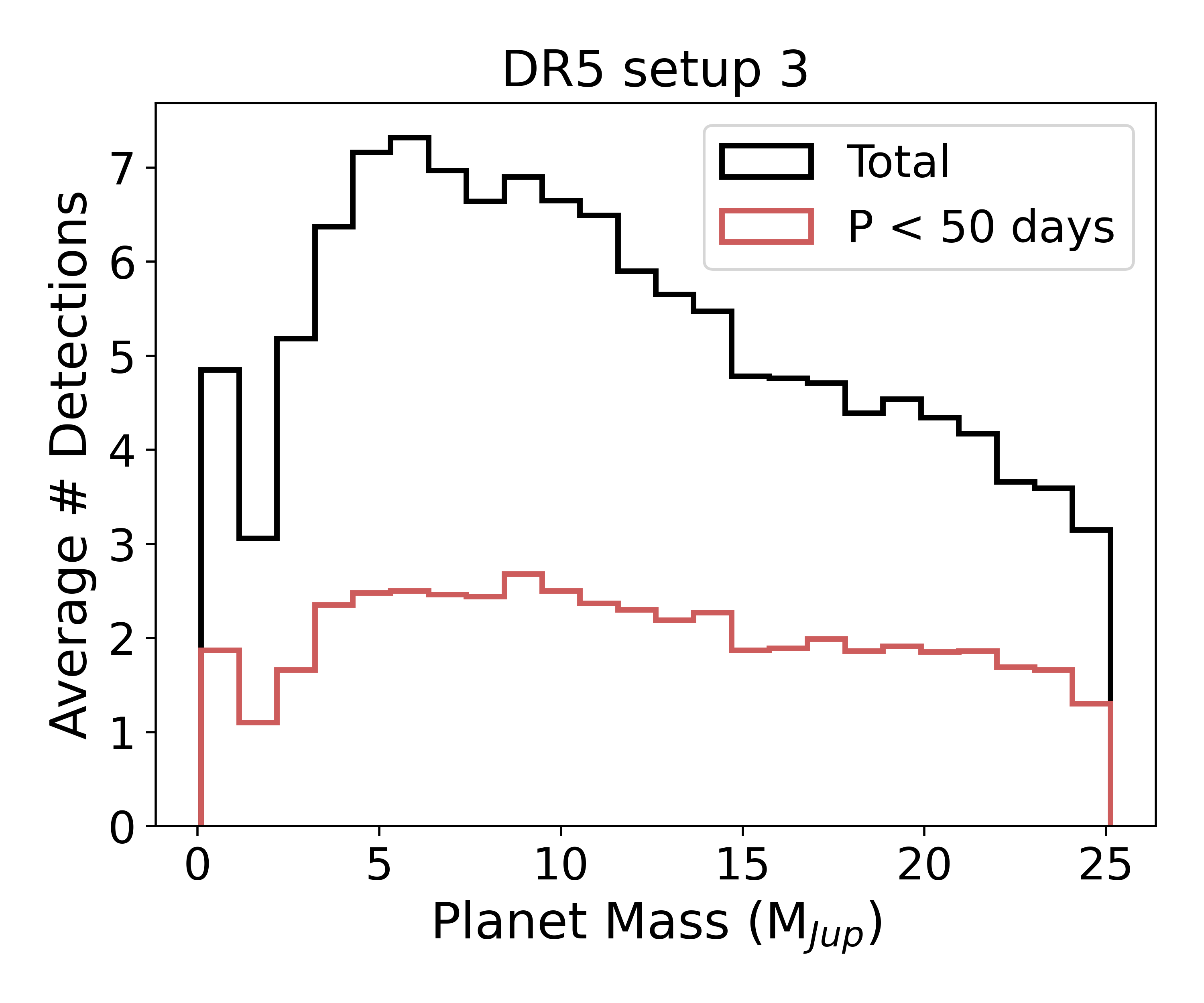}
    \caption{Distributions of detected planet masses of \gaia circumbinary planets in the \edit{S}etups 1, 2, and 3 (top, middle, and bottom). The other distributions are those restricted to binaries with \(P_{\rm bin}<50\,{\rm days}\).}
    \label{fig:Gaia_Mpls}
\end{figure}

Table \ref{tab:GaiaYieldsBintype} provides planet yields within each setup, separating them also for \gaia DR4 and DR5. Both the total yield and that for binaries with \(P_{\rm bin} < 50\,{\rm days}\) are detailed.

\begin{table}
    \centering
    \caption{Circumbinary planet yields for each of the 7 setups in \gaia DR4 and DR5, \edit{the yields for} those orbiting binaries with orbital periods shorter than 50 days \edit{are also shown}.}
    \begin{tabular}{lcccc}
    \hline
    Setup & Total & Total & \(P_{\rm bin}<50\) days  & \(P_{\rm bin}<50\) days \\
     & (DR4) & (DR5) & (DR4)  & (DR5) \\
    \hline
    1 & \editt{100} & \editt{276} & \editt{47} & \editt{106} \\
    2 & \editt{142} & \editt{394} & \editt{67} & \editt{153} \\
    3 & \editt{46} & \editt{127} & \editt{22} & \editt{49} \\
    4 & \editt{222} & \editt{564} & \editt{23} & \editt{42} \\
    5 & \editt{161} & \editt{423} & \editt{35} & \editt{75} \\
    6 & \editt{51} & \editt{152} & \editt{38} & \editt{101} \\
    7 & \editt{76} & \editt{220} & \editt{37} & \editt{88} \\ 
    \hline
    \end{tabular}
    \label{tab:GaiaYieldsBintype}
\end{table}

First, we examine how assumptions behind the mass distribution affect the yields. As expected \edit{(since the SNR increases with larger planet mass)}, increasing the amount of massive planets in the population leads to a clear increase in the yield, and this increase applies as well for short-period binaries as long-period. The planet mass distributions for \edit{S}etups 1, 2, and 3 (the experiment in varying mass distribution) are shown in Figure \ref{fig:Gaia_Mpls}. \edit{Across these three setups, most detected planets are super-Jupiters, and the shape of the distribution does not change much, peaking between \(5-10\,{\rm M_{Jup}}\)}. One difference which does appear is that when more Saturn-mass planets \edit{and fewer super-Jupiters} are injected (i.e. \edit{S}etup 3) there is a growing second peak from some of them being detected \editt{(and fewer larger mass planets being present comparatively}, however even in \edit{S}etup 3 this only gives an expected \(\sim 4\) detections in \gaia DR5. These low-mass detections are planets that happened to be injected around the nearest binaries to \edit{E}arth \edit{(see Figure \ref{fig:SNRexamples})}.

\begin{figure*}
    \centering
    \includegraphics[width=0.8\columnwidth]{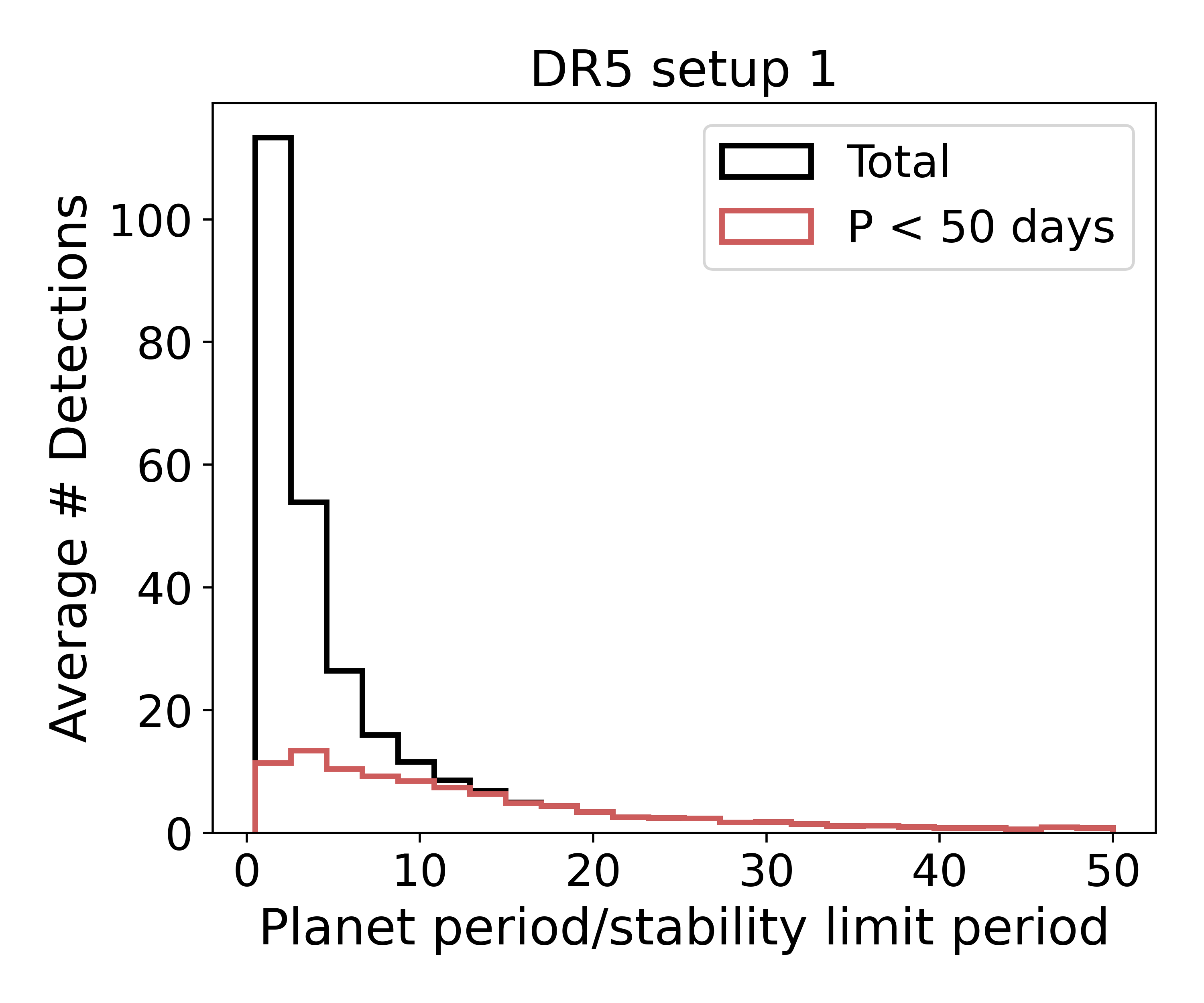}
    \includegraphics[width=0.8\columnwidth]{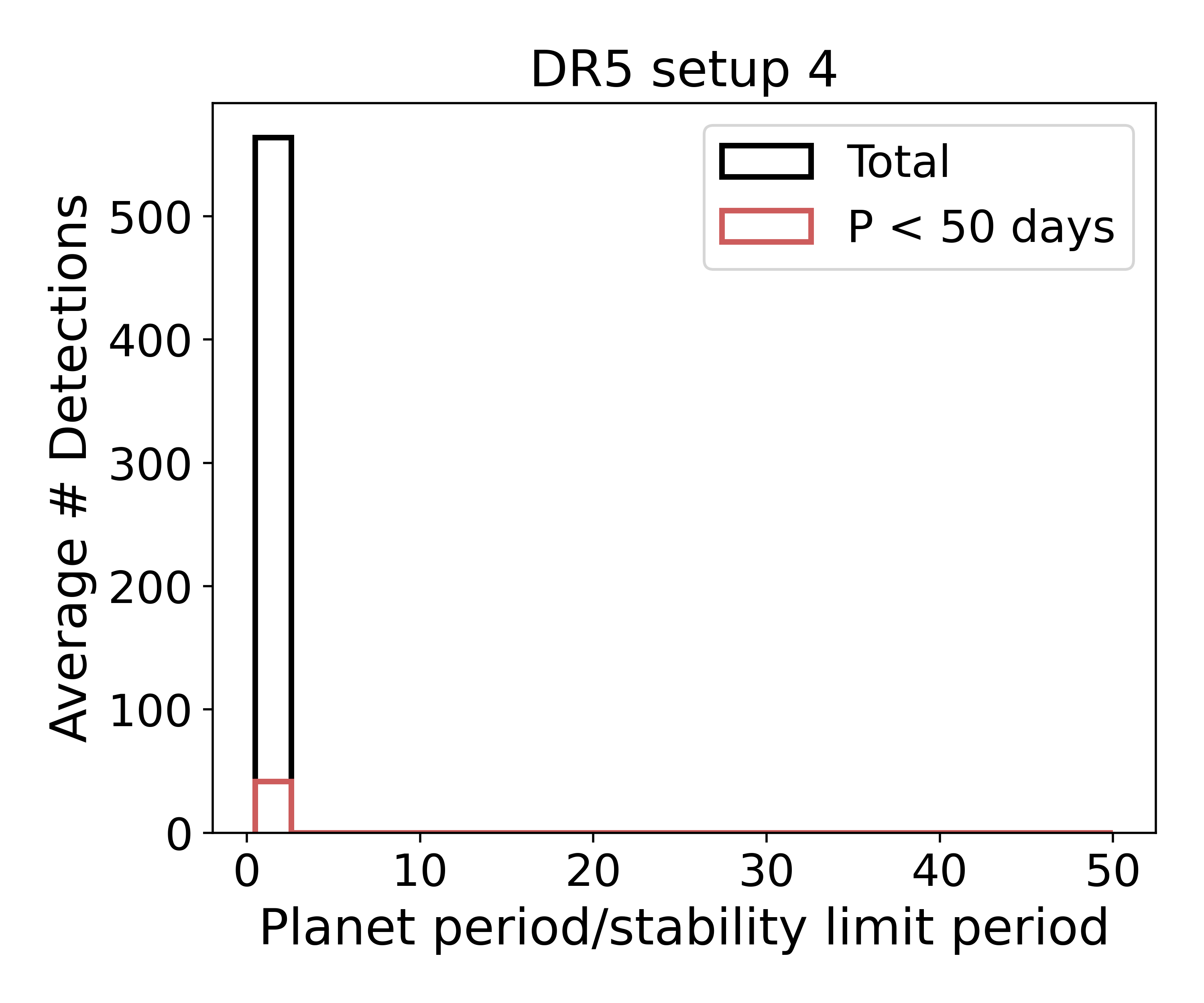}\\
    \includegraphics[width=0.8\columnwidth]{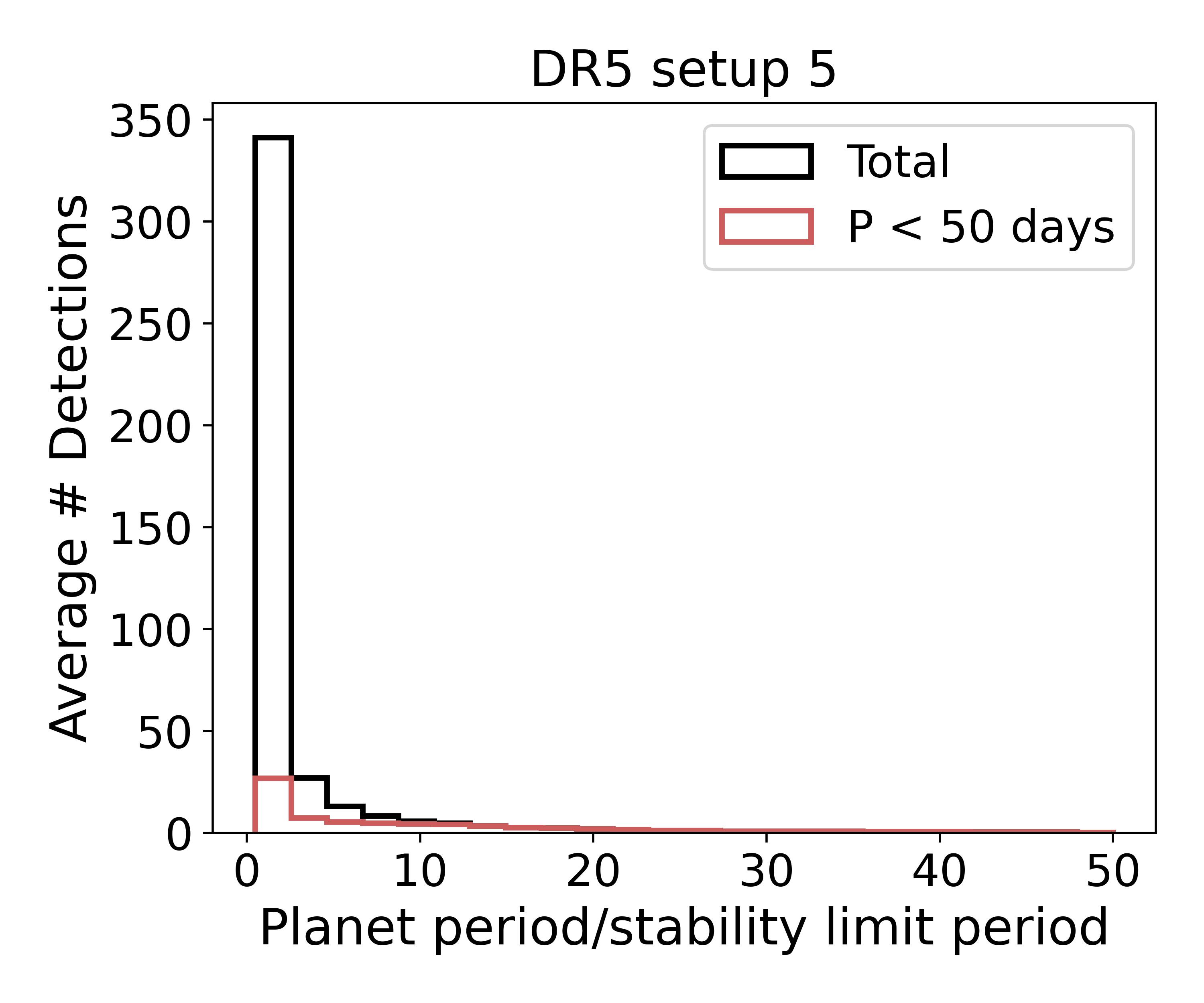}
    \includegraphics[width=0.8\columnwidth]{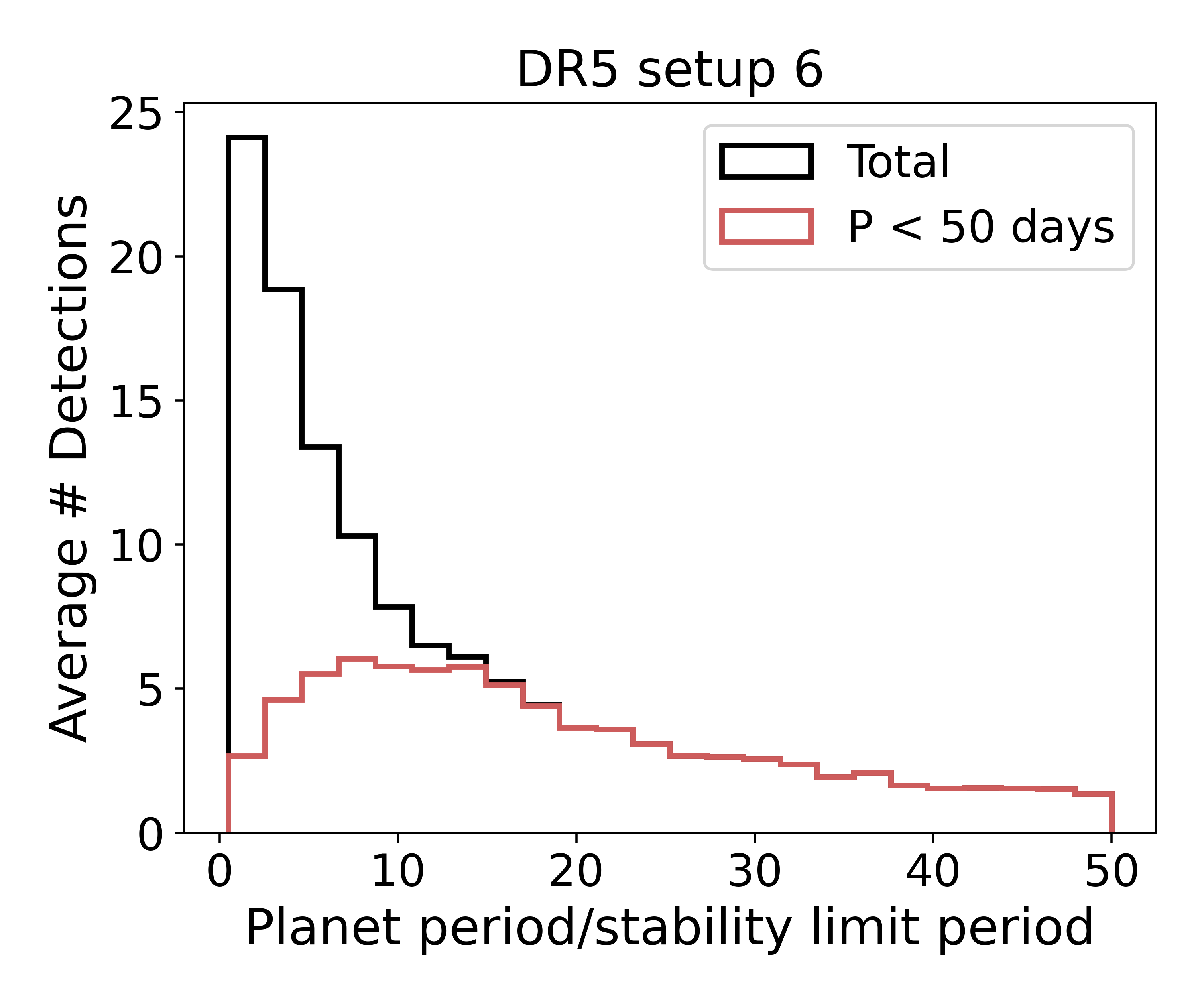}
    \caption{Distributions of planet's scaled orbital period for predicted \gaia circumbinary planet detections in the \edit{S}etups 1, 4, 5, and 6 (top-left,top-right,bottom-left,bottom-right). The other distributions are those restricted to binaries with \(P_{\rm bin}<50\,{\rm days}\).}
    \label{fig:Gaia_Pstabs}
\end{figure*}

Now we turn to examining how the circumbinary planets' orbital separation distribution affects yields. \edit{While astrometry favours the detection of planets on longer-period orbits,} the more planets are injected on short-periods (relative to the binary), the higher the yield. The main reason for this is that there are more binaries with longer orbital periods (at least within the period range \gaia is sensitive to, see Figure \ref{fig:gaia_bins} where the binary frequency peaks around \(10^5\) days). In \edit{S}etup 4, where all the planets are placed at the stability limit, this results in a higher rate of detectable planets around the longer period binaries, and since such binaries are both more common and longer orbital periods are also more easily detectable with \gaia, it results in a higher yield. \gaia therefore has an observing bias towards finding circumbinary planets nearer the stability limit, but it is a bias inherent to the binary population as well as which binaries are accessible astrometrically. Table \ref{tab:GaiaYieldsBintype} demonstrates that while shifting the planets' period distribution nearer the binary (e.g. comparing \edit{S}etup 1 to \edit{S}etup 4) drastically increases the total yield, it has the opposite effect on the yield of planets orbiting the short-period binaries. This bias is also visible in Figure \ref{fig:Gaia_Pstabs}, which plots the distributions of scaled orbital period for \edit{S}etups 1, 4, 5 and 6. The bottom-right panel of Figure \ref{fig:Gaia_Pstabs} displays the results from \edit{S}etup 6 where the inject \(a_{\rm sc}\) distribution is Uniform. The orbital periods of the planets detected orbiting binaries with \(P_{\rm bin}<50\,{\rm days}\) is closer to Uniform, whereas those detected orbiting astrometric binaries is strongly biased towards the stability limit.

\begin{figure}
    \centering
    \includegraphics[width=0.8\columnwidth]{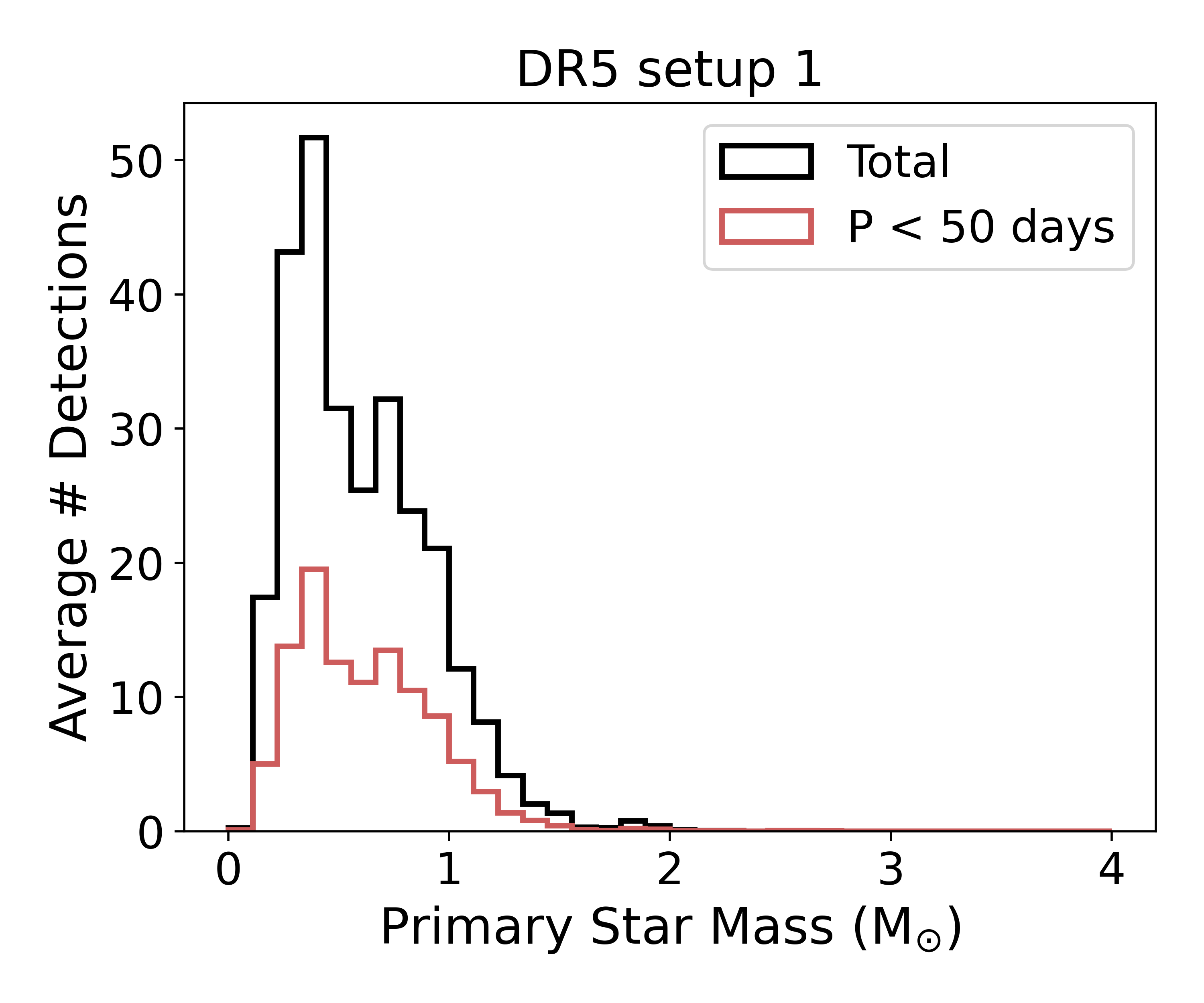}
    \includegraphics[width=0.8\columnwidth]{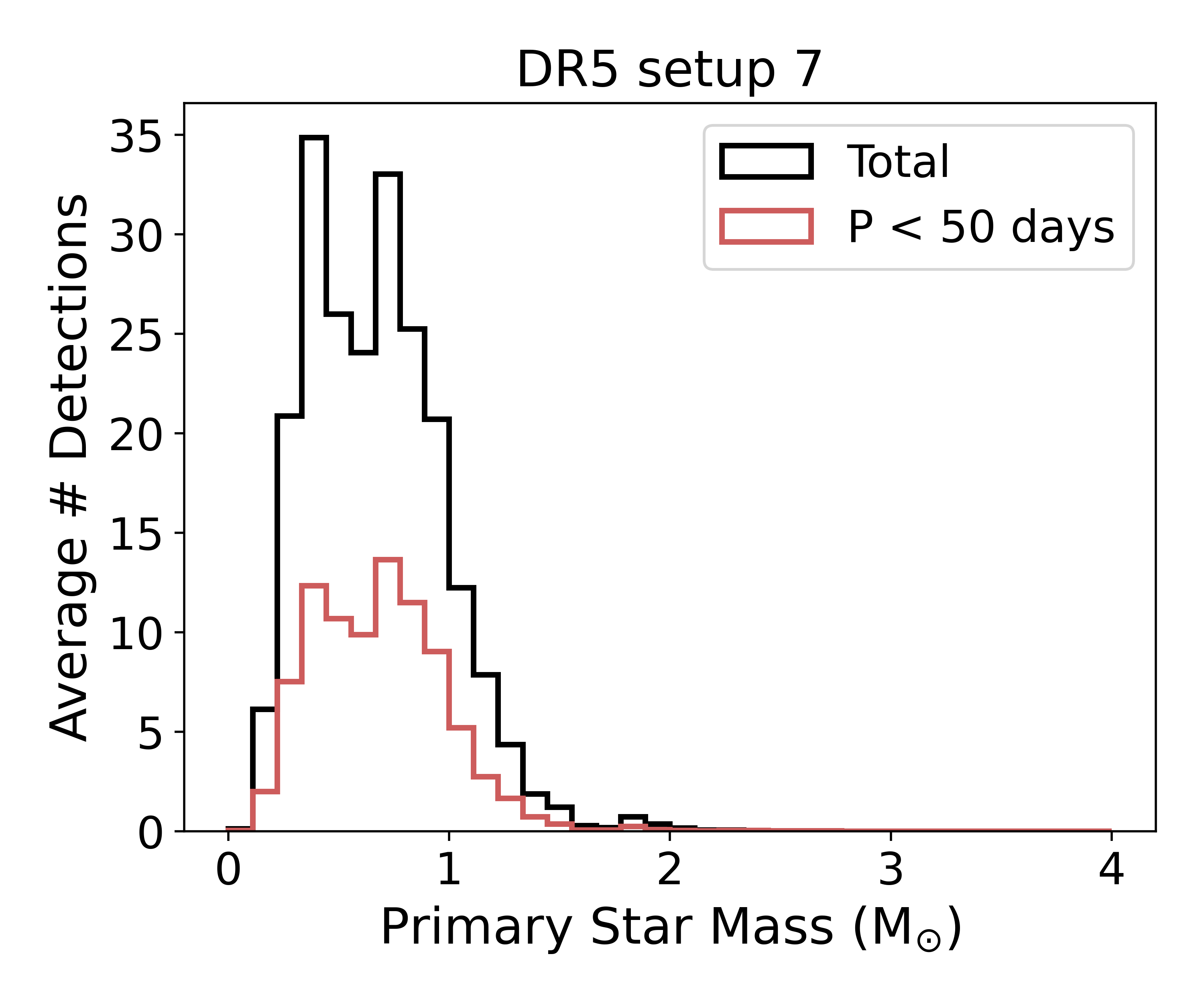}
    \caption{Distributions of the primary star mass for the systems in which circumbinary planets are detected in the fiducial \edit{S}etup 1 (top) and in \edit{S}etup 7 (bottom) where the planet mass scales with mass of the binary star. The other distributions are those restricted to binaries with \(P_{\rm bin}<50\,{\rm days}\).}
    \label{fig:Gaia_Mpris}
\end{figure}

The final \edit{comparison} is \edit{to} \edit{S}etup 7, that in which the masses of the injected planets are scaled linearly with the mass of the inner binary (those around \(M_{\rm bin} = 1\,M_{\odot}\) are kept the same mass). The main difference is displayed in Figure \ref{fig:Gaia_Mpris} which compares the masses of the primary stars in the detected systems. \edit{While in Setup 1 the main peak is for M-dwarf binaries, when scaling the planet mass for Setup 7 the peak for M-dwarfs and FGK-dwarf primaries are comparable in size, the yield orbiting M-dwarfs having been reduced.}

\subsection{Discussion of injection yield}\label{sec:yielddisc}

The injection we perform most similar to that from \citet{sahlmann_gaias_2015} is \edit{S}etup 4, where all the planets are placed near the detection limit. While we obtain a yield of \editt{222} for \gaia DR4, \citet{sahlmann_gaias_2015} obtained a yield of 516, slightly over two times larger. This is likely the effect of our synthetic populations containing fewer high mass planets. \edit{The shape of the mass distributions of detected planets is similar between \citet{sahlmann_gaias_2015} and our results. Their distribution peaks at \(10\,{\rm M_{Jup}}\) where ours peak between \(5-10\,{\rm M_{Jup}}\), another effect of our population containing fewer high-mass planets to begin with. In both cases the rate decreases towards higher mass. While \citet{sahlmann_gaias_2015} chose an upper limit of \(30\,{\rm M_{Jup}}\) and we chose \(25\,{\rm M_{Jup}}\), this difference is not significant enough to explain the differences between the yields; to meaningfully increase our yield, the upper-mass limit would need to be significantly higher than \(25\,{\rm M_{Jup}}\).}

While it is possible to claim the results from the fiducial \edit{S}etup 1 as the predicted yield of circumbinary planets from \gaia, and we do investigate the results of \edit{S}etup 1 below, we will not make a strong claim of this being "the prediction" from our work. The circumbinary planet population is not well enough known to say which of the various setups is most likely. The most important result from this set of tests is that the different setups led to large variations in the yield and in the way the yield manifested around the binaries of different orbital periods. This range of outcomes strongly suggests that the results from \gaia will be able to constrain the population distributions of circumbinary planets, especially the massive planets \(M>{\rm M_{Jup}}\) and planets orbiting longer-period binaries. Even in the most pessimistic injection \gaia DR4 still yields more than 40 circumbinary planets, a comparable number to even the most optimistic count of currently known circumbinary exoplanets.

\begin{figure}
    \centering
    \includegraphics[width=\columnwidth]{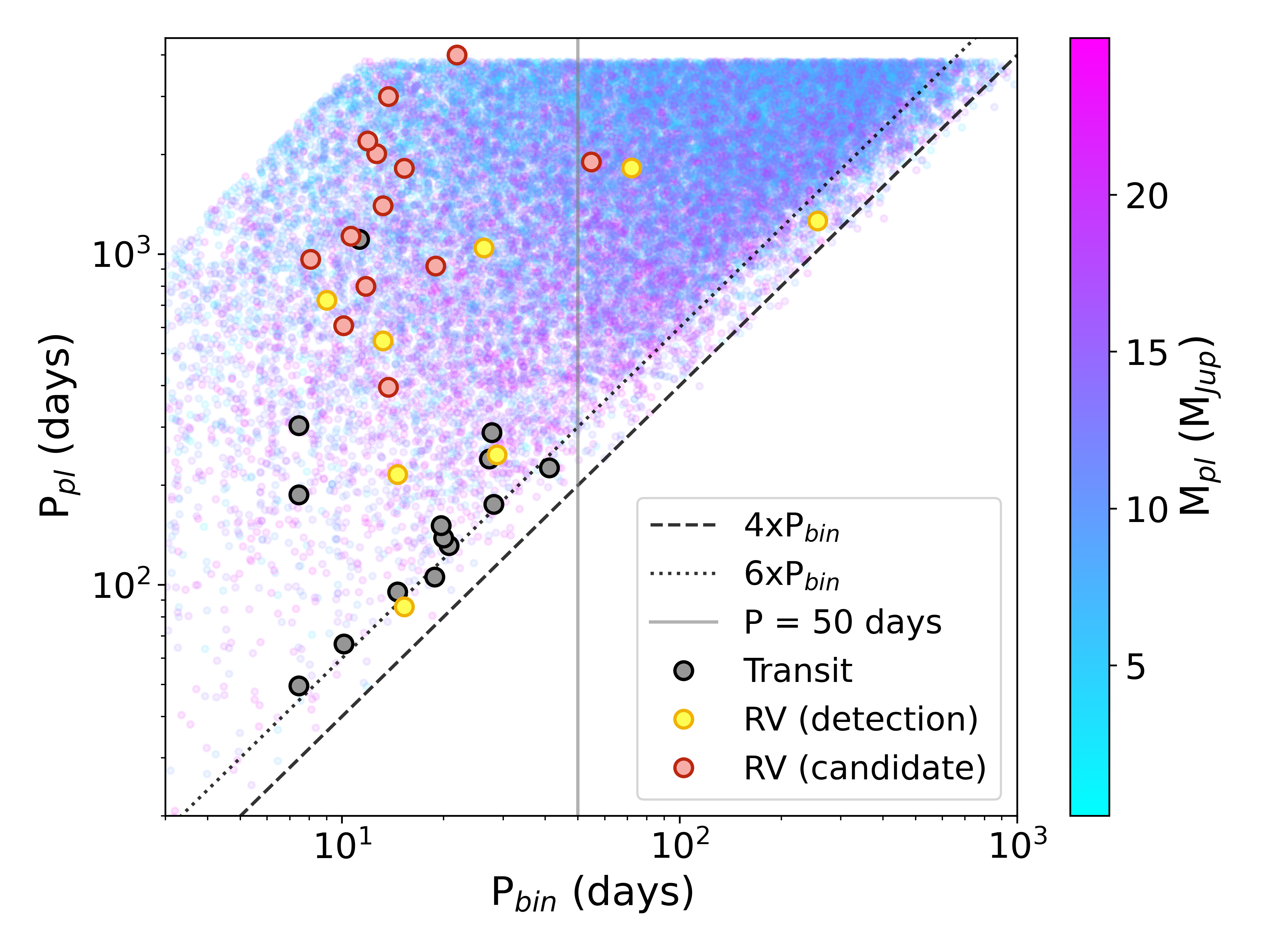}
    \caption{Binary and Planet periods for the detected planets in the synthetic yield of \edit{S}etup 1 (coloured by planet mass) alongside known circumbinary planets detected by transit or radial velocities. Dashed and dotted lines show the approximate stability threshold at \(4\times P_{\rm bin}\) and approximate location of the "pile-up" from the transiting systems at \(6\times P_{\rm bin}\). The vertical lines shows the 50 day marker that we use to distinguish the type of binaries most studied by the transit and radial velocity samples.}
    \label{fig:PP_plot}
\end{figure}

Figure \ref{fig:PP_plot} shows the resulting distribution of planets yielded from \edit{S}etup 1 in \(P_{\rm bin}\) vs \(P_{\rm pl}\), coloured by \(M_{\rm pl}\). Note that the points in Figure \ref{fig:PP_plot} represent all simulated systems in the 100 iterations of \edit{S}etup 1; the true yield would be much less densely populated. For comparison, Figure \ref{fig:PP_plot} also includes circumbinary planets known from transits and radial velocities. The radial velocity detections and candidates are those from the BEBOP survey \citep[as laid out in][]{baycroft_tools_2025} as well as HD202206 \citep[][see discussion of this system in Section \ref{sec:known_ms}]{benedict_hd_2017}. We see that most of the known planets orbit a binary with orbital period \(P_{\rm bin} < 50\,{\rm days}\). The "pile-up" of transiting circumbinary planets around \(6\times P_{\rm bin}\) is also visible.

The majority of the planets detected are, as mentioned above, longer period planets orbiting longer period binaries, with the distribution decreasing as either of these is decreased. The hard limit in Figure \ref{fig:PP_plot} at \(P_{\rm pl} \approx 4000\,{\rm days}\) is due to the limited \gaia timespan. The empty triangle in the upper-left corner is due to the injected distribution of \(a_{\rm sc}\) being bounded by \(a_{\rm sc} < 20\), therefore not allowing for very long period planets around very short period binaries. While this does potentially lower the yield estimates slightly, this is a small distinction and we prefer to keep a consistent \(a_{\rm sc}\) across different binary periods. On top of this, in the sample of eclipsing binaries from \textit{Kepler} there appears to be a dearth of transiting circumbinary planet orbiting binaries with \(P_{\rm bin} \lesssim 8 \,{\rm days}\). The reasons for this are not fully understood though a few hypotheses have been posited \citep{munoz_survival_2015,martin_no_2015,xu_disruption_2016,mogan_concealing_2025,farhat_capture_2025}. It is possible that planets do exist orbiting the shortest period binaries, just at long orbital periods and/or on misaligned orbits which \gaia could detect, helping to resolve this open question.

The majority of detections from our synthetic population correspond to planets different from most of the known population from transits or radial velocities, i.e. the binaries they orbit have a longer orbital period. Three known circumbinary planets from radial velocities have \(P_{\rm bin} > 50\,{\rm days}\), but also some of those found by microlensing \citep[e.g.][]{kuang_ogle-2019-blg-1470labc_2022,han_ogle-2023-blg-0836l_2024} appear to be in this regime, though since orbital periods are not measured we do not show them in Figure \ref{fig:PP_plot}. Since the injected mass and period distributions are based on a population of planets orbiting binaries with \(P_{\rm bin} < 50\,{\rm days}\), it is likely that - despite some planets certainly existing in this regime - extrapolating this to those with \(P_{\rm bin} > 50\,{\rm days}\) is not correct. However, the fact that this region is where \gaia has the best sensitivity to a circumbinary planet population means that \gaia will lead towards answering the question of whether the population of circumbinary planets depends on binary orbital period. \edit{I}f there happens to be some mechanism which results in a drop in circumbinary planet occurrence at \(P_{\rm bin} > 50\,{\rm days}\), these results show that \gaia would still detect a significant number of planets in the regime of \(P_{\rm bin} < 50\,{\rm days}\) which is at the moment better understood (see Table \ref{tab:GaiaYieldsBintype}). \edit{Furthermore, since \gaia is expected to detect many planet orbiting these longer period binaries if the populations are similar, if \gaia does not detect these then this will have strong implications that there is some mechanism (possibly differences in formation paths) causing such a dearth of planets.}

\edit{In Appendix \ref{sec:SNRorbcov}, we investigate the impact of our cuts/limits in SNR, orbital coverage, and distance. The conclusions are that while the choices for these cuts are justified, the true yield of \gaia will likely be higher than we have predicted. Extending to greater distances would increase the yield of high-mass planets (\(\gtrsim 10\,{\rm M_{Jup}}\)), and decreasing the SNR threshold that of low-mass planets (\(\lesssim 10\,{\rm M_{Jup}}\)). Our choices of cuts to be conservative and include only confident detections, with the caveat that the true population of circumbinary planets is not well known, means that our predictions should be interpreted as lower-bounds on the expected yield from \gaia.}

\subsection{Results on known main-sequence binaries}\label{sec:known_ms}

Using the detection criteria above, we now assess whether \gaia will be sensitive to any of the known circumbinary planets orbiting main-sequence binaries. 

\gaia will not be sensitive to any of the transiting circumbinary planets from Kepler or TESS. The planets cause too small an astrometric signal and are too distant from the Solar system, which exacerbates this. The closest (by a large margin) is the Kepler-16 system, at \(\approx 75\) pc. Despite that, the planet in this system, Kepler-16b, is expected to only produce an SNR of \(\approx 1\) in \gaia DR4. Kepler-1660b \citep{goldberg_5mjup_2023} as well as the other candidates proposed to explain dynamically-induced eclipse timing variations in the Kepler data \citep{borkovits_comprehensive_2016}, are also unfortunately too distant to be detectable by \gaia.

Of the planets detected by radial velocities, two will be accessible to \gaia. 

The first, HD 202206 \citep{correia_coralie_2005}, is a system with a dichotomy. The radial velocity solution could be interpreted as a two-planet system with an outer Jupiter and an inner low-mass brown dwarf \(\sim 18 {\,\rm M_{Jup}}\), or as a circumbinary low mass brown dwarf \citep{correia_coralie_2005}. A dynamical analysis by \citet{couetdic_dynamical_2010} found that the circumbinary brown dwarf solution is unstable, limiting the inclination of the system \(> 15^{\circ}\) with respect to the plane of the sky. However, later astrometric measurements using the Hubble fine-guidance sensor by \citet{benedict_hd_2017} found a solution implying the orbits are inclined at \(10.9\pm0.8^{\circ}\) and \(7.7\pm1.1^{\circ}\) to the line of sight, in contradiction with the dynamical analysis. The \gaia DR4 SNR for the inner/outer bodies will be \(\approx 80/\approx70\) if both orbits are edge\edit{-}on compared to the line-of sight. If the correct solution is that proposed by \citet{benedict_hd_2017}, then the SNR will be \(\approx 400/\approx250\). \gaia will therefore be able to weigh in on this system, resolving the dichotomy once and for all.

The second known system detectable with \gaia is BEBOP-4 \citep{triaud_bebop_2025}. This is a circumbinary brown dwarf of mass \(\approx 20 {\, \rm M_{Jup}}\). The brown dwarf companion ought to have a \gaia SNR of \(\approx 40\) using DR4 alone. Other circumbinary planets detected with radial velocities so far \citep{standing_radial-velocity_2023,baycroft_progress_2024,baycroft_bebop_2025,baycroft_tools_2025} should not be detectable in \gaia data unless their orbital inclination is nearly parallel to the plane of the sky. If this were the case, which is highly unlikely, it would mean all of them were much more massive, likely brown dwarfs.

\subsection{Results on post-common-envelope binaries}

\subsubsection{Previously claimed planets}\label{sec:PCEB_planets}

Circumbinary planets and planetary-mass objects have been claimed to orbit post-common-envelope binaries based on eclipse timing variations. These are for the most part controversial, their existence being called into question. The planetary models typically do not predict future data well, and often imply multiple planets in rapidly unstable orbital configurations \citep{horner_detailed_2013,pulley_eclipse_2022}. Various magnetic field-based mechanisms have been posited as potential other explanations for the eclipse timing variations \citep[e.g.][]{applegate_mechanism_1992,lanza_internal_2020}. As stated in \citet{sahlmann_gaias_2015,baycroft_new_2023} \gaia astrometry should have the ability to help resolve the question of their existence for a handful systems such as HW Vir \citep{beuermann_quest_2012}. 

We have collated and tabulated all such claimed planets orbiting eclipsing post-common-envelope binaries that we were able to locate in the literature. This is shown in Table \ref{tab:PCEBs}. A large portion of this collection is based on a table compiled by \citet{basturk_eclipse_2023}, which we supplemented with other systems. Only systems where the third body is possibly below stellar mass (\(\lesssim 80 {\,\rm M_{Jup}}\)) are included, but others exist with a more massive tertiary object. Table \ref{tab:PCEBs} shows the parameters for each system, which we use to calculate the \gaia \SNR\,of each proposed candidate planet. Uncertainties are not provided as in many cases the mass or period of the candidate planet is uncertain, varying, or just an estimate. 

\begin{figure}
    \centering
    \includegraphics[width=0.98\columnwidth]{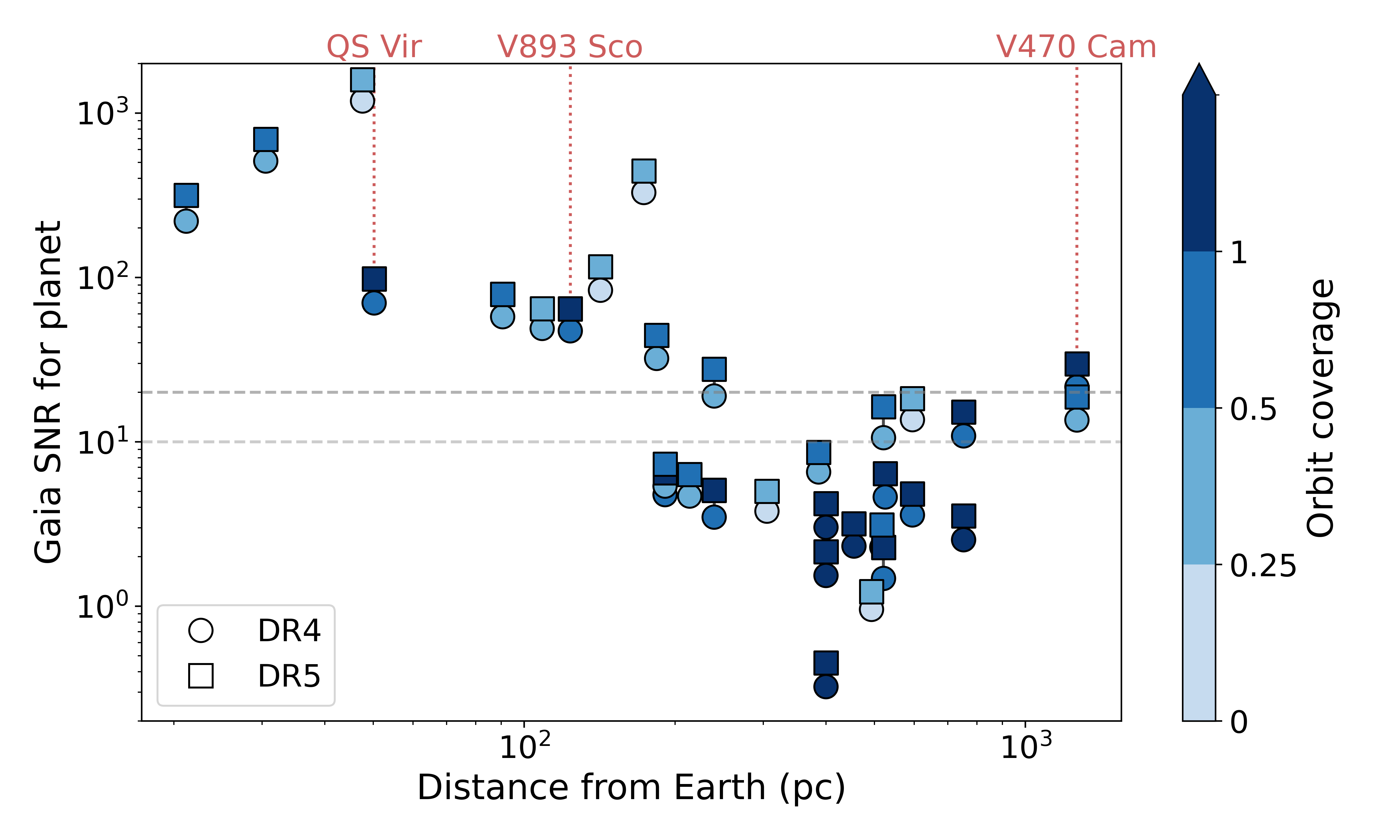}
    \caption{\gaia \SNR\,vs distance \editt{from earth} for the claimed planets orbiting post-common-envelope binaries shown in Table \ref{tab:PCEBs}. Each has two markers showing the \SNR\,for \gaia DR4 and DR5, with the \editt{colour} of the marker showing how much of the orbit of the claimed planet would have been covered. Dashed lines show the strong and weak detection thresholds of 20 and 10.}
    \label{fig:PCEB_planets}
\end{figure}

The \gaia SNR is calculated as described in Section \ref{sec:methods} for both \gaia DR4 and DR5 for each claimed planet. With these being known systems, we use the \gaia Observation Forecast Tool\footnote{\href{https://gaia.esac.esa.int/gost/index.jsp}{https://gaia.esac.esa.int/gost/index.jsp}} (GOST) to predict the number of observations with which we calculate the expected \SNR; we reduce the predicted number of observations by \(20\%\) to account for potential data loss (as recommended in GOST). The fraction of the orbit that will be covered by the dataspan of DR4 and DR5 is also calculated. The \SNR\,and coverage are shown in Table \ref{tab:PCEBs} for \gaia DR5. This information is also displayed for DR4 and DR5 in Figure \ref{fig:PCEB_planets} along with the distance to each system. Figure \ref{fig:PCEB_planets} reveals the expected trend for \gaia data that the systems with the highest amplitude signals are those that are nearest to the Solar system, with sensitivity decreasing as distance increases. It is also clear that, if they actually exist, a significant number of these claimed post-common-envelope candidate circumbinary planets would  produce large signal in \gaia data, but not many of these will have a full orbit covered. 

\begin{table*}
    \centering
    \caption{Table of post-common-envelope binaries with planetary-mass companions claimed from eclipse timing variations. Those in bold are the 11 that meet the relaxed criteria for detection with \gaia. \(^{(1)}\) There is evidence against its existence from \citet{hardy_first_2015} and \citet{baycroft_new_2023}}
    \begin{tabular}{llrrrrrr}
    \hline
    System name & Publication & M\(_{\rm bin}\) & Distance & M\(_{\rm pl}\) & P\(_{\rm pl}\) & \gaia DR5 SNR & \gaia DR5 Coverage \\
     &  & [\(M_{\odot}\)] & [pc] & [\(M_{\rm  Jup}\)] & [days] &  &  \\
    \hline
    HW Vir & \cite{baycroft_new_2023} & 0.54 & 173 & 17.8 & 15000 & 444.1 & 0.26 \\
    \textbf{DE CVn} & \cite{han_cvn_2018} & 0.92 & 31 & 11.5 & 4100 & 691.6 & 0.93 \\
    NN Ser & \cite{ozdonmez_investigation_2023} & 0.65 & 521 & 2.6 & 3000 & 2.3 & 1.28 \\
    \textbf{NN Ser} & \cite{ozdonmez_investigation_2023} & 0.65 & 521 & 10.5 & 7300 & 16.4 & 0.52 \\
    \textbf{RR Cae} & \cite{rattanamala_eclipse_2023} & 0.62 & 21 & 3.1 & 6000 & 315.8 & 0.64 \\
    2M1938+4603/Kepler-451 & \cite{esmer_detection_2022} & 0.6 & 400 & 1.9 & 43 & 0.5 & 89.13 \\
    2M1938+4603/Kepler-451 & \cite{esmer_detection_2022} & 0.6 & 400 & 2.0 & 410 & 2.1 & 9.35 \\
    2M1938+4603/Kepler-451 & \cite{esmer_detection_2022} & 0.6 & 400 & 1.7 & 1460 & 4.2 & 2.62 \\
    NSVS 14256825/V1828 Aql & \cite{wolf_possible_2021} & 0.53 & 753 & 3.0 & 1280 & 3.5 & 2.99 \\
    \textbf{NSVS 14256825/V1828 Aql} & \cite{wolf_possible_2021} & 0.53 & 753 & 8.4 & 2500 & 15.2 & 1.53 \\
    NY Vir & \cite{esmer_testing_2023} & 0.6 & 595 & 2.1 & 3200 & 4.8 & 1.2 \\
    NY Vir & \cite{esmer_testing_2023} & 0.6 & 595 & 4.2 & 8400 & 18.3 & 0.46 \\
    DP Leo & \cite{basturk_eclipse_2023} & 0.7 & 305 & 6.3 & 10500 & 5.0 & 0.36 \\
    HU Aqr & \cite{qian_detection_2011} & 1.08 & 191 & 6.3 & 2370 & 6.5 & 1.62 \\
    HU Aqr & \cite{qian_detection_2011} & 1.08 & 191 & 4.7 & 4360 & 7.3 & 0.88 \\
    \textbf{QS Vir} & \cite{basturk_eclipse_2023} & 1.28 & 50 & 6.7 & 2870 & 98.0 & 1.34 \\
    V808 Aur & \cite{leichty_eccentric_2024} & 0.95 & 214 & 7.1 & 4100 & 6.3 & 0.93 \\
    HT Cas & \cite{han_orbital_2023} & 0.7 & 142 & 13.6 & 11000 & 116.3 & 0.35 \\
    \textbf{V470 Cam} & \cite{sale_eclipse_2020} & 0.61 & 1267 & 28.3 & 2870 & 29.8 & 1.34 \\
    \textbf{V470 Cam} & \cite{sale_eclipse_2020}  & 0.61 & 1267 & 12.4 & 4840 & 18.8 & 0.79 \\
    KIC 10544976 & \cite{basturk_eclipse_2023} & 1.0 & 517 & 13.6 & 6000 & 3.1 & 0.64 \\
    \textbf{OY Car} & \cite{basturk_eclipse_2023} & 0.75 & 91 & 8.9 & 5110 & 79.0 & 0.75 \\
    V2051 Oph & \cite{basturk_eclipse_2023} & 0.93 & 109 & 7.6 & 7880 & 64.5 & 0.49 \\
    \textbf{V893 Sco} & \cite{basturk_eclipse_2023} & 0.88 & 124 & 10. & 3720 & 64.4 & 1.03 \\
    DV Uma & \cite{basturk_eclipse_2023} & 1.3 & 387 & 27.2 & 6420 & 8.6 & 0.6 \\
    FBS 1531+381 & \cite{basturk_eclipse_2023} & 0.49 & 525 & 1.5 & 3830 & 6.4 & 1.0 \\
    \textbf{SDSS J143547.87+373338.5} & \cite{wolf_possible_2021} & 0.71 & 184 & 17.8 & 4770 & 44.5 & 0.8 \\
    GK Vir & \cite{basturk_eclipse_2023} & 0.62 & 493 & 1.0 & 8760 & 1.2 & 0.44 \\
    UZ For & \cite{basturk_eclipse_2023} & 0.84 & 239 & 3.6 & 2120 & 5.1 & 1.81 \\
    \textbf{UZ For} & \cite{basturk_eclipse_2023} & 0.84 & 239 & 10.5 & 5400 & 27.7 & 0.71 \\
    KIC 3853259 & \cite{borkovits_comprehensive_2016} & 2.0 & 455 & 10.5 & 330 & 3.2 & 11.61 \\
    V471 Tau & \cite{guinan_best_2001}\(^{(1)}\) & 1.61 & 48 & 41.9 & 11130 & 1594.6 & 0.34 \\
    \hline
    \end{tabular}
    \label{tab:PCEBs}
\end{table*}

The restriction that the SNR cut should be 20 and that one whole orbit is covered (as used otherwise) can in principle be relaxed since these planet candidates have known/predicted orbits which can be tested against. \editt{T}hree claimed planets would meet the strict criteria in \gaia DR5, \editt{these are: QS Vir \citep{qian_giant_2010}, V470 Cam \citep{sale_eclipse_2020}, and V893 Sco \citep{basturk_eclipse_2023}, these are labelled in Figure \ref{fig:PCEB_planets}.} 11 claimed planets would meet \editt{more} relaxed criteria of \({\rm SNR}>10\) and an orbital coverage above \(0.5\). These 11 planets are highlighted in bold in Table \ref{tab:PCEBs}. The most strongly constrained system will be QS Vir. It is unclear exactly how well \gaia astrometry will constrain partial orbits, especially in the presence of existing eclipse timing data; this should be investigated before or during the application of \gaia data to these systems. \editt{\gaia will therefore confirm or refute at least three of these planets, potentially weighing in on more.}

\subsubsection{Sensitivity to a population}

These claimed planets are orbiting post-common-envelope binaries that have had eclipses measured over long timespan (typically in the 10s of years). More post-common-envelope binaries exist where planets have not been claimed, or that do not eclipse. We analyse the sensitivity of \gaia to planets around a sample of post-common-envelope binaries. The catalogue of potential post-common-envelope binaries is taken from \citet{kruckow_catalog_2021}. All binaries in this catalogue have orbital periods \(P_{\rm bin} \leq 100\) days, and are therefore amenable for characterisation and to be searched for circumbinary companions with \gaia. The original 848 from the catalogue are filtered to only conserve those with Gmag values \editt{and \gaia DR3 IDs} recorded in \textit{Simbad}\footnote{\href{https://simbad.u-strasbg.fr/simbad/sim-fid}{https://simbad.u-strasbg.fr/simbad/sim-fid}} and with measured masses for primary and secondary, \edit{only a value for the masses is needed, even if there are no uncertainties}. The filtering leaves \editt{426} systems.

\begin{figure}
    \centering
    \includegraphics[width=\columnwidth]{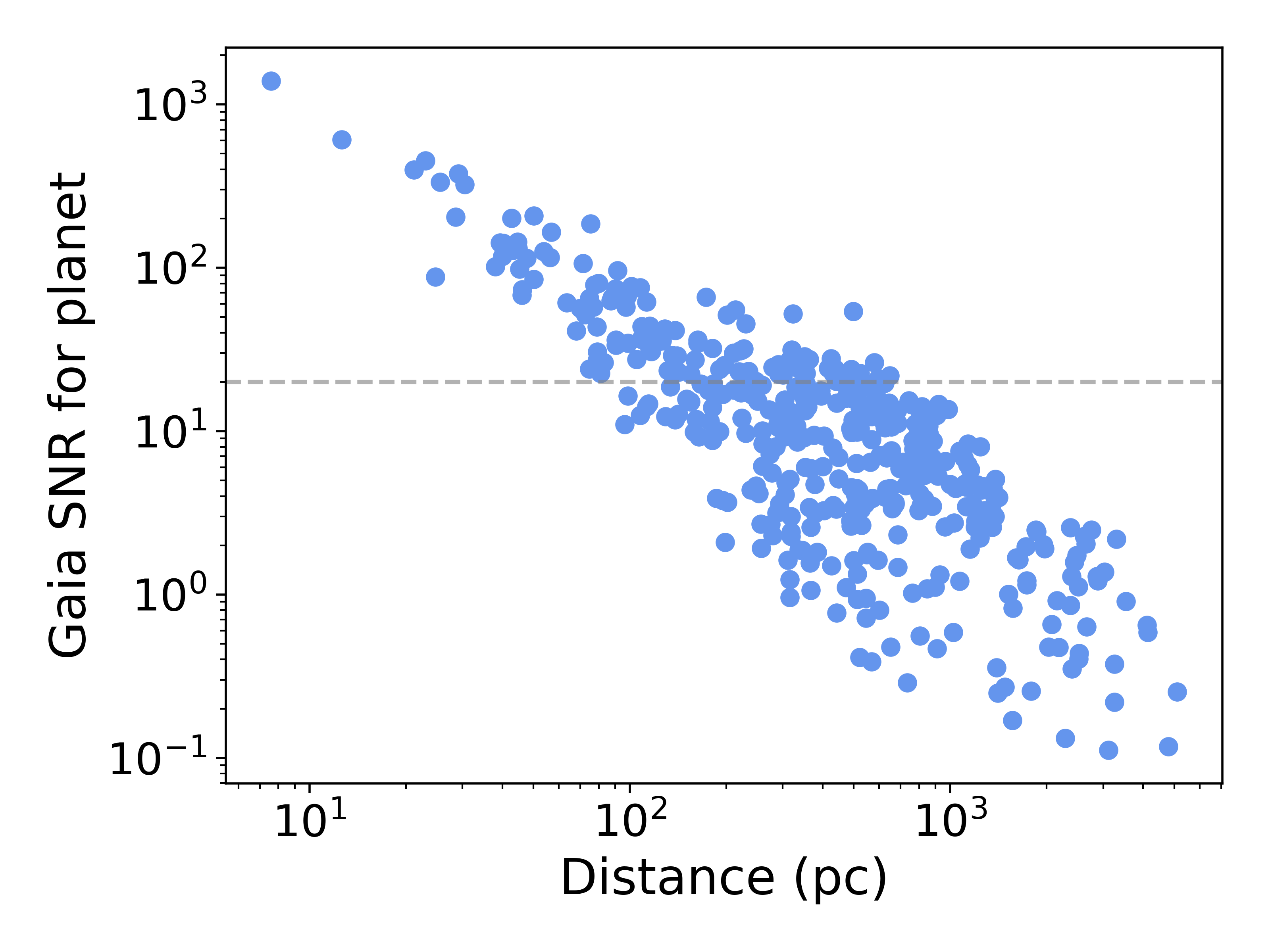}
    \caption{The \gaia \SNR\,that would be achieved by a \(6.5\,M_{\rm Jup}\) planet on a 3000 day orbit for each system in the post-common-envelope binary catalogue. Horizontal dashed line shows the detection sensitivity threshold of \({\rm SNR}=20\).}
    \label{fig:PCEB_single}
\end{figure}

\begin{figure}
    \centering
    \includegraphics[width=0.9\columnwidth]{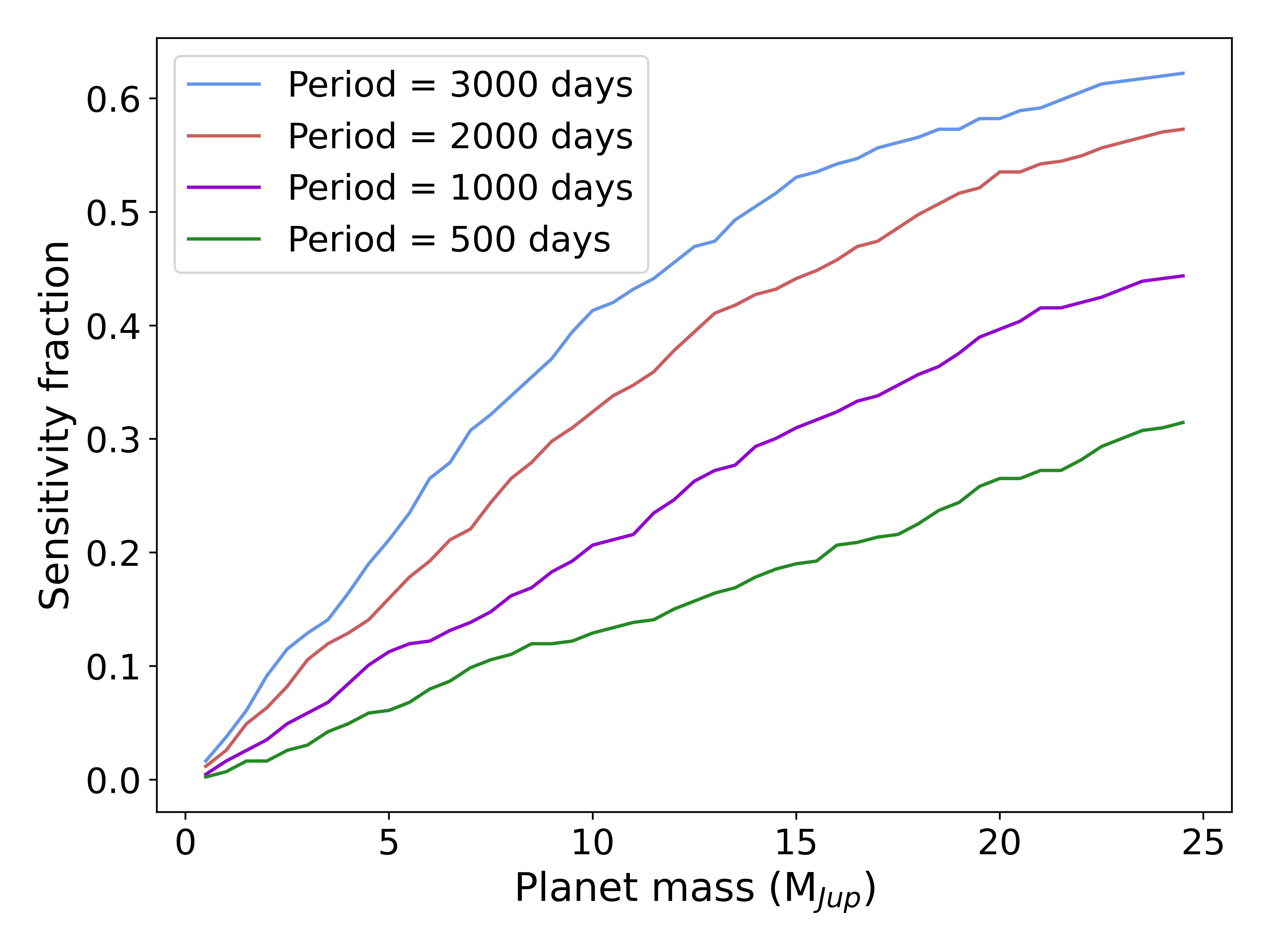}
    \includegraphics[width=0.92\columnwidth]{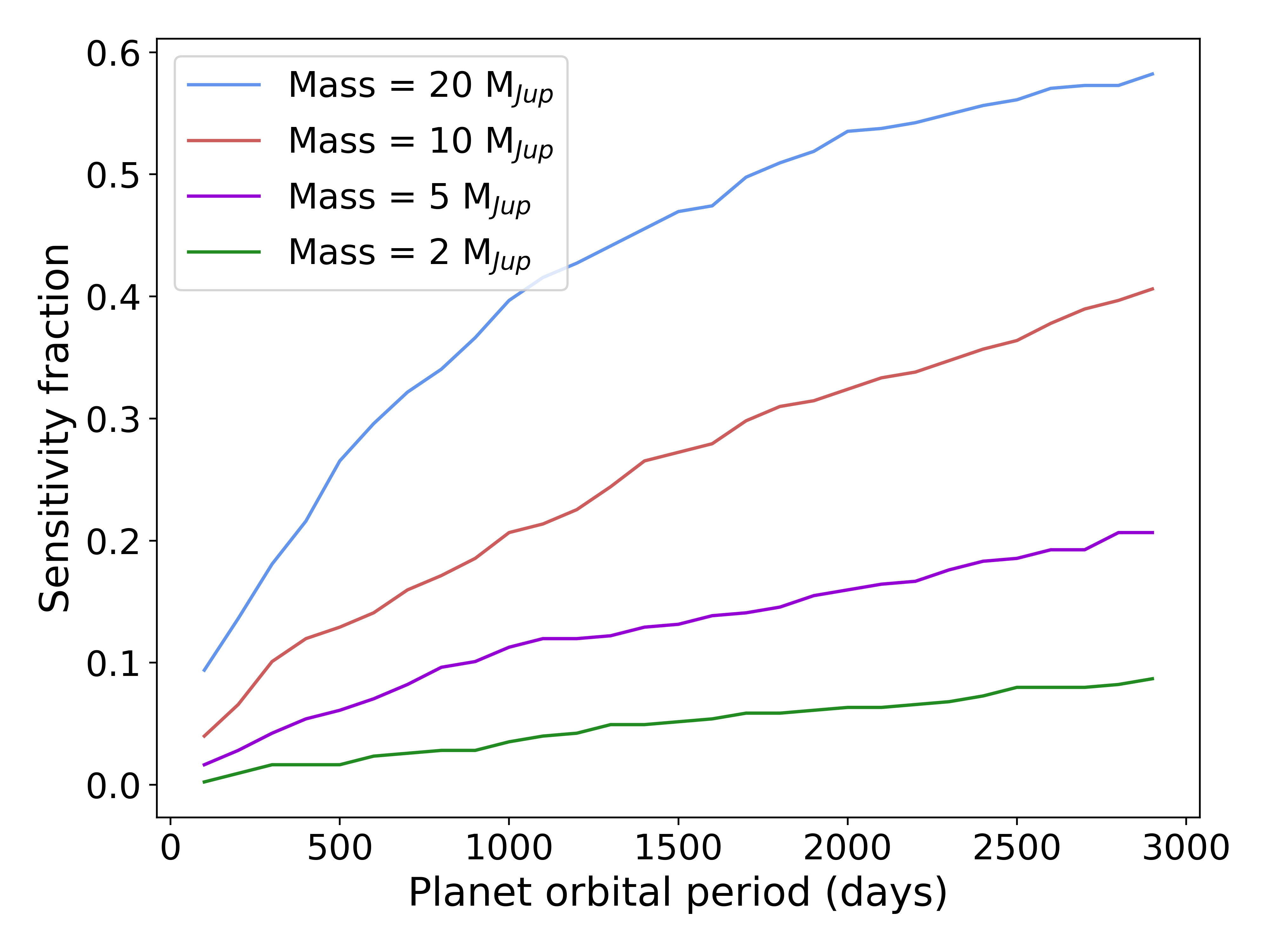}
    \caption{\gaia DR5 sensitivity to planets of varying mass and orbital period for the catalogue of post-common-envelop binaries. Top: range of masses for 4 different periods. Bottom: range of periods for 4 different masses.}
    \label{fig:PCEB_catalogue}
\end{figure}

Figure \ref{fig:PCEB_single} graphically presents the \gaia \SNR\,that would be achieved by a \(6.5\,M_{\rm Jup}\) planet on a 3000 day orbit for each system in the catalogue. These are the approximate parameters of the claimed planet orbiting QS Vir (see Table \ref{tab:PCEBs}). In this case, there would be sensitivity to such a planet in \editt{119} out of the \editt{426} systems (\(28\%\)). Sensitivity here is calculated strictly as earlier \editt{since this is a blind sample without candidate planets}, with a \SNR\,cut of 20. Repeating this calculation for a series of different masses and orbital periods produces Figure \ref{fig:PCEB_catalogue}. Only periods less than the timespan of \gaia DR5 are considered.

The results in Figure \ref{fig:PCEB_catalogue} reveal that a large population of planetary-mass circumbinary objects existing orbiting post-common-envelope binaries would be detectable in this sample. Therefore \gaia and in particular \gaia DR5 will be able to constrain the population of high-mass planets orbiting post-common-envelope binaries, and assess how common systems such as many of those claimed planets in Table \ref{tab:PCEBs} actually are.

If planets are found to truly exist orbiting post-common-envelope binaries, there is then the question of whether these are planets that survived the evolution of the binary star, or planets that formed in a second generation disc \citep{mustill_main-sequence_2013}. The masses of the claimed companions seen in Table \ref{tab:PCEBs} are very large, mostly in the brown dwarf regime. Based on \citet{triaud_bebop_2025}, circumbinary brown dwarf companions do exist orbiting main-sequence binaries, but it is difficult to say yet whether in high enough numbers to produce the full population claimed orbiting post-common-envelope binaries. \gaia will likely constrain the occurrence of circumbinary brown dwarfs orbiting main-sequence binaries and enable this comparison to be made. As seen above, \gaia DR5 will be sensitive to \edit{between 1/10 and} 1/3 of the claimed planets orbiting post-common-envelope which \edit{may be} enough to test how reliable the eclipse timing variations are at detecting planets. \edit{If enough of these 11 can be confirmed or refuted, this could provide} conclusions on the overall reliability of eclipse timing variation method for post-common envelope binaries.

\section{Conclusions}\label{sec:conclusions}

The next \gaia data releases (DR4 and DR5) will have a powerful impact on our understanding of circumbinary planets. A volume-limited analysis of binaries within 200pc can be expected to yield in the 10s to 100s of circumbinary planets in \gaia. \edit{With conservative constraints we find our yields to likely be lower limits}. We can compare these \edit{approximate yields} to the predicted yields of planets orbiting single-stars, though it is worth noting that the studies for single stars did not cut at 200pc as we did, and \edit{since} we have used very strict criteria for detectability, so this comparison may not be completely fair. Compared to the projected numbers of thousands and tens of thousands of exoplanet discoveries around single stars for DR4 and DR5, respectively,
\citep[e.g.][]{perryman_astrometric_2014,lammers_exoplanet_2025}, the yield we produce may appear small. However, so little is known of the properties of the circumbinary exoplanet population, that we expect \gaia \edit{may} alter the view we have of these types of systems. The population of circumbinary planets that \gaia will detect orbiting main-sequence binaries is expected to constrain the distribution of mass and orbital separation of circumbinary planets\edit{, and how this manifests around binaries of different orbital periods}.

\gaia will provide additional insight into many known or candidate circumbinary systems. In particular, \gaia will be able to confirm or refute \edit{some} of the claimed exoplanets orbiting post-common-envelope binaries, and will assess whether such planets exist orbiting other post-common-envelope binaries with no currently proposed planetary companions.

\section*{Acknowledgements}

The research leading to these results is supported by a grant from the ERC/UKRI Frontier Research Guarantee programme (EP/Z000327/1/CandY).
This work has made use of data from the European Space Agency (ESA) mission
{\it Gaia} (\url{https://www.cosmos.esa.int/gaia}), processed by the {\it Gaia}
Data Processing and Analysis Consortium (DPAC,
\url{https://www.cosmos.esa.int/web/gaia/dpac/consortium}). Funding for the DPAC
has been provided by national institutions, in particular the institutions
participating in the {\it Gaia} Multilateral Agreement.

\section*{Data Availability}

\gaia data can be found at \url{https://www.cosmos.esa.int/web/gaia-users} and the \gaia GOST data at \url{https://gaia.esac.esa.int/gost/}.

This work has made use of the following software: matplotlib \citep{hunter_matplotlib_2007}, numpy \citep{harris_array_2020}, astropy \citep{astropy_collaboration_astropy_2022}, and astroquery \citep{ginsburg_astroquery_2019}



\bibliographystyle{mnras}
\bibliography{GaiaCBPpred} 




\appendix

\section{Astrometric precision}

\begin{figure}
    \centering
    \includegraphics[width=\linewidth]{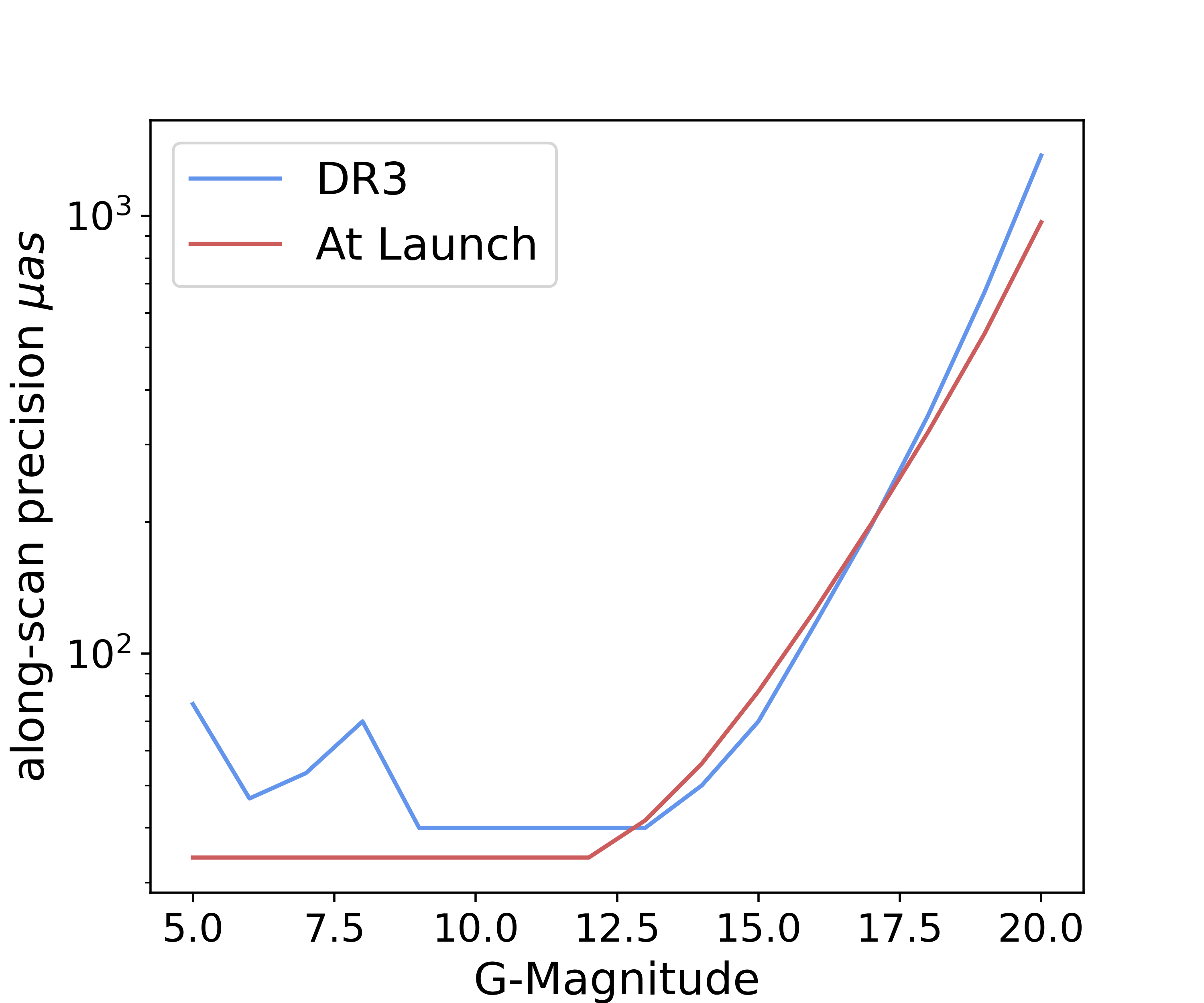}
    \caption{Approximation the the uncertainty curve from DR3 in Fig. 3 of \citet{holl_gaia_2023}, compared to pre-launch prediction following \citet{de_bruijne_gaia_2014}.}
    \label{fig:prec_approx}
\end{figure}

\edit{In Figure \ref{fig:prec_approx} we show the approximation that we use for the \gaia along-scan precision per epoch.}

\section{Impact of SNR and orbital coverage}\label{sec:SNRorbcov}

\edit{In this section we investigate the impact that our choices of limiting to SNR\(\geq 20\); that the orbit has to be fully covered have on the results; and to limit the study to a volume of 200 pc.}

\begin{figure*}
    \centering
    \includegraphics[width=0.8\linewidth]{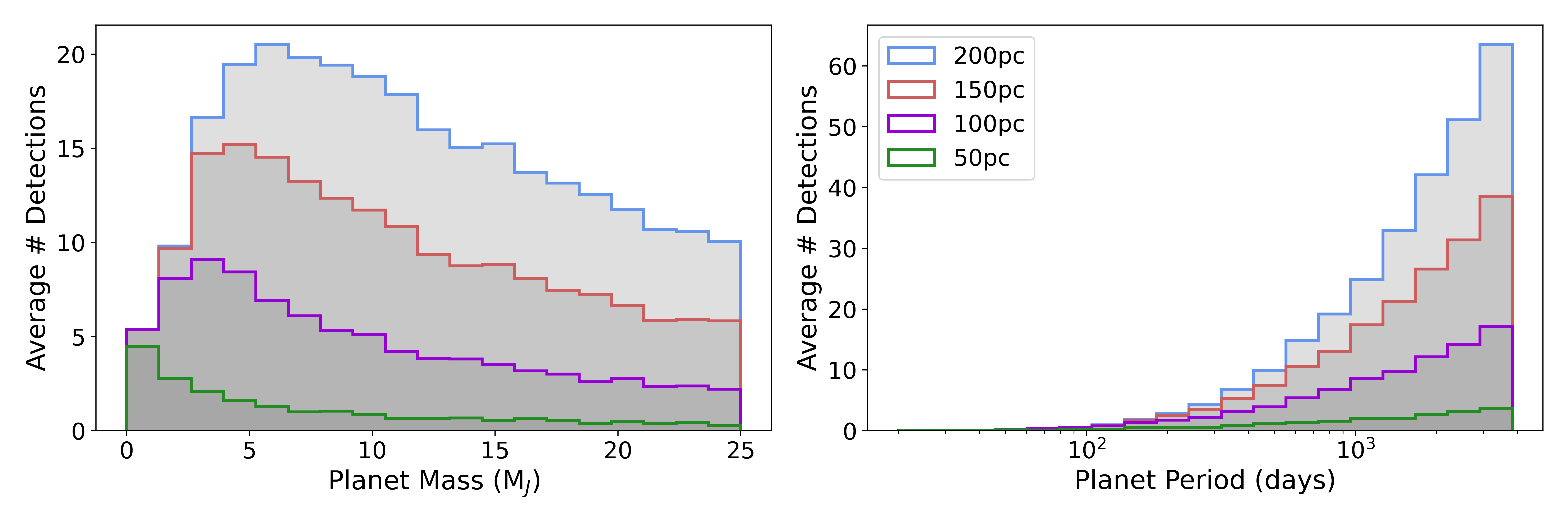}
    \includegraphics[width=0.8\linewidth]{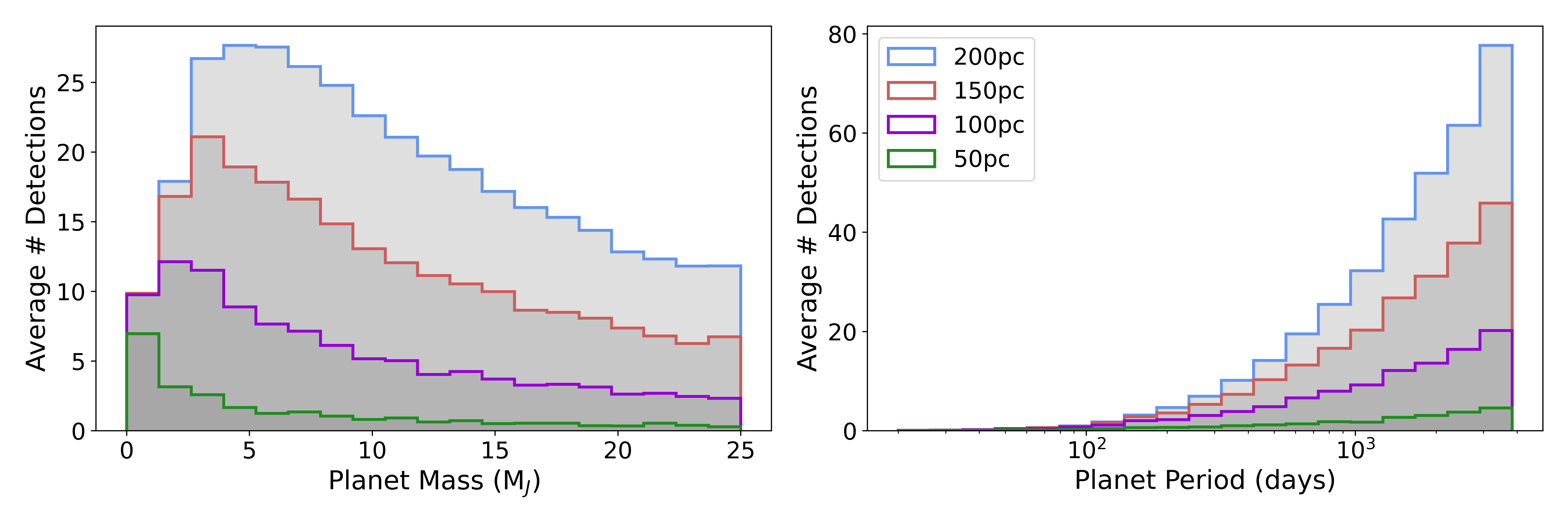}
    \includegraphics[width=0.8\linewidth]{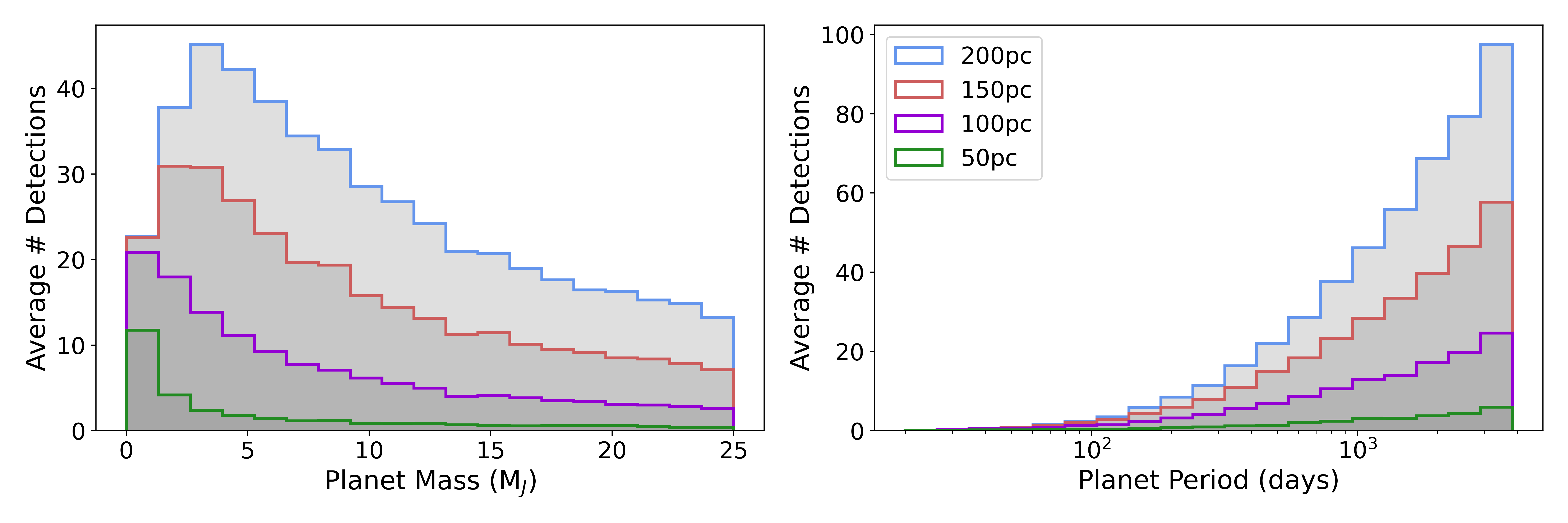}
    \caption{\edit{Distribution of detected planet masses (left) and periods (right) for \gaia DR5, separated into the nested volumes of radius 50-200 pc from Earth. The three versions use different values as the threshold in SNR to class as a detection: top - SNR\(\geq20\), middle - SNR\(\geq15\), bottom - SNR\(\geq10\). This is calculated using Setup 1.}}
    \label{fig:SNRexamples}
\end{figure*}

\begin{figure*}
    \centering
    \includegraphics[width=0.4\linewidth]{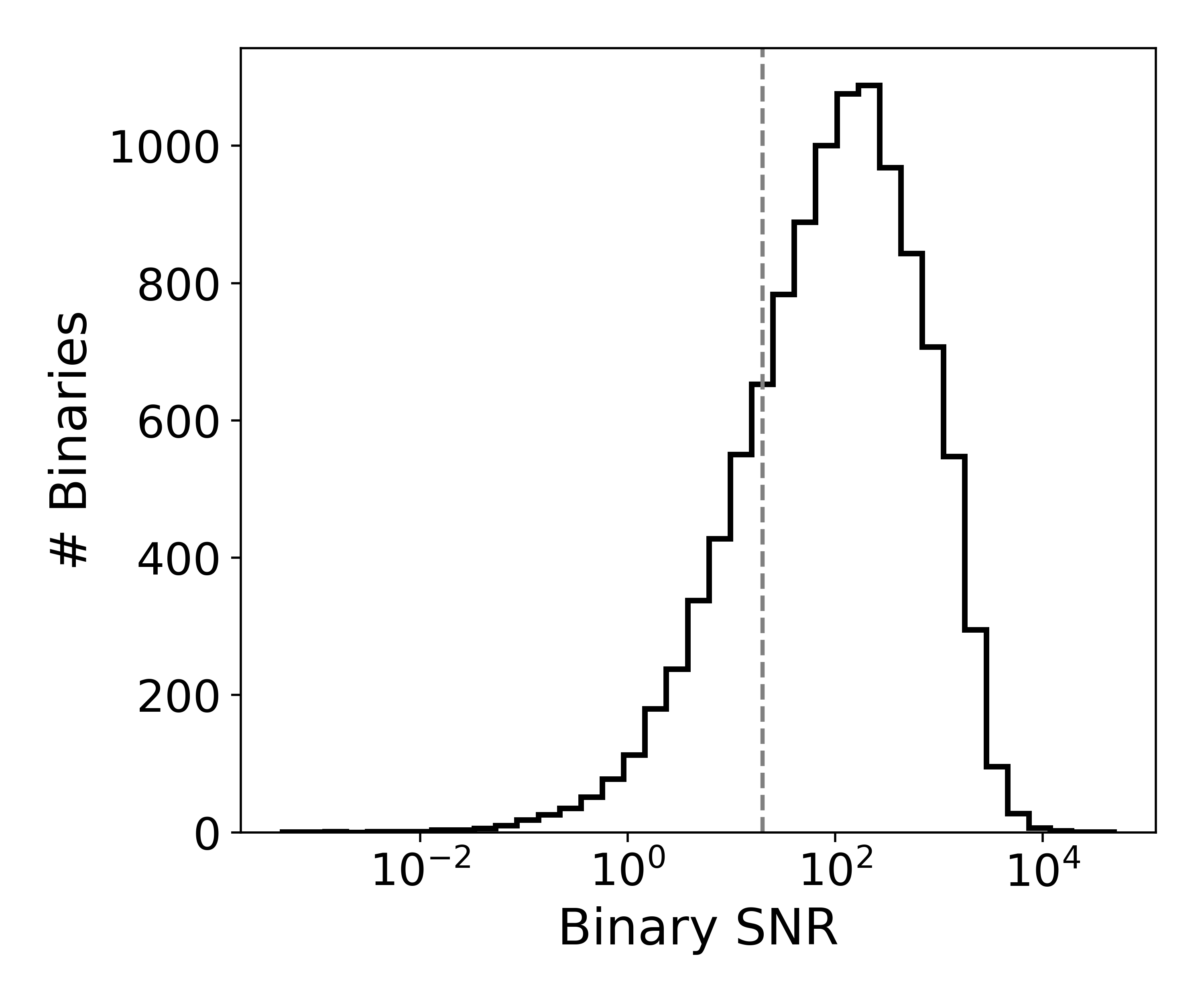}
    \includegraphics[width=0.47\linewidth]{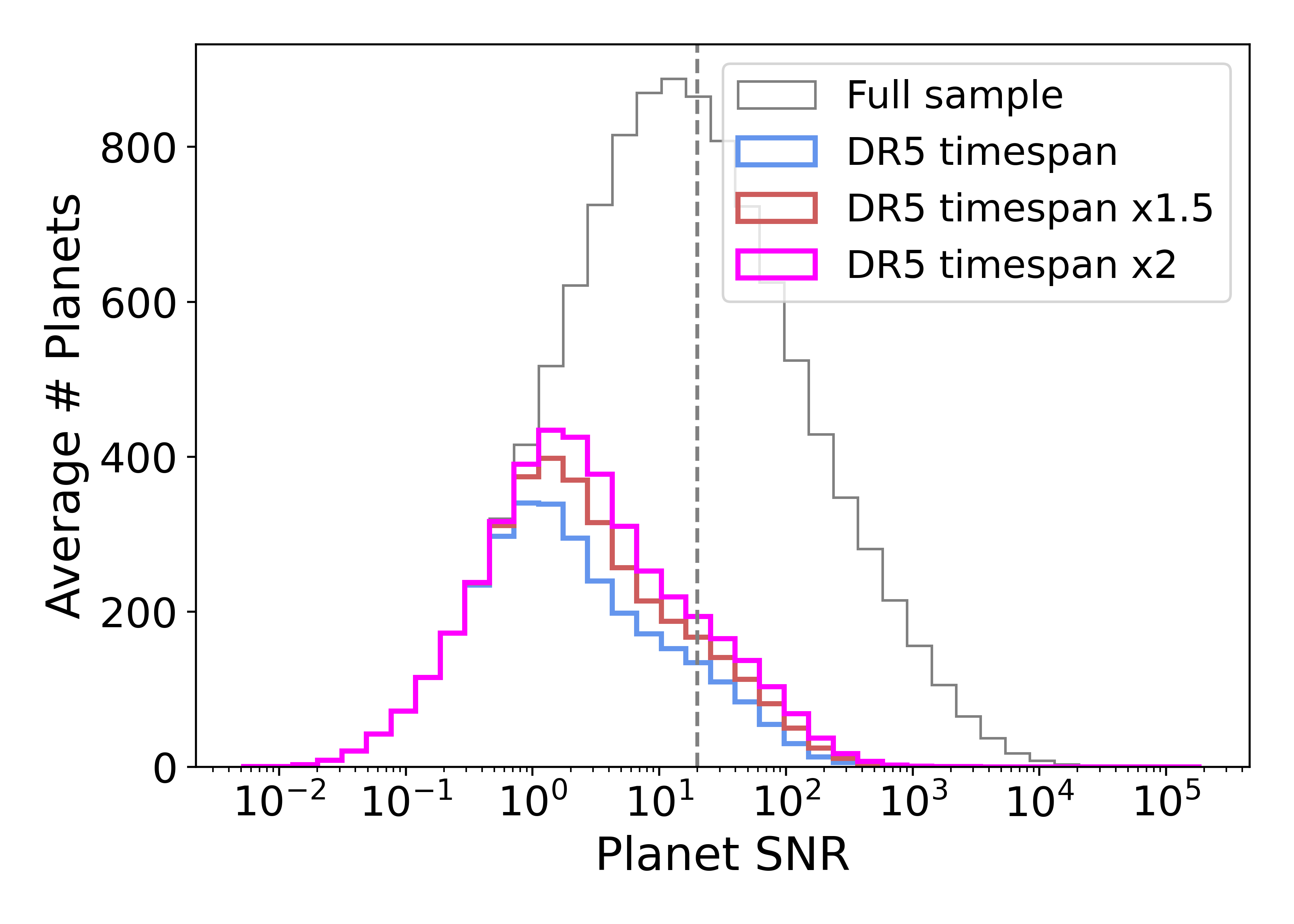}
    \includegraphics[width=0.44\linewidth]{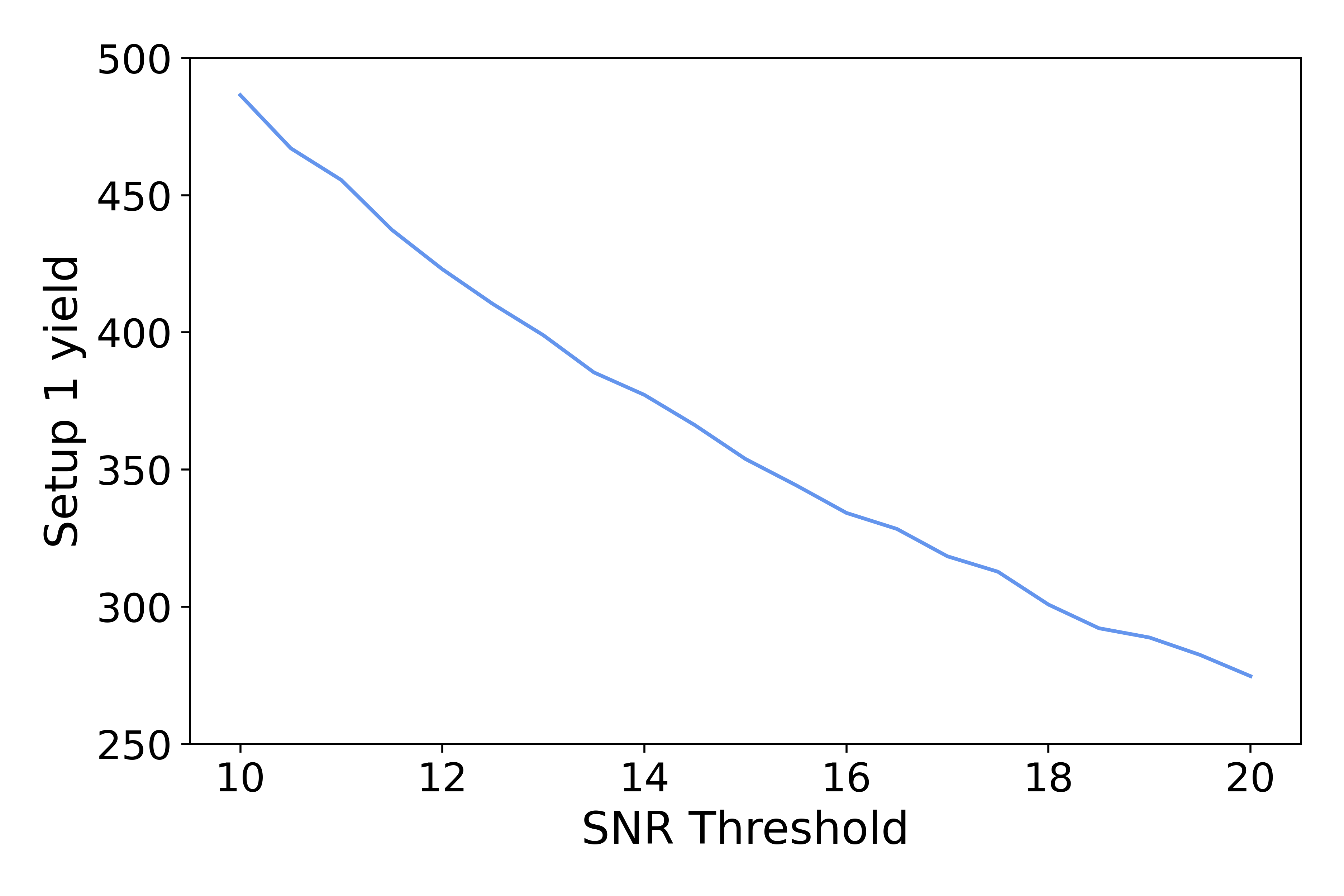}
    \includegraphics[width=0.43\linewidth]{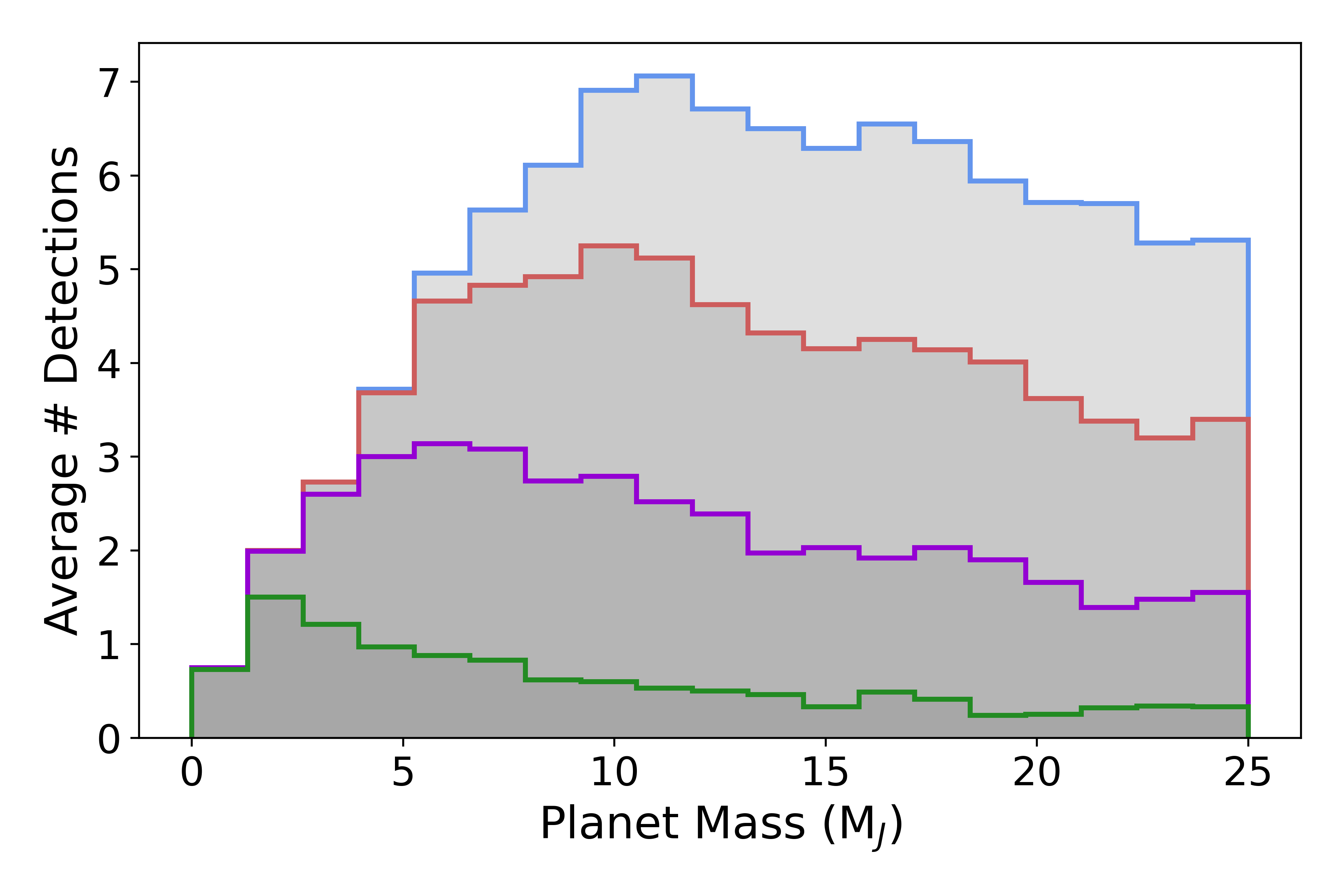}
    \caption{\edit{Top: distribution of SNRs in the sample, for binaries (left) and planets (right), the dashed vertical line shows the threshold of SNR\(=20\). For the planet distribution (top-right) the distributions of those planets whose period is less than the \gaia DR5 timespan (also \(\times1.5\) and \(\times2\)) are shown. Bottom left: Yield in Setup 1 for SNR thresholds between 10-20. Bottom right: distribution of detected planet masses for \gaia DR4, separated into the nested volumes of radius 50-200 pc from Earth.}}
    \label{fig:fourpanel}
\end{figure*}

\edit{First we investigate the impact of the SNR cut. Figure \ref{fig:SNRexamples} shows the mass and period distribution of Setup 1 (separated into distance bins) for different SNR thresholds (10, 15, and 20) for \gaia DR5. The bottom left panel of Figure \ref{fig:fourpanel} shows how the yield varies between thresholds of 10-20. The yield increases with lower SNR threshold, and while the orbital period distribution does not change, the mass distribution peaks at lower values for lower SNR thresholds. This difference is because the semi-major axis (and therefore SNR) depends more strongly on the mass of the companion than its orbital period.}

\edit{Second we investigate the cut in distance. \ref{fig:SNRexamples} is divided into distance bins, and the bottom right panel of Figure \ref{fig:fourpanel} shows the same mass distribution for Setup 1 (with SNR \(=20\) cut) but for the \gaia DR4 yield. These figures show that within the 200 pc cut, the low mass planets are only found in the nearest systems, and in more distant systems, only the more massive planets are detectable.  This is because the angular semi-major axis (and therefore SNR) decreases with greater distance. This does mean that extending to a larger volume will increase the yield, but only for the high-mass end, for example in \gaia DR4, only planets with \(M\gtrsim10\,{\rm M_{Jup}}\) would be detected by expanding the search volume (see Figure \ref{fig:fourpanel}). Therefore while in the main results we chose to keep the 200pc cut to allow for comparison with \citet{sahlmann_gaias_2015}, the true yield should be higher than we predict, with more high mass planets.}

\edit{Third we investigate the impact of allowing for orbital periods longer than the timespan of data. The top right panel of Figure \ref{fig:fourpanel} shows the distribution of SNRs for all the planets in the sample as well as for those with orbital periods below the timespan and below \(1.5\times\) and \(2\times\) the timespan of \gaia DR5. We see that the vast majority of planets with large SNR are at significantly longer orbital periods than the timespan. The definition of SNR compares the amplitude of the astrometric signal to the astrometric uncertainties, in our SNR calculation we are assuming that the whole orbit is covered by the data such that the amplitude of the detected signal is equivalent to that of the planet. This assumption is only sensible if the whole orbit, or close to the whole orbit, has been covered. Therefore while including orbits up to \(1.5\times\)(\(2\times\)) the timespan of DR5 increases the yield from \editt{276} to \editt{395(502)}, these numbers are overestimates.}

\edit{Overall relaxing our strict cuts would certainly increase the yields we predict, but these yields would have more uncertainty attached to them due to the uncertain and orbital-parameter-dependent effect on detectability of lower SNR signals and longer period orbits. We therefore chose to keep the yields as predicted but refer to these as lower bounds on the expected yields for each simulated circumbinary planet population.}

\section{Investigating the use of a lower binary fraction}\label{sec:fewerbinaries}

\begin{figure*}
    \centering
    \includegraphics[width=0.33\linewidth]{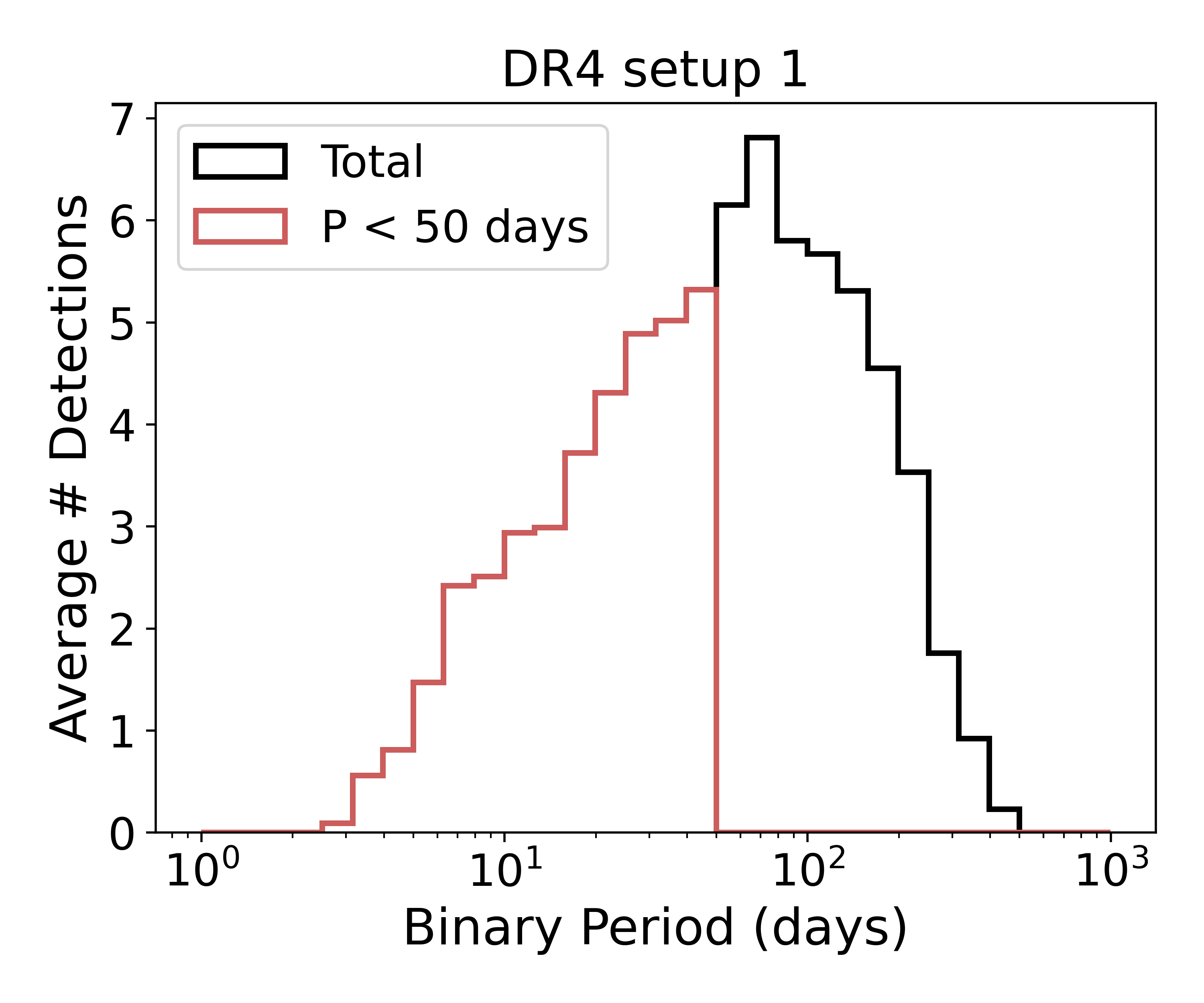}
    \includegraphics[width=0.33\linewidth]{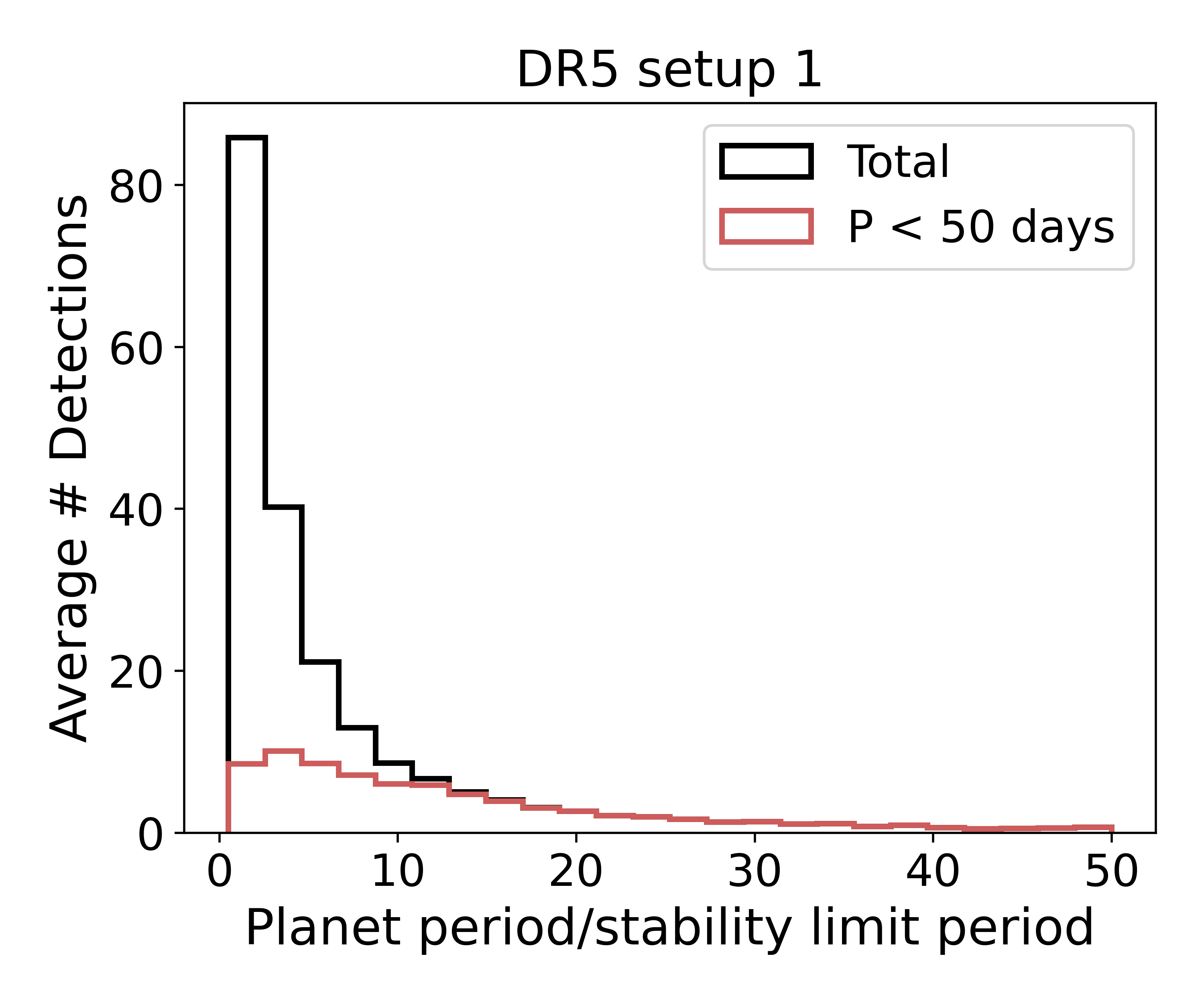}
    \includegraphics[width=0.33\linewidth]{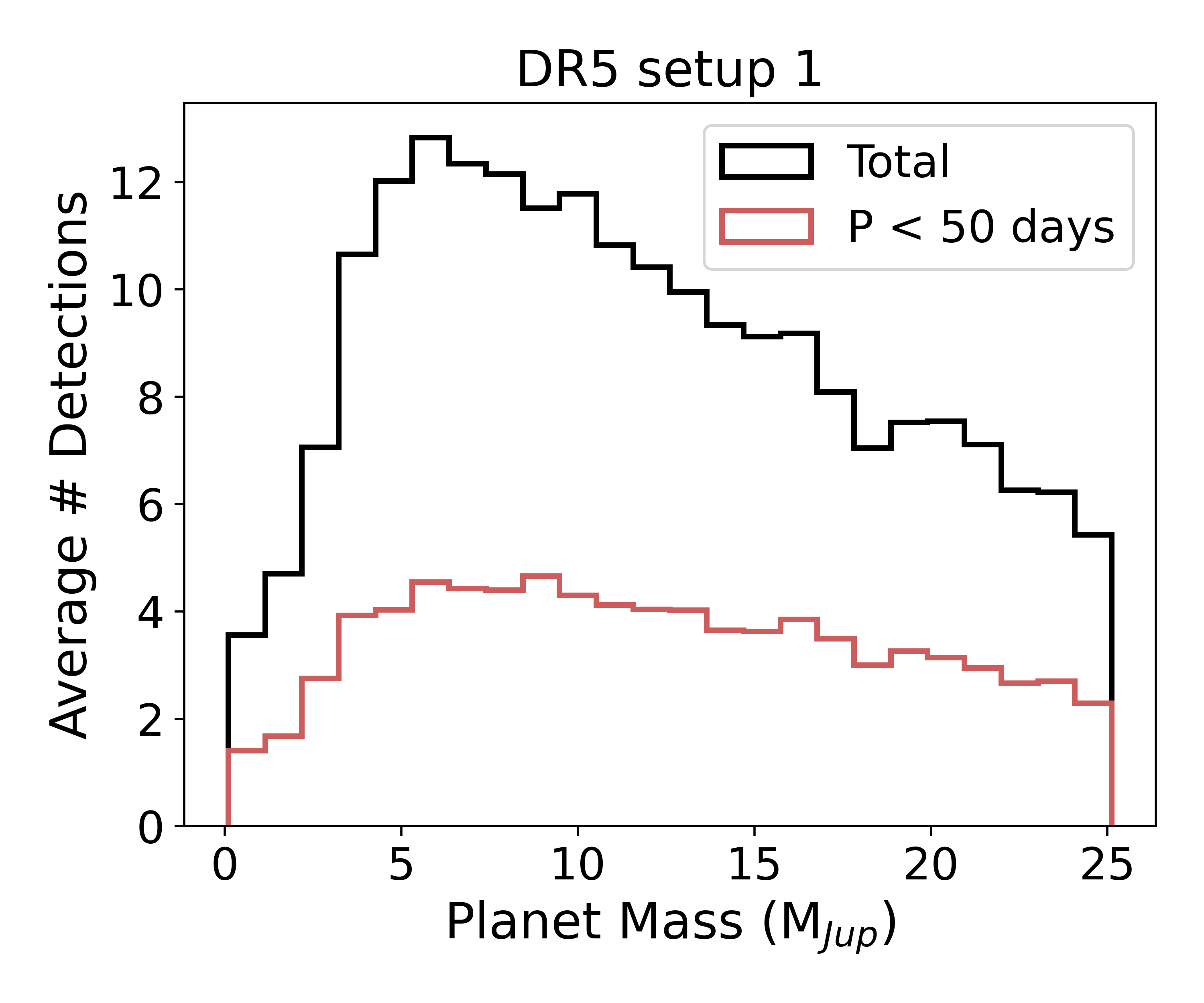}
    \includegraphics[width=0.33\linewidth]{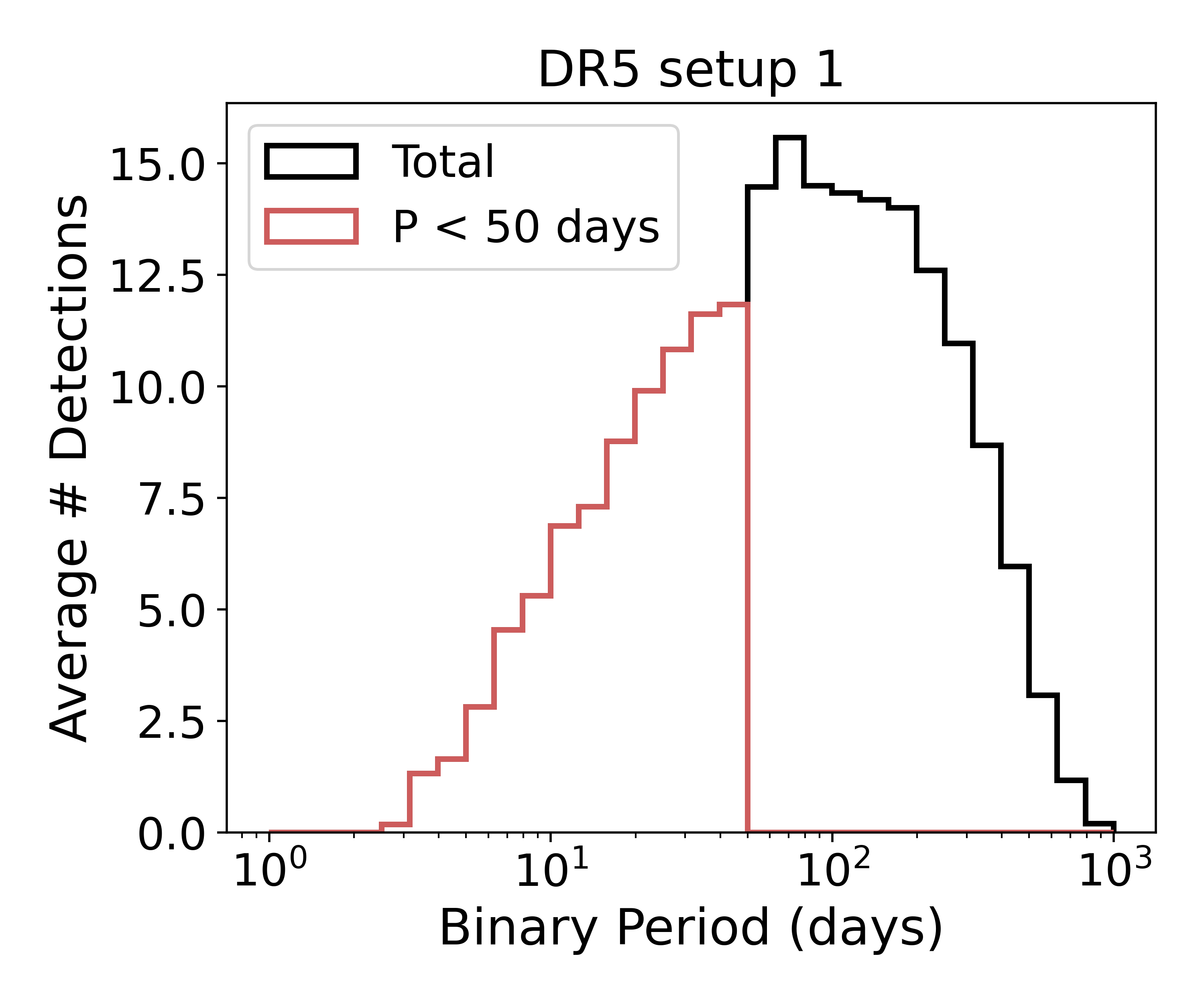}
    \includegraphics[width=0.33\linewidth]{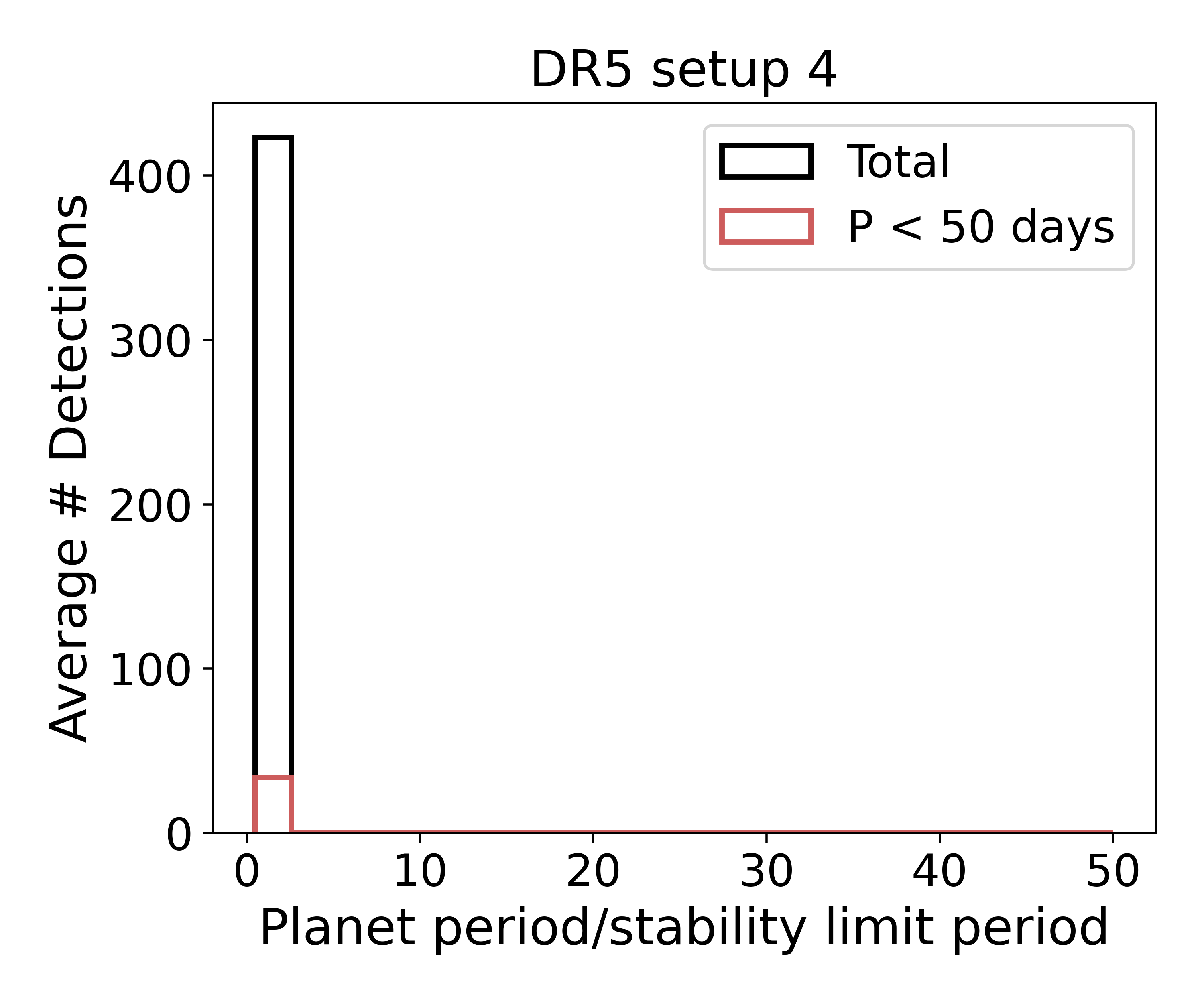}
    \includegraphics[width=0.33\linewidth]{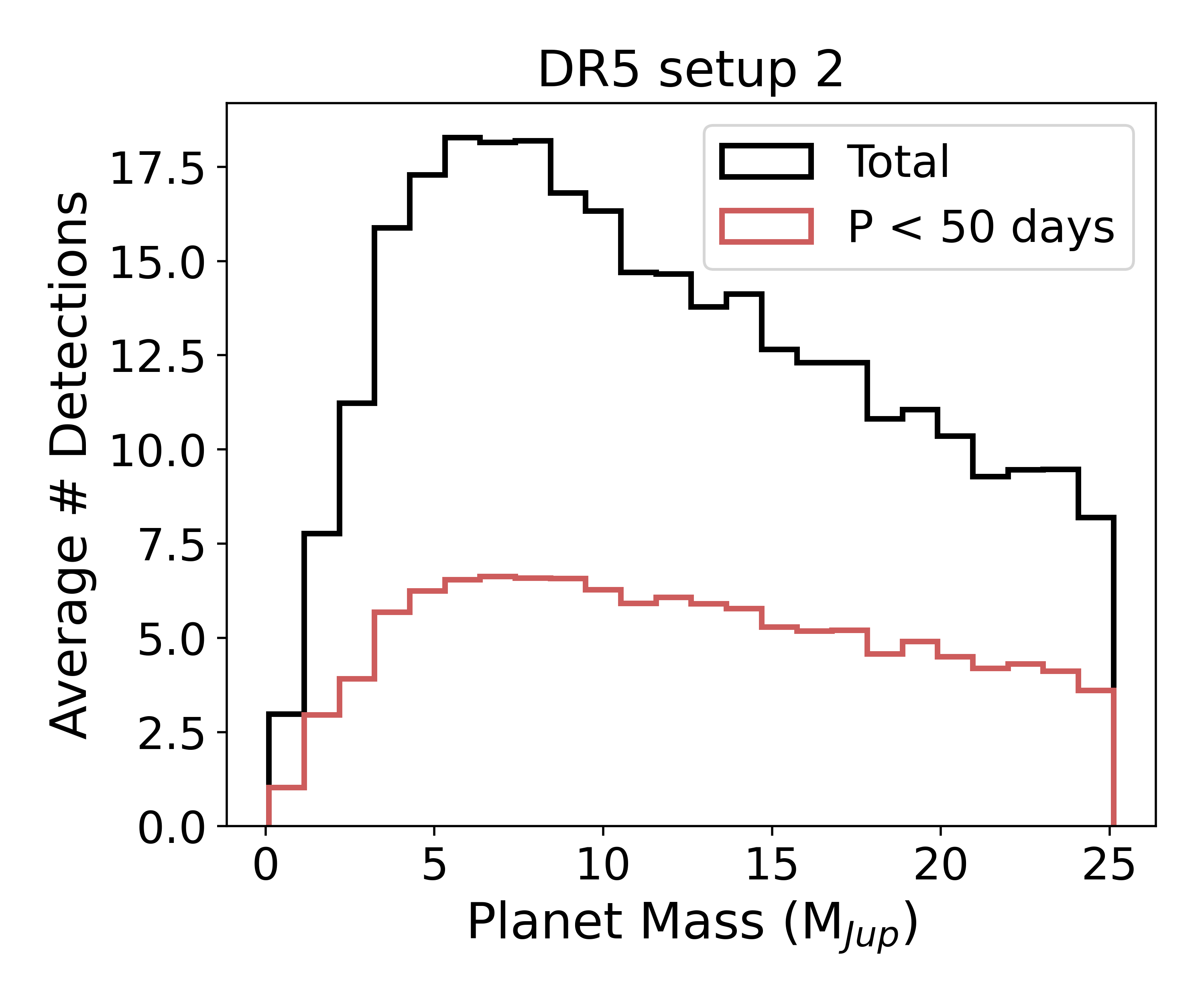}
    \includegraphics[width=0.33\linewidth]{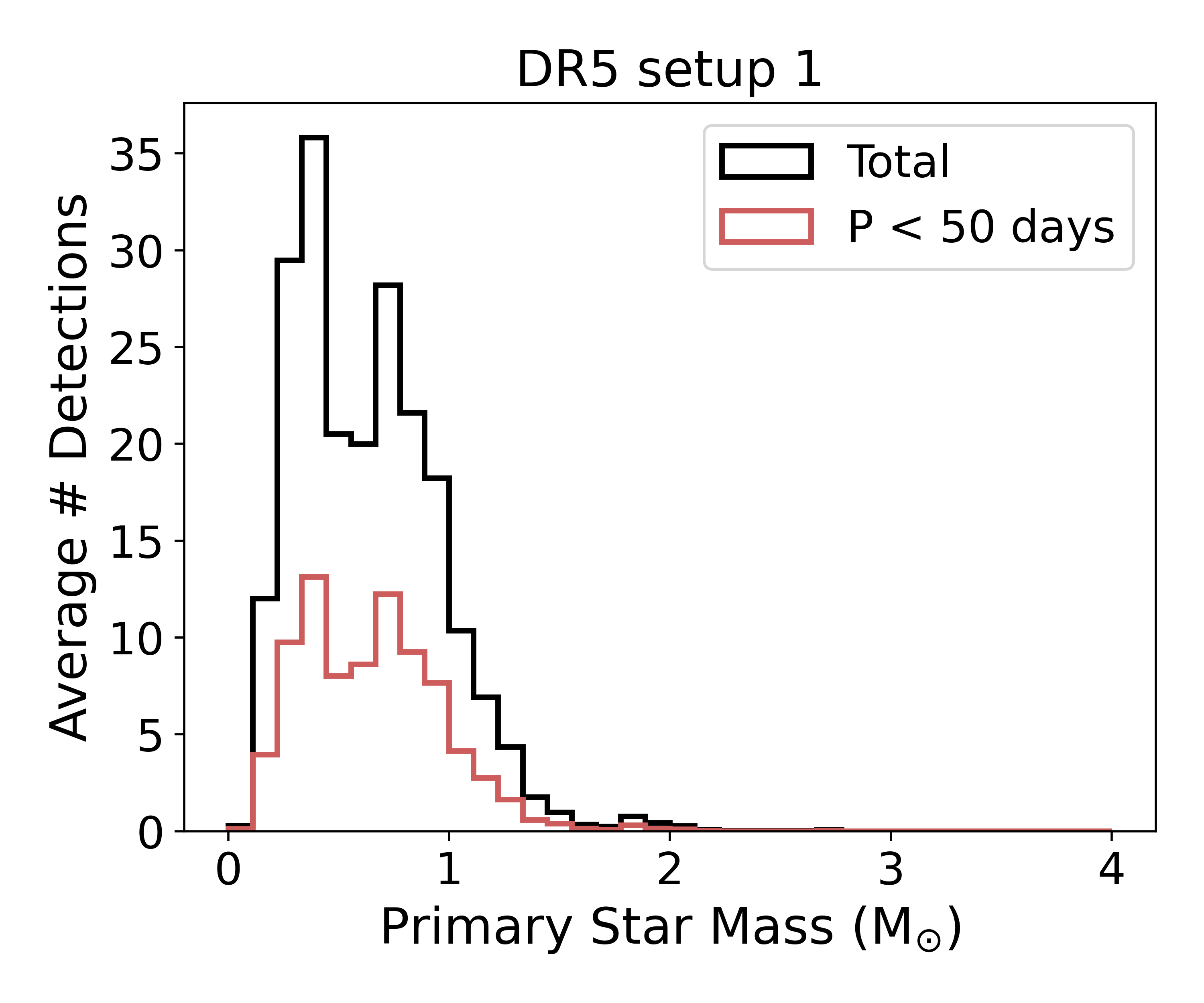}
    \includegraphics[width=0.33\linewidth]{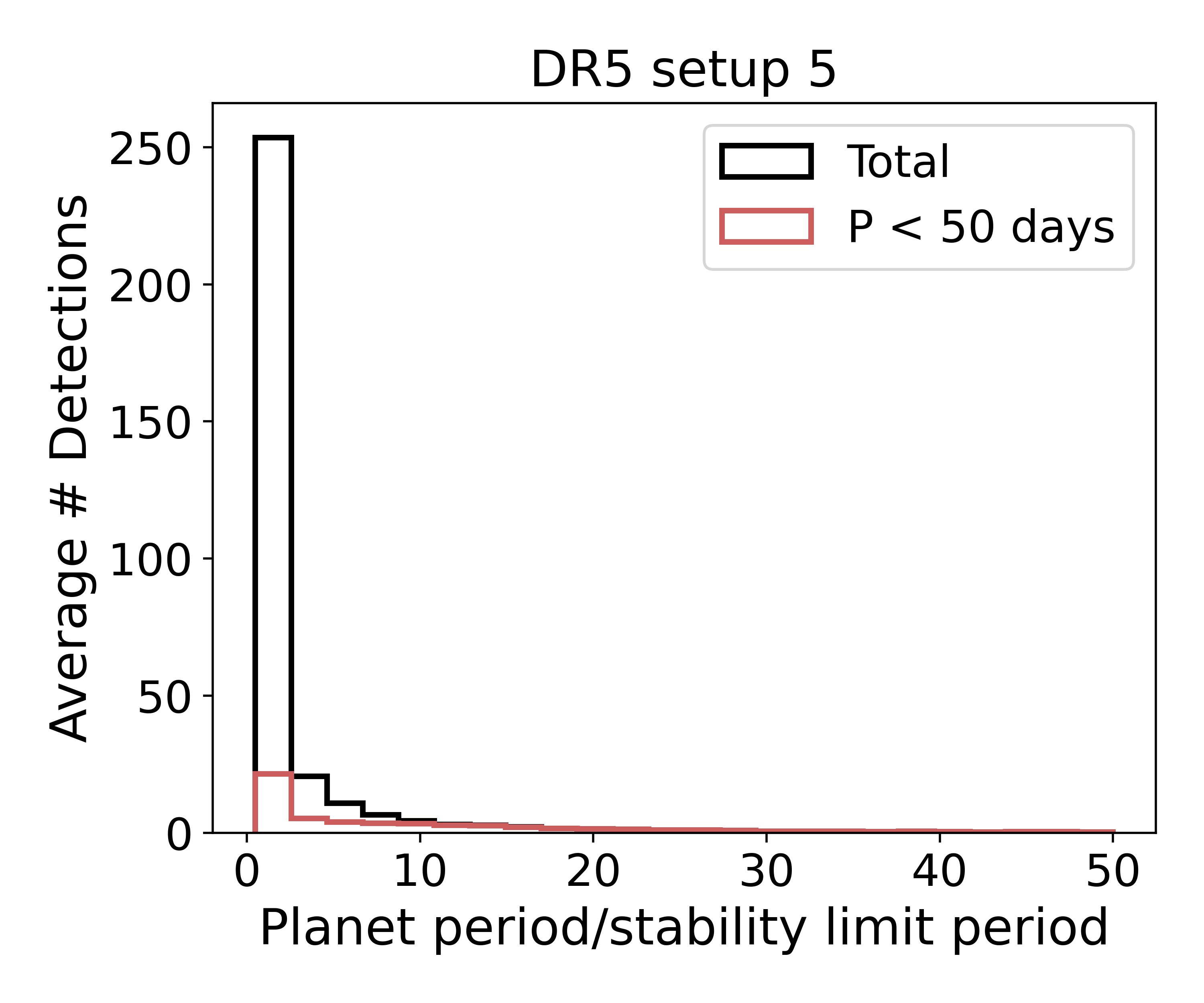}
    \includegraphics[width=0.33\linewidth]{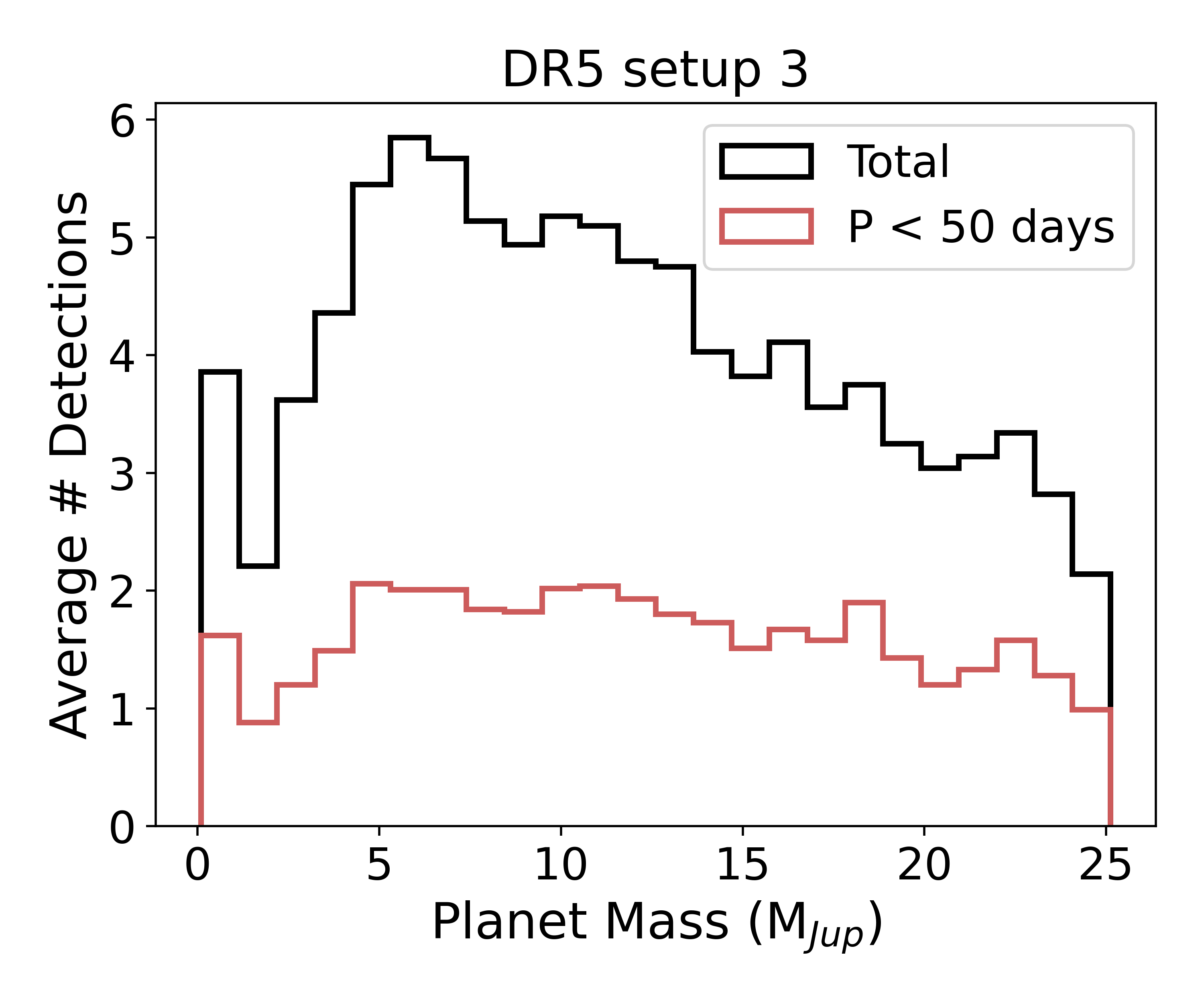}
    \includegraphics[width=0.33\linewidth]{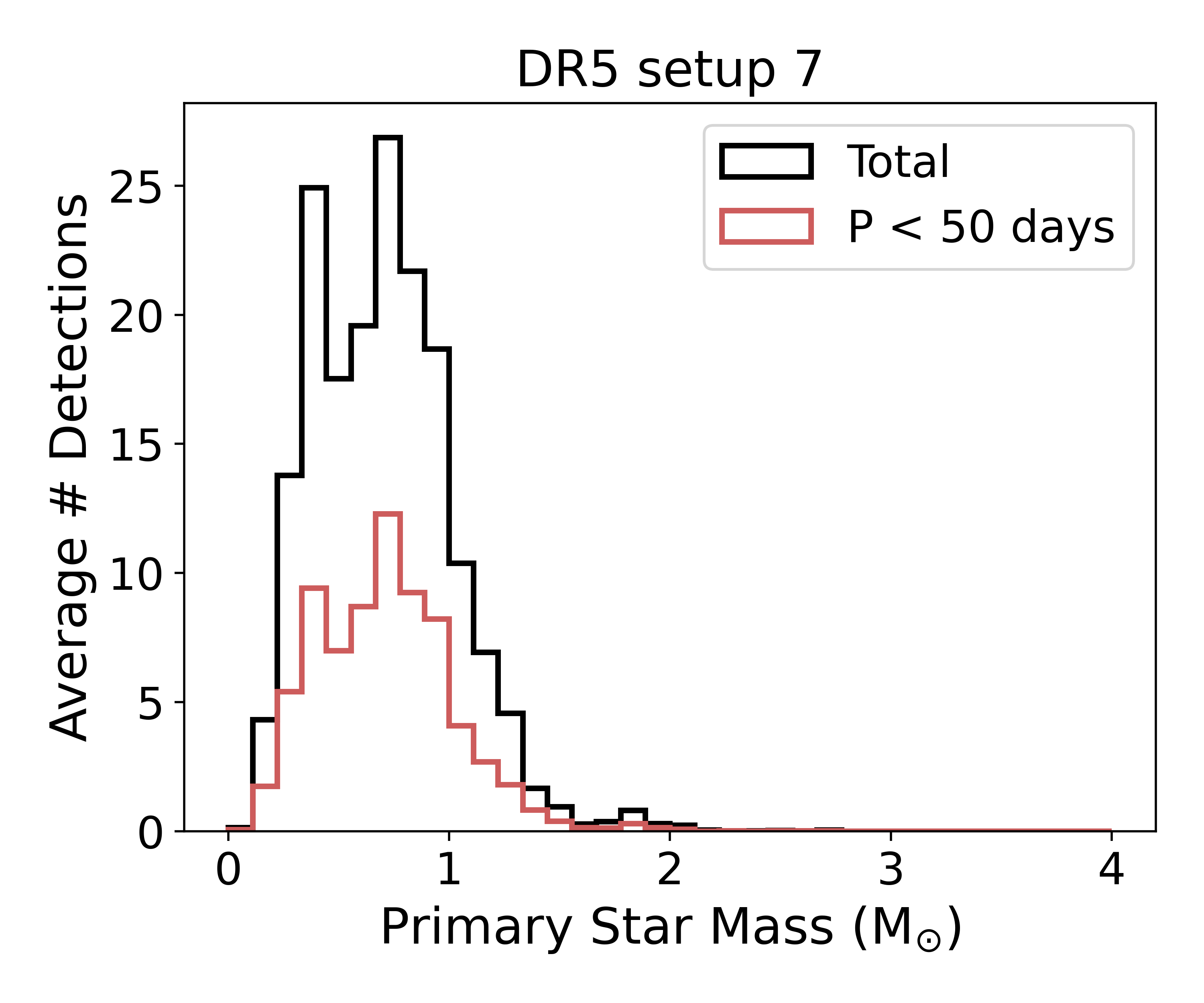}
    \includegraphics[width=0.33\linewidth]{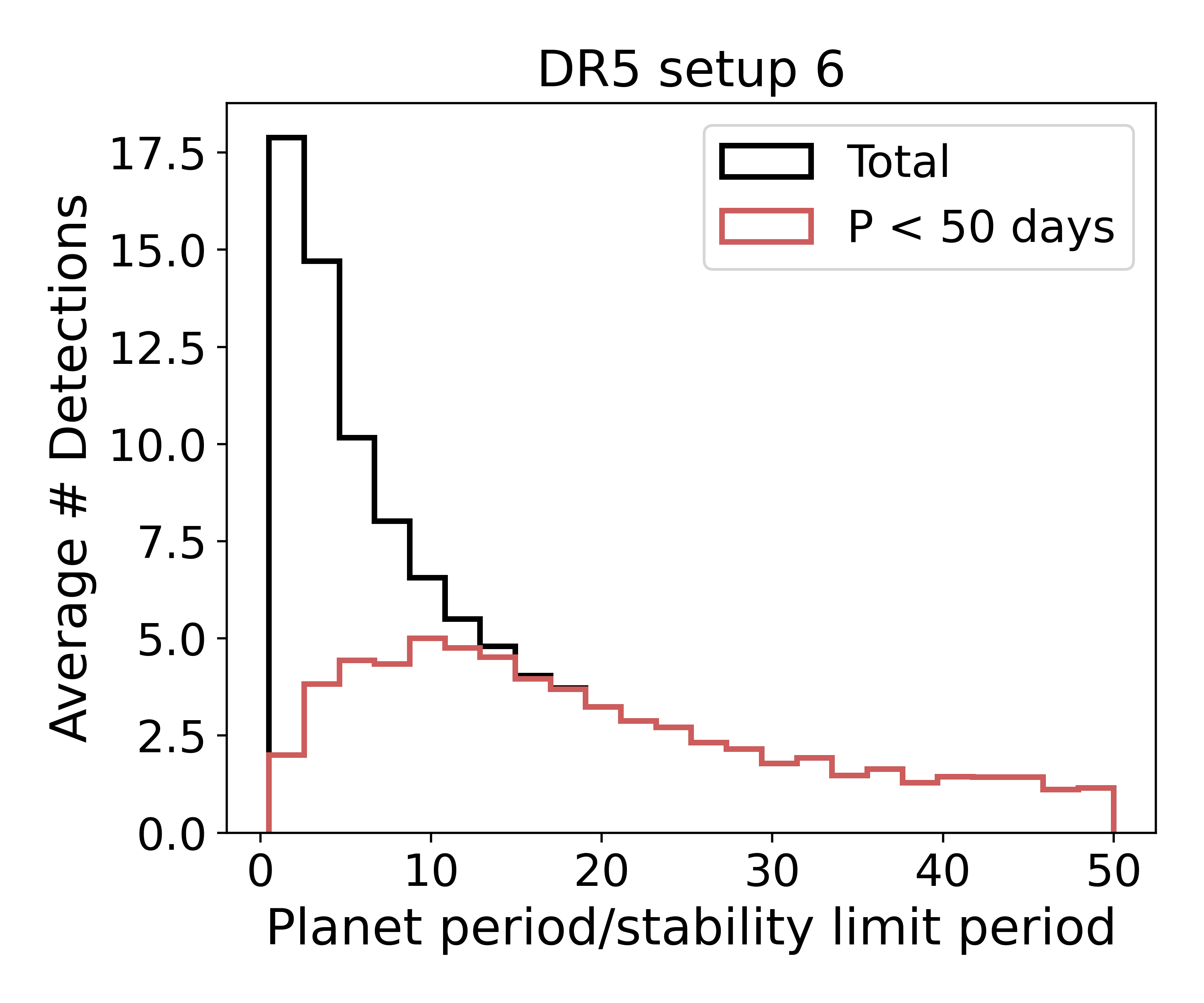}
    \caption{\edit{Distributions of various parameters for the detected planets in the alternate yield calculation with lower binary fraction for FGKM primaries. Distribution for binaries with \(P\leq50\,{\rm days}\) is also shown.}}
    \label{fig:alternate}
\end{figure*}

\edit{We investigate the effect of changing the binary fraction used in generating the binary sample (see Section \ref{sec:synthpop}). We test an alternate binary sample which is identical to that presented in the main text, but with reduced binary fractions for FGKM stars. The rationale behind this test is that in the main analysis we use the lower-bound on multiplicity from \citet{raghavan_survey_2010} as the binary fraction, but that this leaves in some systems that would have tertiary stars. We now use a binary rate of \(40\%\) for F-type primaries, \(35\%\) for G-type primaries, \(30\%\) for K-type primaries, \(20\%\) for M-type primaries. The resulting yields from this are compared to those from the main analysis in Table \ref{tab:GaiaYieldsBintype_appendix}. The shapes of the parameter distributions (see Figure \ref{fig:alternate}) are similar to those in the main body, with the primary mass distribution slightly shifted in the favour of FGK-type over M-type in comparison.}

\edit{The yield numbers in Table \ref{tab:GaiaYieldsBintype_appendix} are for \gaia DR5. Comparing this alternate sample to the original sample we see a decrease in the yield to between \(75-80\%\) of the original value.} 

\begin{table}
    \centering
    \caption{\edit{Circumbinary planet yields for each of the 7 setups in \gaia DR5, the yields for those orbiting binaries with orbital periods longer and shorter than 50 days are also shown. This is shown for both the calculation presented in the main body, and the one using an alternate smaller sample of binaries.}}
    \begin{tabular}{lcccc}
    \hline
    Setup & Total & Total & \(P_{\rm bin}<50\) days  & \(P_{\rm bin}<50\) days \\
     & (Original) & (Alternate) & (Original)  & (Alternate) \\
    \hline
    1 & 276 & 213 & 106 & 83 \\
    2 & 394 & 306 & 153 & 122 \\
    3 & 127 & 98 & 49 & 39 \\
    4 & 564 & 423 & 42 & 34 \\
    5 & 423 & 318 & 75 & 59 \\
    6 & 152 & 121 & 101 & 82 \\
    7 & 220 & 174 & 88 & 73 \\
    \hline
    \end{tabular}
    \label{tab:GaiaYieldsBintype_appendix}
\end{table}

\bsp	
\label{lastpage}
\end{document}